\documentclass[journal=jacsat,manuscript=article]{achemso}
\usepackage{achemso}
\setkeys{acs}{maxauthors = 0}

\usepackage[version=3]{mhchem}
\usepackage[english]{babel}
\usepackage{filecontents}
\usepackage{makecell}
\usepackage{graphicx}
\usepackage{courier}
\usepackage{amssymb,amsmath,amsthm,mathrsfs,comment,afterpage,accents}
\usepackage{mathtools}
\usepackage{braket}
\usepackage{mathrsfs}
\usepackage{courier}
\usepackage{color}
\usepackage{subdepth}
\makeatletter
\def\@dotsep{4.5}
\makeatother
\usepackage{dcolumn}
\usepackage{bm}
\renewcommand\vec\mathbf
\newcommand\mat\mathbf

\newcommand{\insertnew}[1]{{\textcolor{black} {#1}}}
\newcommand{\replacewith}[2]{{\textcolor{red}{}}{\textcolor{black}{#2}}}
\newcommand{\iterate}[1]{{\textcolor{black} {#1}}}
\newcommand{\beginsupplement}{%
        \setcounter{table}{0}
        \renewcommand{\thetable}{A\arabic{table}}%
        \setcounter{figure}{0}
        \renewcommand{\thefigure}{A\arabic{figure}}%
        \setcounter{equation}{0}
        \renewcommand{\theequation}{A\arabic{equation}}%
}
\SectionNumbersOn

\title{Distinguishing Artificial and Essential Symmetry Breaking in a Single Determinant: 
Approach and Application to the \ce{C60}, \ce{C36}, and \ce{C20}
Fullerenes
}
\author{Joonho Lee}
\email{linusjoonho@gmail.com}
\author{Martin Head-Gordon}
\email{mhg@cchem.berkeley.edu}
\affiliation{
Department of Chemistry, University of California, Berkeley, California 94720, USA
Chemical Sciences Division, Lawrence Berkeley National Laboratory, Berkeley, California 94720, USA
}
\begin{document}
\newpage
\maketitle
\begin{abstract}
We present a thorough analysis of symmetry breaking observed in Hartree-Fock (HF) solutions of fullerenes \ce{C60}, \ce{C36}, and \ce{C20} in order to characterize the nature of electron correlation in them. Our analysis is based on (1) the critical regularization strength to restore symmetry breaking in the recently developed regularized orbital optimized second-order M{\o}ller-Plesset perturbation theory ($\kappa$-OOMP2), (2) singlet-triplet gaps from various MP2 methods, and (3) natural orbital occupation numbers from restricted coupled-cluster with singles and doubles (RCCSD) and coupled-cluster valence bond with singles and doubles (CCVB-SD). Based on these three independent probes, we conclude that \ce{C36} (D$_\text{6h}$) exhibits genuine strong correlation and symmetry breaking whereas 
\ce{C60} exhibits {\it artificial} HF symmetry breaking and is {not strongly correlated}.
Investigating the critical regularization strength, we discuss strong correlation in \ce{C20} at the Jahn-Teller distorted geometries (C$_\text{2h}$, D$_\text{2h}$, C$_\text{i}$, and D$_\text{3h}$) and the I$_\text{h}$ geometry. Only \ce{C20} (I$_\text{h}$) was found to be strongly correlated while others exhibit {\it artificial} HF symmetry breaking. \iterate{This analysis highlights a useful feature of the recommended $\kappa$-OOMP2 method. It is an electronic structure method that describes dynamic correlation, and attenuates strong correlation in MP2 towards zero by regularization. Therefore, $\kappa$-OOMP2 will exhibit symmetry breaking in its reference determinant only when correlation is strong (i.e., essential symmetry breaking). Artificial symmetry breaking (arising in HF due to neglect of dynamic correlation) is thus removed in $\kappa$-OOMP2.}
\end{abstract}
\newpage
\section{Introduction}
\begin{figure}[h!]
\includegraphics[scale=0.45]{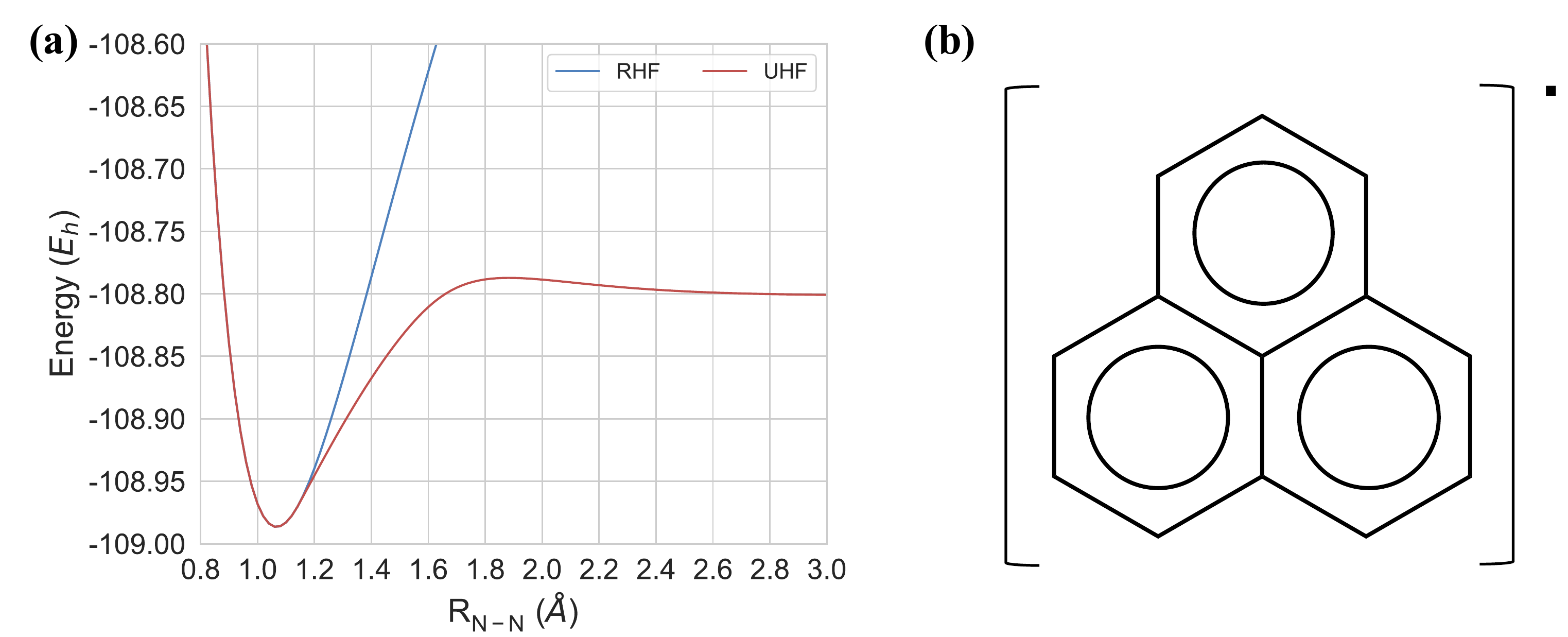}
\caption{
(a) Bond dissociation of \ce{N2} in the cc-pVTZ basis set and (b) a phenalenyl radical.
In (a), RHF stands for spin-restricted Hartree-Fock and UHF stands for spin-unrestricted Hartree-Fock.
The Coulson-Fischer point is located at 1.16 \AA.
}
\label{fig:mhg1}
\end{figure}

\replacewith{A common}{The conventional} wisdom in quantum chemistry is that spin symmetry-breaking is often necessary for describing strongly correlated systems that are beyond the scope of single-reference wavefunctions with spin-restricted orbitals.\cite{Pulay1988,Bofill1989}
\insertnew{A familiar example, illustrated in Figure \ref{fig:mhg1} (a) is the spin-polarization that occurs in homolytic bond-breaking at the Coulson-Fischer point,\cite{Coulson1949} leading to spin-unrestricted fragments at separation. We can say that this is {\it essential symmetry-breaking}, because without it, the character of a one-determinant wavefunction is fundamentally wrong (it has ionic character at dissociation). This is analogous to the fact that the exact wavefunction in this recoupling regime has {\it essential} multiconfigurational character, and is not qualitatively similar to a single determinant of spin-restricted orbitals. In other words, essential symmetry-breaking yields qualitatively better energies because it captures some aspects of electron correlation relative to a spin-restricted determinant.}

However, {\it artificial symmetry breaking} has appeared in numerous systems including open-shell systems and polyaromatic hydrocarbons. \cite{Jackels1976,Davidson1983,Andrews1991,Ayala1998,McLean1985,Sherrill1999,Crawford2000,Paldus2007}
It is ``artificial'' in the sense that
these systems are not strongly correlated and a single Slater determinant wavefuction with restricted orbitals should be a faithful representation. 
However, employing unrestricted Hartree-Fock (UHF) wavefunctions often yields significantly spin-contaminated results, which then generally give poor energetics. 
\insertnew{
An example is the doublet phenalenyl radical, shown in Figure \ref{fig:mhg1} (b), which is quite stable, with known solution chemistry.\cite{Small2004, Small2005} Yet UHF/6-31G* calculations lead to $\langle \hat{S}^2\rangle$ = 2.08 rather than 0.75,\cite{Lochan2007} even though there is no obvious  ``essential'' electron correlation effect that is captured by this extensive spin symmetry-breaking. Rather we might say that this ``artificial'' symmetry-breaking is lowering the energy by recovering a bit of the dynamical correlation.
}

It is \iterate{another difficult} {\it symmetry dilemma} \iterate{as to how to} distinguish a genuine symmetry breaking from an artificial one. \cite{Lykos1963} At the mean-field level, spontaneous symmetry breaking occurs to lower the mean-field electronic energy. Broken symmetry solutions are variationally preferred due to the very limited form of a single deteminant wavefunction. However, subsequent correlated wavefunction calculations \insertnew{(particularly those that include perturbative corrections)} on top of broken symmetry orbitals often yield qualitatively incorrect energetics and properties.\cite{Farnell1983,Nobes1987,Gill1988,Jensen1990,Andrews1991,Yamanaka1994,Ayala1998,Crawford2000}

To mitigate this problem, it is often preferred to employ approximate Br{\"u}ckner orbitals.\cite{Meyer1976, Shavitt1976, Dykstra1977, Handy1989} The exact Br{\"u}ckner orbitals may be obtained from an exact wavefunction by enforcing the zero singles constraint (i.e. the defining property of Br{\"u}ckner orbitals).\cite{Brueckner1954,Nesbet1958} Having an exact wavefunction is not a realistic assumption so, in practice, one may utilize coupled-cluster doubles (CCD) to obtain so-called Br{\"u}ckner doubles (BD).\cite{Handy1989} Our group proposed a variational formulation of BD, called orbital-optimized doubles (OD), which optimizes orbitals in the presence of the CCD correlation energy.\cite{Sherrill1998,Krylov1998} 

One major drawback of OD (or BD) is that its computational complexity scales sextically with system size. A more economical way to obtain approximate Br{\"u}ckner orbitals with fifth-order complexity is to use orbital-optimized second-order M{\o}ller-Plesset theory (OOMP2).\cite{Lochan2007} OOMP2 has been successfully applied to shed light on artificial spin-symmetry breaking problems.\cite{Stuck2011,Kurlancheek2012} It also improves energetics of radical-involving systems drastically compared to regular MP2.\cite{Neese2009, Kossmann2010} 

Although OOMP2 has shown its \replacewith{strength}{utility} in the aforementioned examples, \replacewith{there are two disturbing aspects of it}{it also has two disturbing features}.\cite{Stuck2013,Razban2017} First, its potential energy surface often exhibits a first-order derivative discontinuity even though the electronic energy is minimized with respect to orbitals \insertnew{which should give only a second-order derivative discontinuity}. This is a consequence of the disappearance of Coulson-Fischer points.\cite{Coulson1949} Second, it often finds a divergent energy solution (i.e. $E=-\infty$) by preferring a unphysical zero-gap restricted solution. We argue that these two drawbacks were satisfactorily solved and thoroughly analyzed in our previous work where we employed an orbital energy dependent regularizer.\cite{Lee2018}

Fullerenes have \replacewith{drawn lots of}{attracted great} attention in \replacewith{the}{}physical chemistry \replacewith{community}{}and interested readers are referred to ref. \citenum{Fowler2007}, an excellent textbook which summarizes the history of fullerenes in physical chemistry, and references therein. 
Starting from the smallest fullerene \ce{C20},\cite{Prinzbach2000} fullerenes are made solely of carbon and exhibit uncommon cage geometries.
These extraordinary features of fullerenes are interesting on their own.
What is surprising from an electronic structure standpoint is that many of these fullerenes exhibit complex generalized Hartree-Fock (cGHF) solutions as \replacewith{suggested}{discovered} by Jim{\'e}nez-Hoyos et al.\cite{Jimenez-Hoyos2014} These \insertnew{symmetry-broken} cGHF solutions were \replacewith{then}{}interpreted as an indicator of polyradicaloid character of fullerenes. This is unexpected because experimentally synthesized fullerenes are quite stable and thus these stable ones are likely closed-shell \insertnew{in character}.

The most striking conclusion of Jim{\'e}nez-Hoyos and co-workers' work is that buckminsterfullerene, \ce{C60}, is polyradicaloid because of the existence of a cGHF solution. This contradicts with our group's previous attempt at characterizing the electron correlation of \ce{C60}.
Our group discovered a restricted (R) to unrestricted (U) instability in \ce{C60} at the HF level.\cite{Stuck2011} We carefully assessed the nature of spin-symmetry breaking in comparison with \ce{C36} which is known to be strongly correlated. \insertnew{For \ce{C60}} RMP2 yielded a more reasonable single-triplet gap than UMP2 and scaled opposite spin OOMP2 restored the spin-symmetry. \replacewith{Based on these}{For these reasons}, our conclusion \replacewith{in the work}{}was that \ce{C60} is not strongly correlated and should be described as a closed-shell molecule.\replacewith{, directly contradicting Jim{\'e}nez-Hoyos and co-workers' claim.}{} 

In this work, we address this controversy on symmetry breaking of \ce{C60} and characterize the electron correlation of related molecules (i.e., \ce{C20} and \ce{C36}) using the recently developed regularized OOMP2 ($\kappa$-OOMP2)\cite{Lee2018} and coupled-cluster valence bond with singles and doubles (CCVB-SD) methods.\cite{Small2012, Lee2017}


\section{Theory}
We will use $i,j,k,l,\cdot\cdot\cdot$ to index occupied orbitals, $a,b,c,d,\cdot\cdot\cdot$ to index virtual orbitals, and $p,q,r,s,\cdot\cdot\cdot$ to index either of those two. $n_\alpha$ denotes the number of $\alpha$ electrons, $n_\beta$ denotes the number of $\beta$ electrons, $n_\text{mo}^\sigma$ is the number of molecular orbitals of spin $\sigma$, and $n_\text{so}$ is the number of spin orbitals.
\subsection{Classification of HF Solutions}
Fukutome pioneered \replacewith{in}{}the group-theoretic classification of non-relativistic HF solutions.\cite{Fukutome1981}
There are a total of eight distinct classes based on the symmetry of \insertnew{the electronic} Hamiltonian (complex conjugation ($\hat{K}$), time-reversal ($\hat{\Theta}$), and spin-operators $\hat{S}_\mathbf{n}$ where $\mathbf{n}$ is a collinear axis and $\hat{S}^2$). We will follow the classification by Stuber and Paldus,\cite{Stuber2003} which is identical to Fukutome's classification with a different name \replacewith{to}{for} each class. \replacewith{This includes a total of eight different classes}{The eight different classes are}: restricted HF (RHF), complex RHF (cRHF), unrestricted HF (UHF), complex UHF (cUHF), generalized HF (GHF), and complex GHF (cGHF) along with paired UHF (pUHF) and paired GHF (pGHF). For the purpose of this work, we discuss symmetry breaking of $\hat{K}$, $\hat{S}_\mathbf{n}$, and $\hat{S}^2$ and do not discuss $\hat{\Theta}$. Therefore, there is no distinction between complex solutions and paired solutions since they differ only by time-reversal symmetry. We will thus discuss a total of six classes of HF solutions of each fullerene presented below.
\subsection{Generalized MP2}
\replacewith{In GHF, there is no}{GHF eliminates the} distinction between $\alpha$ and $\beta$ spin orbitals \insertnew{characteristic of UHF}.\cite{Fukutome1981,Yamaki2000,Stuber2003,Jimenez-Hoyos2011, Small2015,Cassam-Chenai2015} \replacewith{Each}{Instead each} electron occupies a spin-orbital \replacewith{whose spin function}{that} can be an arbitrary linear combination of $\alpha$ and $\beta$ \insertnew{basis functions}. We refer \insertnew{to} such an orbital as a generalized spin-orbital (GSO). 
The usual spin-orbital MP2 energy expression reads,
\begin{equation}
E_\text{MP2} = E_0-\frac14\sum_{ijab}\frac{|\langle ij||ab\rangle|^2}{\Delta_{ij}^{ab}},
\label{eq:emp2}
\end{equation}
where \insertnew{$E_0$ is the energy of the reference HF determinant the remainder is the MP2 correlation energy.} 
$\Delta_{ij}^{ab}$ is a non-negative orbital energy gap defined as
\begin{equation}
\Delta_{ij}^{ab} = \epsilon_a+\epsilon_b-\epsilon_i-\epsilon_j,
\end{equation}
and a two-electron integral \replacewith{$\langle ij|ab\rangle$}{$(ia|jb)$} is 
\begin{equation}
(ia|jb) 
= \int_\mathbf{r_1,r_2} \frac{\phi_i(\mathbf{r_1})^* \phi_a(\mathbf{r_1}) \phi_j(\mathbf{r_2})^* \phi_b(\mathbf{r_1})}{||\mathbf{r_1} - \mathbf{r_2}||_2}.
\end{equation}
and an antisymmetrized integral is
\begin{equation}
\langle ij||ab\rangle = (ia|jb) - (ib|ja)
\end{equation}
The evaluation of a two-electron integral with GSOs can be achieved in the following way:
\begin{equation}
(ia|jb) = (i^\alpha a^\alpha|j^\alpha b^\alpha) 
+ (i^\alpha a^\alpha|j^\beta b^\beta) 
+ (i^\beta a^\beta|j^\alpha b^\alpha) 
+ (i^\beta a^\beta|j^\beta b^\beta)
\label{eq:integral}
\end{equation}
where
we used the fact that each GSO, $|\phi_p\rangle$, \replacewith{takes a form of}{has an $\alpha$ and a $\beta$ component}
\begin{equation}
|\phi_p\rangle= 
\left(
  \begin{tabular}{c}
  $\phi_p^\alpha$ \\
  $\phi_p^\beta$ \\
  \end{tabular}
  \right).
  \label{eq:spinor}
\end{equation}
Having expanded out the two-electron integral this way, the implementation of GMP2 is \insertnew{(at least conceptually)} trivial \replacewith{in any existing MP2 programs}{on top of an existing MP2 program}.

The resolution of the identity approximation can be also applied to Eq. \eqref{eq:integral} using \cite{Feyereisen1993,Bernholdt1996}
\begin{equation}
(i^{\sigma_1} a^{\sigma_1}|j^{\sigma_2} b^{\sigma_2}) = \sum_{PQ}(i^{\sigma_1} a^{\sigma_1}|P) (P|Q)^{-1} (Q|j^{\sigma_2} b^{\sigma_2})
\label{eq:contract}
\end{equation}
where $(i^{\sigma_1} a^{\sigma_1}|P)$ represents a 2-electron 3-center Coulomb integral and $(P|Q)$ is the metric matrix of the above decomposition which is a 2-electron 2-center Coulomb integral among auxiliary basis functions.

The scaling of RI-cGMP2 is the same as RI-RMP2 or RI-UMP2 (i.e., $\mathcal O(N^5)$). However, it carries a large prefactor compared to RI-RMP2 or RI-UMP2. The bottleneck of RI-MP2 is forming two-electron integrals in Eq. \eqref{eq:contract}. In the case of RI-RMP2 this step scales $\mathcal O (n_\alpha^2 (n_\text{mo}^\alpha-n_\alpha)^2 n_\text{aux})$ whereas RI-UMP2 scales $\mathcal O ((n_\alpha^2 (n_\text{mo}^\alpha-n_\alpha)^2+n_\beta^2 (n_\text{mo}^\beta-n_\beta)^2+n_\alpha n_\beta (n_\text{mo}^\alpha-n_\alpha)(n_\text{mo}^\beta-n_\beta)) n_\text{aux})$. Assuming $n_\alpha = n_\beta$, RI-UMP2 is three times more expensive than RI-RMP2. In the case of RI-GMP2, we have a scaling of $\mathcal O(n_\text{elec}^2 (n_\text{so} - n_\text{elec})^2 n_\text{aux})$.  Assuming $n_\text{elec} = 2 n_\alpha$ and $n_\text{so} = 2 n_\text{mo}^\alpha$, we conclude that RI-GMP2 is 16 times more expensive than RI-RMP2. RI-cGMP2 carries an extra factor of four due to complex arithmetic operations. Overall, RI-cGMP2 is 64 times more expensive than RI-RMP2. Since we will be studying reasonably large systems, this non-negligible prefactor will limit the applicability of RI-cGMP2 in this study.

We will use the RI approximation throughout this work so we shall drop ``RI" and refer to ``RI-MP2" as ``MP2".

\subsection{Regularized Orbital-Optimized MP2}
When orbital-optimizing Eq. \eqref{eq:emp2}, it is commonly observed that the energy tends to $-\infty$. This divergence is attributed to \replacewith{a small offending}{the development of a small} denominator, \replacewith{$\Delta_{ij}^{ab}$}{$\Delta_{ij}^{ab}\rightarrow 0$ for some $i,j,a,b$}.
\iterate{While simple level shift were first explored\cite{Stuck2013,Razban2017} they proved inadequate.}
To better mitigate this problem, we proposed a regularized MP2 method whose energy reads
\begin{equation}
E_\text{MP2}(\kappa) =
E_0
-\frac14
\sum_{ijab}  \frac{|\langle ij||ab\rangle|^2}
{\Delta_{ij}^{ab}}
\left(
1-e^{-\kappa \Delta_{ij}^{ab}}
\right)^2
\label{eq:kmp2}
\end{equation}
where $\kappa > 0$ is a single empirical parameter that controls the strength of regularization. The exponential damping function ensures that \replacewith{offending}{}small denominators \replacewith{do not}{cannot} contribute to the final energy. Orbital-optimizing Eq. \eqref{eq:kmp2} yields $\kappa$-OOMP2 which we thoroughly analyzed and benchmarked in our previous work.\cite{Lee2018} $\kappa=1.45$ $E_h^{-1}$ was recommended as it appeared to combine favorable recovery of Coulson-Fischer points with good numerical performance. We will also employ its scaled correlation energy variant, $\kappa$-S-OOMP2, where the optimal scaling parameter, $c$, for each $\kappa \in \left[0.05, 4.0\right]$ was determined in our previous work.\cite{Lee2018} 
\subsection{Classification of OOMP2 Solutions}
The classification of OOMP2 solutions can be done in the same way as that of HF solutions. The classification needs to incorporate the MP2 correction to the one-particle density matrix (1PDM) and the first-order correction to the spin expectation values. 

The norm of the imaginary part of \insertnew{the} 1PDM, $\xi$, diagnoses the fundamental complexity of a wavefunction at the one-body level.\cite{Small2015a} This was sufficient for testing the fundamental complexity of OOMP2 solutions studied in this work. In principle, one may consider a more sophisticated diagnostic tool that involves higher order \replacewith{particle}{}density matrices.

The first-order correction to the spin expectation values are needed to compute the first-order correction to a spin covariance matrix $\mathbf{A}$ defined as
\begin{equation}
A_{ij} = \langle \hat{S}_i\hat{S}_j\rangle - \langle \hat{S}_i\rangle \langle \hat{S}_j\rangle
\end{equation}
The nullity of this matrix determines the collinearity of a given wavefunction as noted by Small et al.\cite{Small2015} If there is a zero eigenvalue associated with an eigenvector $\mathbf n$, the wavefunction is collinear along this axis $\mathbf n$. The real part of $\mathbf A$ is a positive semidefinite matrix and the smallest eigenvalue, $\mu_0$, can be used to quantify the non-collinearity of a given wavefunction. This applies to not only non-interacting wavefunctions such as HF states but also correlated states such as MP1 wavefunctions. We present the formula for computing the first-order correction to $\mathbf{A}$ in the Appendix. 

As a side note, during the course of finishing this work, we found that the expression for the first-order correction to $\langle \hat{S}^2 \rangle$ presented in ref. \citenum{Lochan2007} is off by a factor of two and the one in ref. \citenum{Kossmann2010} is off by a factor of four from the correct expression. This is clear from looking at a more general expression given in the Appendix.
\subsection{Coupled-Cluster Methods}

Restricted CC with singles and doubles (RCCSD) generally fails to describe strongly correlated systems and often exhibits non-variationality in such systems.
Recently, we proposed \replacewith{CCVB-SD}{coupled-cluster valence bond with singles and doubles (CCVB-SD)} as a simple alternative that modifies the parametrization of quadruple excitations in RCCSD. \cite{Small2012}
CCVB-SD is better than RCCSD at handling strong correlation as it can describe bond dissociations exactly within a valence active space as long as UHF (or UCCSD) can. This property is inherited from a \replacewith{more crude}{much simpler} correlation model, CCVB.\cite{Small2009, Small2011, Small2014, Small2017, Small2018, Lee2018a}
Furthermore, CCVB-SD was successfully applied to oligoacenes which exhibit emergent strong correlations while RCCSD showed non-variationality.\cite{Lee2017} 
It is clear from our experience that strong correlation is present when we observe qualitative differences between CCVB-SD and RCCSD. This can be shown in terms of working equations. CCVB-SD modifies \insertnew{the} quadruples in RCCSD which plays a crucial role in describing strong correlation. This modification becomes negligible if no substantial connected quadruples are needed and therefore in such cases there is no strong correlation.
Comparing those CC methods will shed light on the electron correlation \replacewith{of}{effects in the} fullerenes.

\section{Probes for Artificial Symmetry Breaking}
\subsection{Probe 1: Symmetry Breaking Landscape from $\kappa$-S-OOMP2}\label{sec:oomp2}
\insertnew{One way to address the problem of distinguishing between essential symmetry breaking (driven by missing static correlation effects) and artificial symmetry breaking (driven by missing dynamical correlation effects) in HF calculations is to apply a method that optimizes the orbitals including only dynamical correlation but not the static or essential correlation. With such an approach, artificial symmetry breaking will be virtually eliminated because dynamic correlation effects are explicitly included and are therefore removed as a driving force for symmetry breaking. Essential symmetry breaking remains because no static correlation is included to handle strong correlation problems. 
}

\begin{figure}[h!]
\includegraphics[scale=0.45]{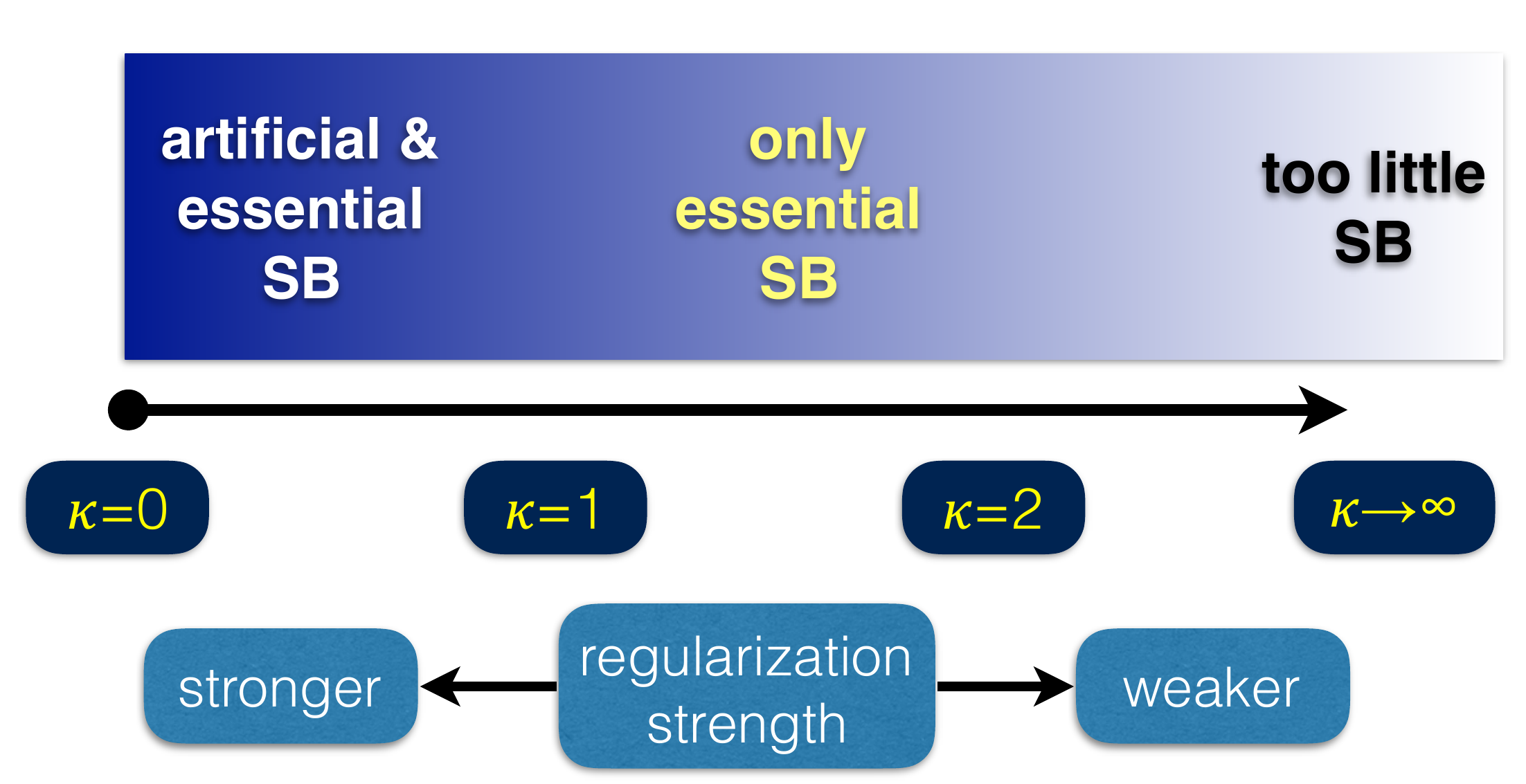}
\caption{
Illustration of artificial and essential symmetry breaking (SB) in $\kappa$-S-OOMP2 as a function of $\kappa$.
$\kappa \in [1.0,2.0]$ exhibits only essential SB.
}
\label{fig:mhg2}
\end{figure}

\insertnew{
           The $\kappa$-S-OOMP2 method is such a theory for reasons that are summarized in Figure \ref{fig:mhg2}. Varying the regularization parameter over the range [0.5, 4.0] for which this method has been parameterized yields a survey of how symmetry breaking depends on regularization strength. Regularization that is weak ($\kappa>2$) defines methods that strongly favor symmetry-restoration, and will often be below the true energy. For example, in bond-breaking, Coulson-Fischer points are pushed to very long bond-lengths, and eventually are even lost. This arises because weak regularization includes part of the exaggerated MP2 description of the (potentially strong) paired correlations associated with small promotion energies. Thus the limit of very weak regularization ($\kappa\rightarrow\infty$) may not even admit essential symmetry breaking for strongly correlated systems.
 }
 
\insertnew{
	By contrast regularization that is strong ($\kappa<1$) will potentially admit both essential and artificial symmetry breaking, because part of the dynamical correlation is also attenuated. This is most clearly seen by considering $\kappa$-OOMP2 without scaling, where the limit as $\kappa\rightarrow0$ recovers HF theory, with all its symmetry breaking characteristics. Using the scaled method, $\kappa$-S-OOMP2, yields methods that have a somewhat lower driving force for symmetry breaking with small $\kappa$. Finally, the intermediate regime, which we view here as roughly $\kappa \in[1.0,2.0]$, presents a transition between strong and weak regularization. The value of $\kappa = 1.45$ which was selected to yield useful accuracy and restore Coulson-Fischer points, as discussed previously,\cite{Lee2018} lies in this region.
	}
	
\insertnew{
            We can use the $\kappa$-dependence of symmetry breaking in $\kappa$-S-OOMP2 to characterize its nature. For an even electron system that exhibits symmetry breaking at the HF level, using $\kappa$-S-OOMP2 at a fixed geometry as a function of $\kappa$ will yield a critical value, $\kappa_c$ above which symmetry restoration is complete. If $\kappa_c$ is large enough (weak regularization, for instance $\kappa_c>2$), so that Coulson-Fischer points are not properly restored in bond-breaking, then we conclude that the HF symmetry breaking was essential in character, since so too is that of $\kappa$-S-OOMP2. On the other hand, if $\kappa_c$ is small enough (strong regularization, for instance $\kappa_c<1$) then we are well into the regime where Coulson-Fischer points exist, and we must therefore conclude that the HF symmetry breaking was artificial in character, because symmetry breaking is readily quenched in $\kappa$-S-OOMP2, even with strong regularization.
}
\subsection{Probe 2: Singlet-Triplet Gap}
\insertnew{A singlet-triplet gap measures the energy cost for breaking a pair of electrons that are singlet coupled. This would be much smaller than a usual single bond energy if two electrons are spatially well separated and singly occupying each orbital. This is the case for biradicaloids where singlet-triplet gaps less than 10 kcal/mol are commonly observed.\cite{Abe2013} On the other hand, if the gap is large then the molecule is best described as a closed-shell molecule. A precise experimental value for this gap is available for \ce{C60} and therefore this is a good probe especially for \ce{C60}.
}
\subsection{Probe 3: Correlated 1PDM}
{We will investigate two quantities that can be obtained from the correlated 1PDM of CC wavefunctions (RCCSD and CCVB-SD).
The first one is natural orbital occupation numbers (NOONs) which are eigenvalues of \iterate{the} 1PDM. For a perfect closed-shell molecule, \iterate{the} NOONs would be either 0.0 (empty) or 2.0 (doubly occupied). For a perfect biradical system, there should be two eigenvalues of 1.0 as well. In general, \iterate{the} NOONs will be distributed between 0.0 and 2.0. A polyradicaloid must have multiple eigenvalues that \iterate{significantly} deviate from 0.0 and 2.0.

The second quantity that we will report is 
Head-Gordon's number of unpaired electrons (NUE),\cite{Head-Gordon2003}
\begin{equation}
\text{NUE} = \sum_i \text{min}(2-n_i, n_i)
\label{eq:nue}
\end{equation}
where $i$ is summed over all natural orbitals and $n_i$ is the occupation number of the $i$-th natural orbital.
This takes the entire spectrum of NOONs and reduces it to a single scalar value that quantifies strong correlation. In the case of a perfect closed-shell molecule, NUE is zero. Molecules with open-shell character will exhibit larger NUE values.
When these quantities show a qualitative difference between RCCSD and CCVB-SD for a given system, we conclude that the system is strongly correlated and vice versa.
}
\section{Results}
We investigated the \ce{C60}, \ce{C36}, and \ce{C20} fullerenes. We studied a total of five different geometries of \ce{C20} whereas only one conformation for other fullerenes was studied. All HF calculations were performed with wavefunction stability analysis to ensure the local stability of solutions. The pertinent cGHF electronic Hessian is provided in the Appendix. All calculations were carried out with a development version of \texttt{Q-Chem}.\cite{Shao2015} All correlation calculations employed the frozen core approximation for the sake of computational efficiency. All the plots were generated with $\texttt{Matplotlib}$ \cite{Hunter2007} and all the molecular figures were generated with $\texttt{Chemcraft}$.\cite{Chemcraft} 

\insertnew{
Obtaining quantitatively accurate answers with CC methods is very computationally intensive for fullerenes so we look for qualitative answers by comparing them to various OOMP2 methods. 
We will employ the STO-3G basis set in order to exaggerate the effect of strong correlation and discuss the implication within this basis set. We only present the CC data for \ce{C60} and \ce{C36} here; the CC data for \ce{C20} showed the same conclusions as the other analyses based on OOMP2 we present below.
}

\subsection{The Nature of Electron Correlation in \ce{C60}}
\begin{figure}[h!]
\includegraphics[scale=0.65]{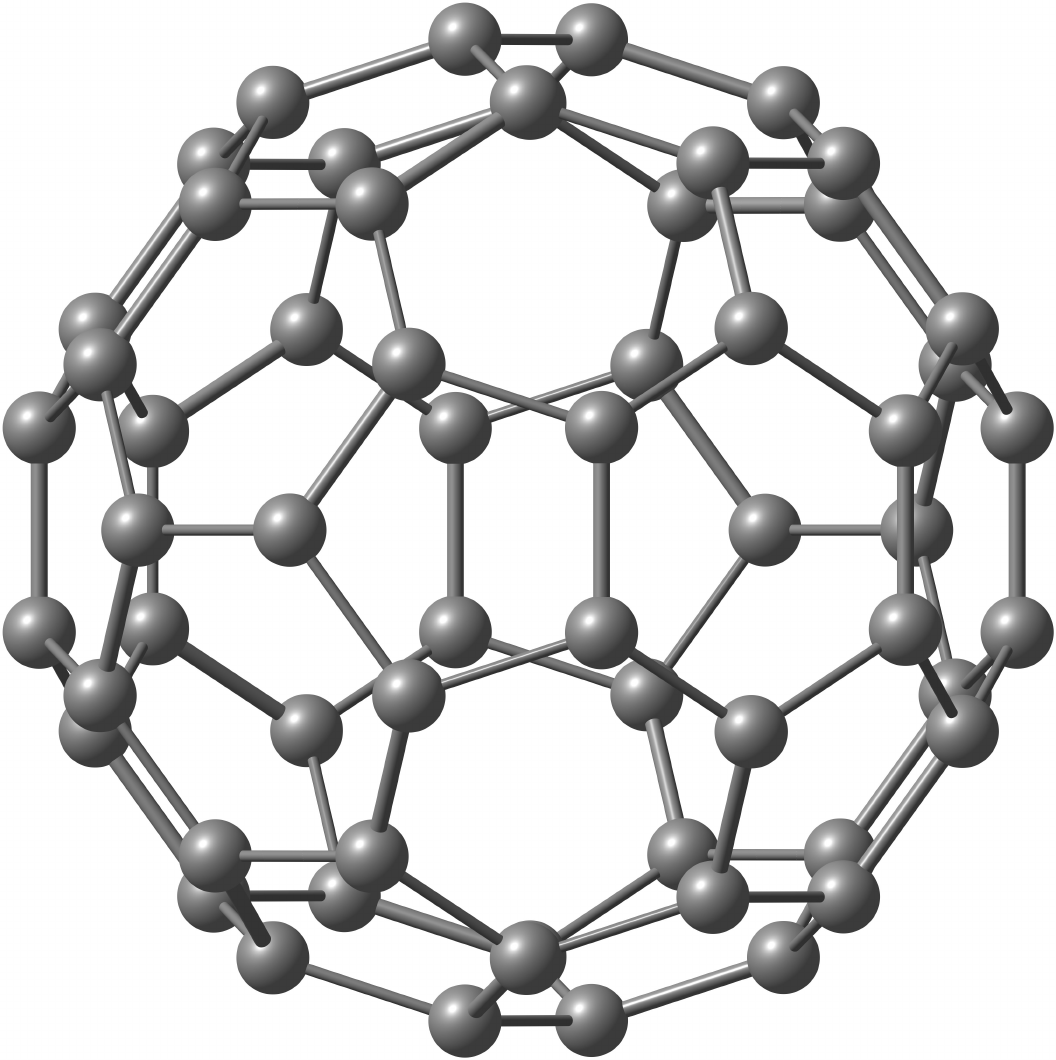}
\caption{The I$_\text{h}$ molecular structure of \ce{C60}.
}
\label{geom:c60}
\end{figure}

\ce{C60} is a well-known electron paramagnetic resonance silent molecule\cite{Paul2002} and its geometry is shown in Figure \ref{geom:c60}. Therefore, it is undoubtedly a molecule with a singlet ground state. Furthermore, its stability has suggested that it is a closed shell molecule.\cite{Tomita2003} This is why it was surprising to observe the existence of R to U symmetry breaking of \ce{C60} at the HF level.\cite{Stuck2011} 
\replacewith{St{\"u}ck and Head-Gordon analyzed this symmetry breaking using SOS-OOMP2 and observed symmetry restoration to restricted orbitals. Furthermore, the experimental singlet-triplet gap (36.95 $\pm$ 0.02 kcal/mol)\cite{Sassara1996} of \ce{C60} was much better reproduced by SOS-OOMP2 and RMP2 compared to UMP2. This led them to conclude that the R to U symmetry breaking in \ce{C60} is {\it artificial} and does not indicate strong correlation.}{
This R to U symmetry breaking was in the end characterized as artificial based on analyses using OOMP2 and the single-triplet gap.
}Later, Jim{\'e}nez-Hoyos et al. found a cGHF solution for \ce{C60} and concluded that \ce{C60} is strongly correlated (or polyradicaloid) based on \replacewith{such a}{this} broken-symmetry HF solution.\cite{Jimenez-Hoyos2014} This was surprising to us because \ce{C60} has been hardly suggested to be polyradicaloid and is also very stable in experiments. 
Therefore, we revisit this problem with $\kappa$-OOMP2 and CC methods and try to determine whether \ce{C60} is strongly correlated.

\subsubsection{Symmetry Breaking Landscape}
Due to limited computational time, we could obtain the symmetry breaking landscape of this molecule only within a minimal basis set, STO-3G.\cite{Hehre1969} We took the cGHF optimized geometry of \ce{C60} from Jim{\'e}nez-Hoyos and co-workers' work. As it is well known, the STO-3G basis set exaggerates strong correlation and facilitates symmetry breaking. The critical $\kappa$ values obtained from minimal basis set calculations would therefore be a good estimate on the upper bound of $\kappa_c$.

\begin{figure}[h!]
\includegraphics[scale=0.6]{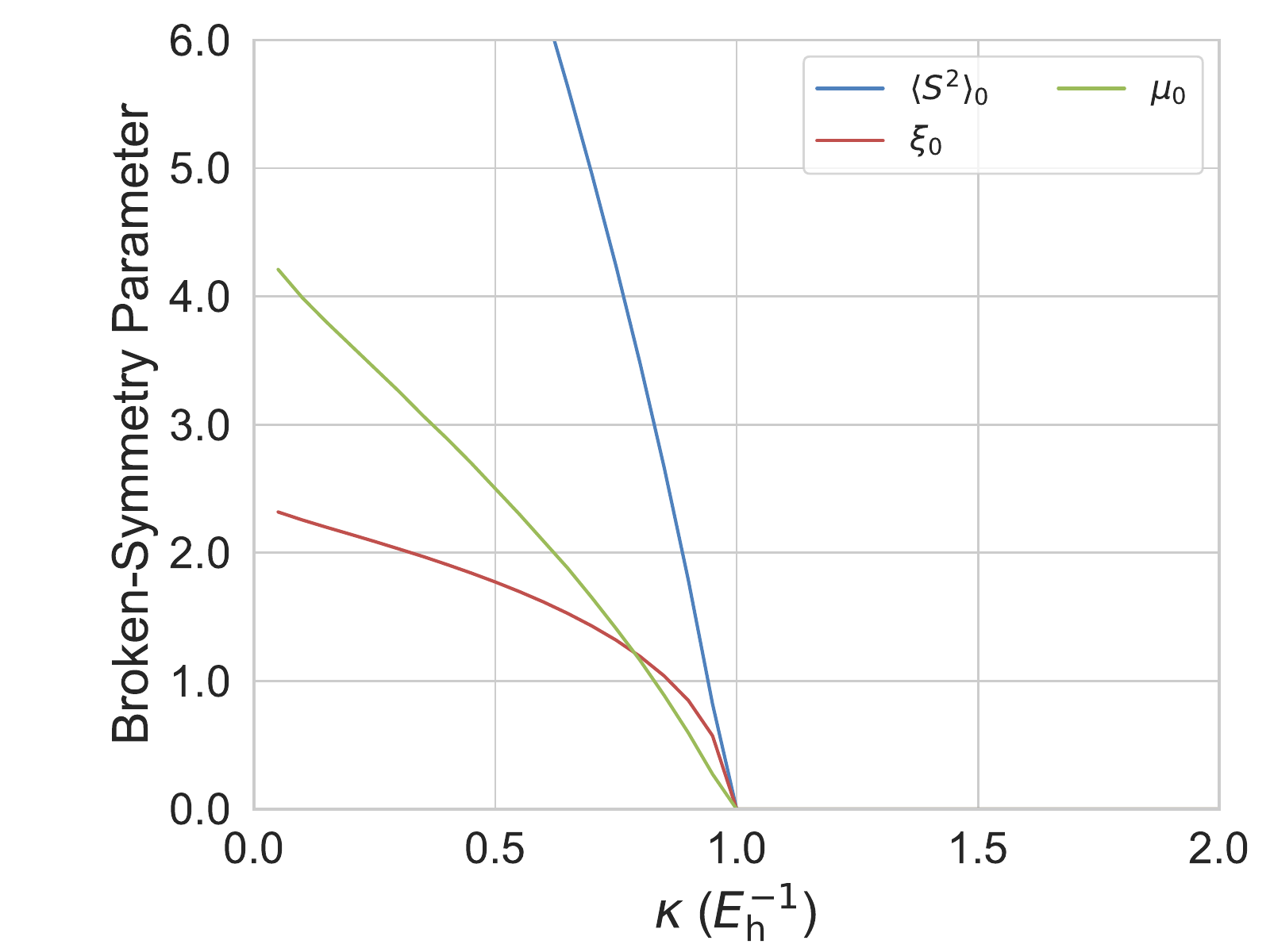}
\caption{\label{fig:c60}
Measures of symmetry breaking ($\langle S^2\rangle_0$, $\xi_0$, and $\mu_0$) as a function of the regularization strength $\kappa$ for \ce{C60} (I$_\text{h}$). It should be noted that these plots are obtained with a scaled correlation energy variant of $\kappa$-OOMP2 (i.e., $\kappa$-S-OOMP2). These quantities do not include correlation corrections: \insertnew{in other words, the plots characterize symmetry-breaking/restoration in the reference determinant}.
}
\end{figure}

In Figure \ref{fig:c60}, we present the symmetry breaking landscape of \ce{C60}. For $\kappa < 1.0$, $\langle S^2\rangle_0$, $\xi_0$, and $\mu_0$ exhibit a steep increase as $\kappa$ decreases. Eventually, the curve reaches the cGHF solution at the limit of $\kappa\rightarrow 0$. 
$\kappa = 1.0$ is enough to restore all symmetries that were broken at the HF level.
\replacewith{This critical value is far from the suggested optimal value}{This critical value is not in the range of essential symmetry breaking illustrated in Section \ref{sec:oomp2}.} Therefore we conclude that the symmetry breaking at the HF level is {\it artificial}. We also confirmed the symmetry restoration at $\kappa = 1.45$ with $\kappa$-OOMP2 within the VDZ (i.e., cc-pVDZ\cite{Dunning1989} without polarization) basis set. Therefore, we believe that the basis set incompleteness error will not affect this qualitative conclusion.
\subsubsection{Singlet-Triplet Gap}
\begin{table}[h!]
  \centering
  \begin{tabular}{c|r|r|r}\hline
Method & \multicolumn{1}{c|}{$\Delta E_\text{S-T}$} & $\langle \hat{S}^2\rangle_{M_S = 0}$ & $\langle \hat{S}^2\rangle_{M_S = 1}$\\\Xhline{3\arrayrulewidth}
RHF & 65.63 & 0.000 & 2.000\\\hline
RMP2 & 50.19 & 0.000 & 2.000\\\hline
SCS-RMP2 & 52.21 & 0.000 & 2.000\\\hline
SOS-RMP2 & 53.22 & 0.000 & 2.000\\\Xhline{3\arrayrulewidth}
UHF & 45.23 & 6.708 & 8.560\\\hline
UMP2 & 77.39 &  5.566 & 7.401\\\Xhline{3\arrayrulewidth}
UOOMP2 & 19.48 & 0.000 & 2.043\\\hline
SCS-UOOMP2 & 28.53  & 0.000 & 2.002 \\\hline
SOS-UOOMP2 & 35.72 & 0.000 & 1.995\\\hline
$\kappa$-UOOMP2 & 49.23 & 0.000 & 2.002\\\Xhline{3\arrayrulewidth}
Experiment\cite{Sassara1996} & 36.95 $\pm$ 0.02 &  & 
  \end{tabular}
  \caption{
The singlet-triplet gap $\Delta E_\text{S-T} $ (kcal/mol) of \ce{C60} from various methods.
The expectation values of $\langle \hat{S}^2\rangle$ for  $M_S=0$ and $M_S=1$ states are presented as well. 
Note that these values include correlation corrections to $\langle \hat{S}^2\rangle$.
  }
  \label{tab:c60}
\end{table}

In Table \ref{tab:c60}, we present the singlet-triplet gap of \ce{C60} computed with various MP2 methods. Here, we used the cc-pVDZ basis set\cite{Dunning1989} along with the corresponding auxiliary basis set.\cite{Weigend2002} For OO methods, we performed orbital optimization starting from stable UHF solutions. 
The results presented in Table \ref{tab:c60} generally agree with what St{\"u}ck and Head-Gordon reported.\cite{Stuck2011}

The singlet-triplet gap predicted by HF is better with UHF (45.23 kcal/mol) than with RHF (65.63 kcal/mol) in comparison to the experimental value (36.95 $\pm$ 0.02 kcal/mol). 
UHF exhibits striking spin-contamination and this is improved to a small extent with UMP2. The singlet-triplet gap of UMP2 is much worse than that of UHF, going from 45.23 kcal/mol to 77.39 kcal/mol whereas RMP2 improves the gap by 15 kcal/mol compared to RHF. 
It is clear that a better reference for subsequent correlation calculations is RHF not UHF. We also compare RMP2 with spin-component scaled MP2 \cite{Grimme2003}(SCS-MP2) and scaled opposite-spin MP2\cite{Jung2004} (SOS-MP2). The singlet-triplet gap is quite insensitive to the choice of scaling factors.

OOMP2 methods successfully remove heavy spin-contamination observed at the HF level. While in terms of $\langle \hat{S}^2 \rangle$ unregularized OOMP2 and its scaled variants are desirable, a striking underestimation of the gap is alarming. In particular, compared to their non-OO variants these gaps are severely underestimated. We suspect that this is due to overcorrelating the $M_S=1$ state via OO. 
Unlike the non-OO variants, unregularized (i.e., $\kappa\rightarrow\infty$) OO methods are sensitive to the scaling parameters.
We observe that the gap from SOS-UOOMP2 is only off by 1 kcal/mol from the experimental value. 
This is likely a fortuitous outcome.

It is interesting that this unphysical overcorrelation of the triplet state seems to be successfully resolved with $\kappa$-UOOMP2. $\kappa$-UOOMP2 yields more or less the same singlet-triplet gap as that of RMP2. 
$\kappa$-UOOMP2 exhibits an error of 12 kcal/mol which is likely due to the limited treatment of electron correlation. It will be interesting to resolve this remaining error using higher order perturbation theory or, perhaps, coupled-cluster methods.
\subsubsection{Correlated 1PDM}
\begin{figure}[h!]
\includegraphics[scale=0.50]{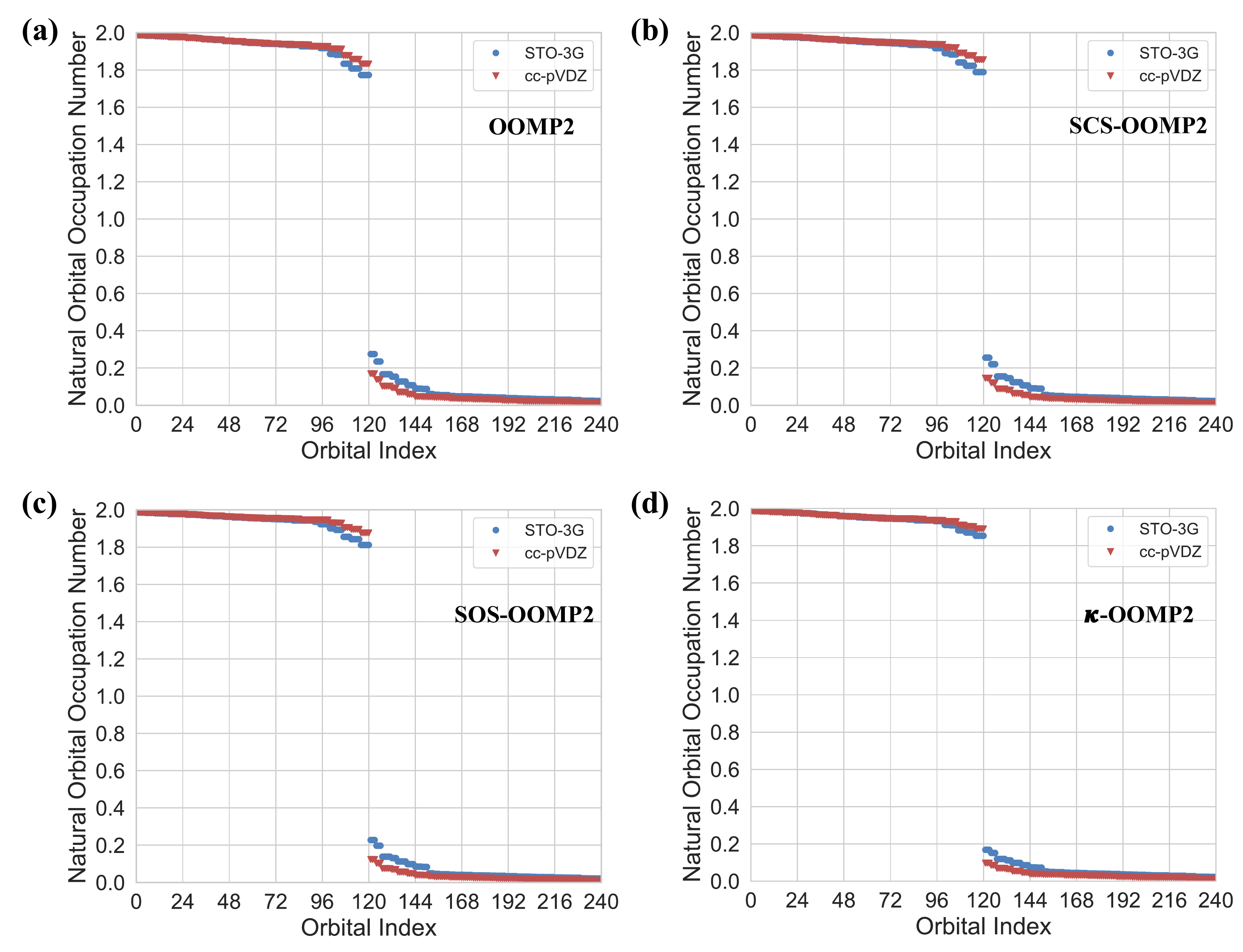}
\caption{Natural orbital occupation numbers of \ce{C60} within a valence active space from (a) OOMP2, (b) SCS-OOMP2, (c) SOS-OOMP2, and (d) $\kappa$-OOMP2. Note that there were no unrestricted solutions found for any of these methods.
}
\label{fig:c60noon1}
\end{figure}
\insertnew{We then obtain NOONs from OOMP2 methods and analyze them.
We will discuss three unregularized OOMP2 methods, OOMP2, SCS-OOMP2,\cite{Neese2009} and SOS-OOMP2\cite{Lochan2007} along with $\kappa$-OOMP2.
}
In Figure \ref{fig:c60noon1}, different OOMP2 methods exhibit more or less identical NOONs.
We could not find unrestricted solutions for any of these methods.
This reflects the simplicity of the electronic structure of \ce{C60}\replacewith{compared to that of \ce{C36}}{.} A slight reduction in open-shell character is observed in $\kappa$-OOMP2 compared to other OOMP2 methods.
These NOONs are far from the usual values for orbitals with open-shell character in strongly correlated wavefunctions (i.e., values between 1.30 and 0.70). 

\begin{figure}[h!]
\includegraphics[scale=0.50]{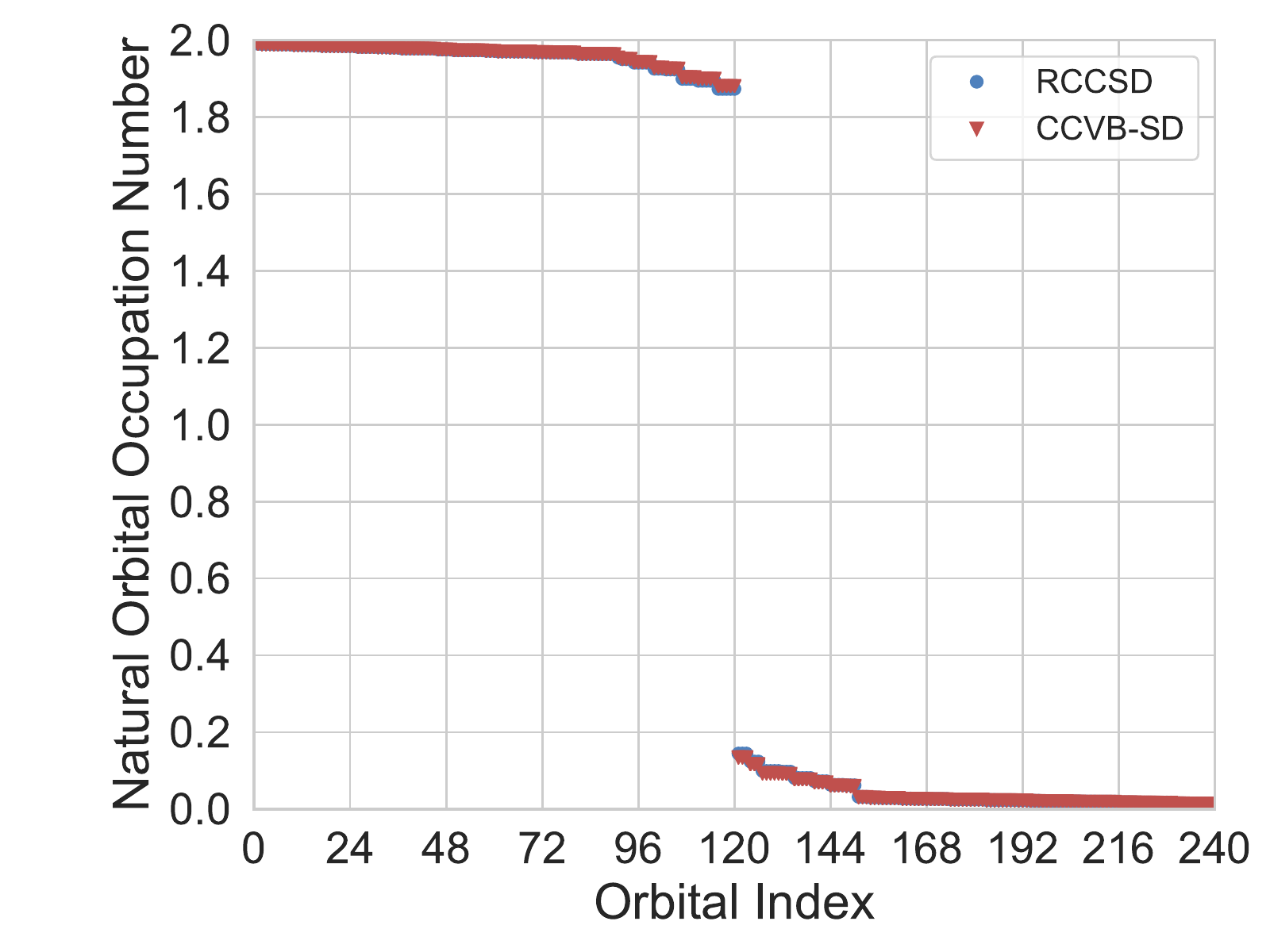}
\caption{Natural orbital occupation numbers of \ce{C60} within the minimal basis set, STO-3G, from RCCSD and CCVB-SD.
Two sets of data points are very close to each other so the blue circles lie right below the red triangles.
}
\label{fig:c60noon2}
\end{figure}

Likewise, the NOONs from CC methods presented in Figure \ref{fig:c60noon2} strongly suggest that this molecule is {\it not strongly correlated}. RCCSD and CCVB-SD exhibit visually identical distributions. Indeed, the NOONs of the highest occupied NO (HONO) and the lowest unoccupied NO (LUNO) are
(1.87, 0.14) for both methods. These values are comparable to naphthalene's NOONs computed with CCVB-SD. Evidently, naphthalene is not a strongly correlated system, which implies that neither is \ce{C60}.

\begin{table}[h!]
  \centering
  \begin{tabular}{c|c|c}\hline
Method & STO-3G & cc-pVDZ\\\hline
OOMP2 & 16.07 (0.27) & 13.99 (0.23)\\\hline
SCS-OOMP2 & 15.32 (0.26) & 12.59 (0.21)\\\hline
SOS-OOMP2 & 13.82 (0.23) & 10.99 (0.18)\\\hline
$\kappa$-OOMP2 & 12.93 (0.22) & 11.61 (0.19)\\\hline
RCCSD & 9.48 (0.16) & \\\hline
CCVB-SD & 9.19 (0.15) & 
  \end{tabular}
  \caption{
  Number of unpaired electrons (NUE) of \ce{C60} computed from various methods.
  The numbers in parentheses are NUE per carbon atom.
  }
  \label{tab:nue2}
\end{table}

In Table \ref{tab:nue2}, we present NUEs (Eq. \eqref{eq:nue}) of \ce{C60} computed by various methods.
\replacewith{Compared to Table \ref{tab:nue1}, all the NUEs are larger with \ce{C60} than with \ce{C36}.
However, this is simply due to the fact that there are more electrons in \ce{C60}.
With a proper normalization (i.e., NUE per C atom), it is clear that \ce{C60} exhibits less open-shell character than \ce{C36}.}
With a larger basis set (cc-pVDZ), NUEs are smaller than those of STO-3G. 
\insertnew{This reflects the reduction in strong correlation with the increase in the basis set size.}
We once again observe almost no differences between RCCSD and CCVB-SD.
\insertnew{
Overall, there are about 0.20 unpaired electrons per C atom in \ce{C60}. As each C atom has four valence electrons, this amounts only 5$\%$ of the total number of electrons. Therefore, the polyradicaloid character in \ce{C60} is only marginal from the global electronic structure viewpoint.}
\replacewith{In this section, we have studied the electron correlation of \ce{C36} and \ce{C60} in terms of natural orbital occupation number using various OOMP2 methods and CC methods. Based on unbiased analyses of the results, we conclude that \ce{C36} is strongly correlated (i.e., a singlet biradicaloid in particular) and \ce{C60} is not strongly correlated. In order to properly describe \ce{C36} with $\kappa$-OOMP2, unrestriction was necessary whereas \ce{C60} did not require any symmetry breaking. Therefore, the conclusion about \ce{C60} in the study of Jim{\'e}nez-Hoyos and co-workers is likely {\it an artifact} caused by Hartree-Fock.}{}

Based on these analyses, we conclude that \ce{C60} should be considered a closed-shell molecule and {\it not strongly correlated}.
\iterate{Thus the very interesting cGHF solution reported previously\cite{Jimenez-Hoyos2014} should be considered as an artificial rather than an essential symmetry breaking.}
\insertnew{This conclusion will be also supported by comparing with our next molecule, \ce{C36} which is a well-known biradicaloid.}

\subsection{The Nature of Electron Correlation in \ce{C36}}

\begin{figure}[h!]
\includegraphics[scale=0.65]{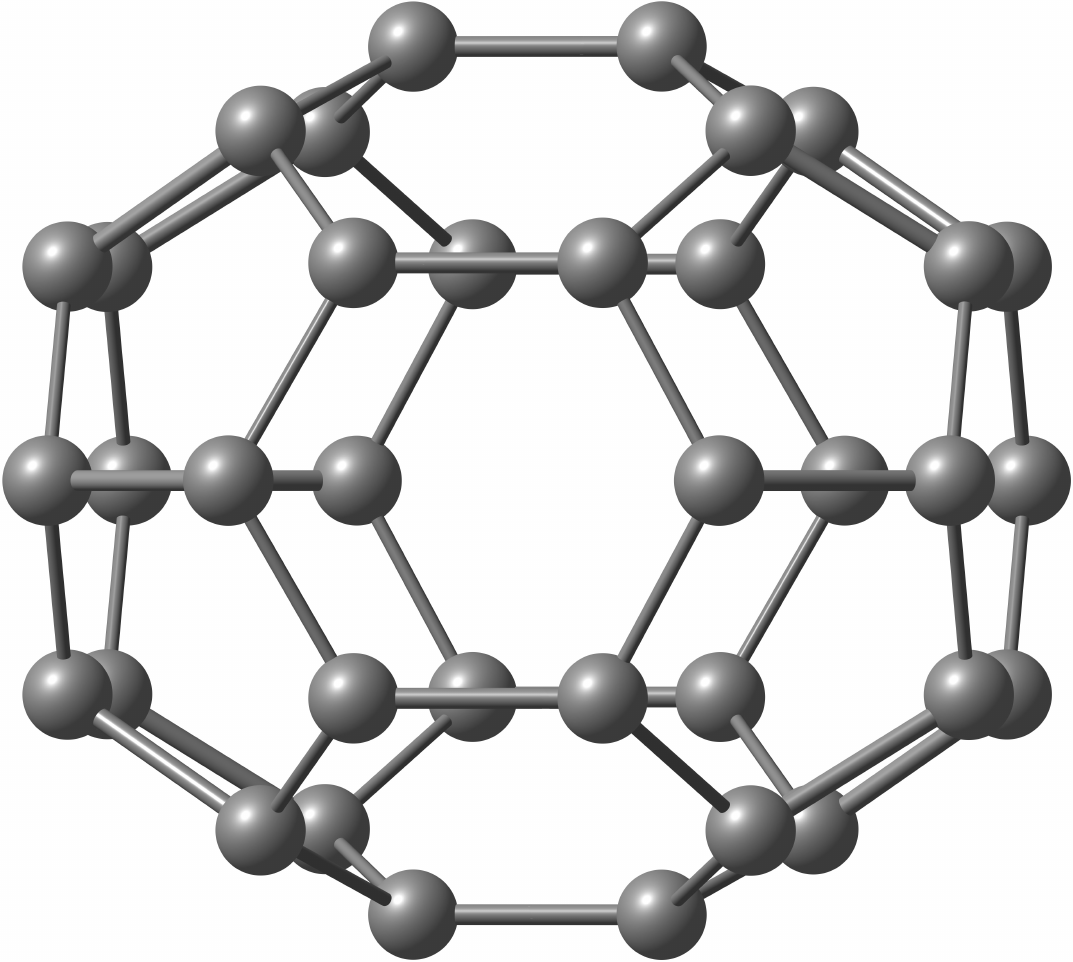}
\caption{The D$_\text{6h}$ molecular structure of \ce{C36}.
}
\label{geom:c36}
\end{figure}

The D$_\text{6h}$ structure of \ce{C36} shown in Figure \ref{geom:c36} has been known to exhibit strong biradicaloid character.\cite{Varganov2002} This has been of great interest both experimentally\cite{Piskoti1998} and theoretically.\cite{Fowler1999,Fowler1999a,Jagadeesh1999,Ito2000,Slanina2000,Lan-FengYuan2000,Varganov2002,Paulus2003} The D$_\text{6h}$ structure is supported by the $^{13}$C NMR spectrum exhibiting a single peak.\cite{Piskoti1998} Different computational methods suggested different structures and even different multiplicities. There is no doubt that \ce{C36} exhibits complex electronic structure and is a strongly correlated system.

St{\"u}ck and Head-Gordon studied this prototypical strongly correlated system using unregularized SOS-OOMP2 before. Fortunately, unregularized OOMP2 did not diverge even with this substantial biradicaloid character. What was not explicitly mentioned in this previous work is that OOMP2 in fact yields a restricted solution when starting from an unrestricted solution. Since unregularized OOMP2 unphysically prefers restricted solutions (closely connected to its singularity problem), this is not unexpected. This also suggests that whether unregularized OOMP2 restores broken symmetry is not a definitive probe to characterize strong electron correlation in a given system \iterate{(it is the $\kappa\rightarrow\infty$ limit shown in Fig. \ref{fig:mhg2})}. We will see how $\kappa$-OOMP2 resolves this artifact and can be used to probe strong correlation in this system.

We obtained the D$_\text{6h}$ molecular structure of \ce{C36} via geometry optimization with restricted density functional calculations using the BLYP exchange-correlation functional\cite{Becke1988,Lee1988} and the 6-31G(d) basis set.\cite{Hariharan1973} The geometry optimization was performed with a D$_\text{6h}$ geometric constraint so the optimization was not allowed to break this spatial symmetry. We do not think that this geometry is quantitatively accurate but for the purpose of this work, it should be sufficient.

\begin{figure}[h!]
\includegraphics[scale=0.6]{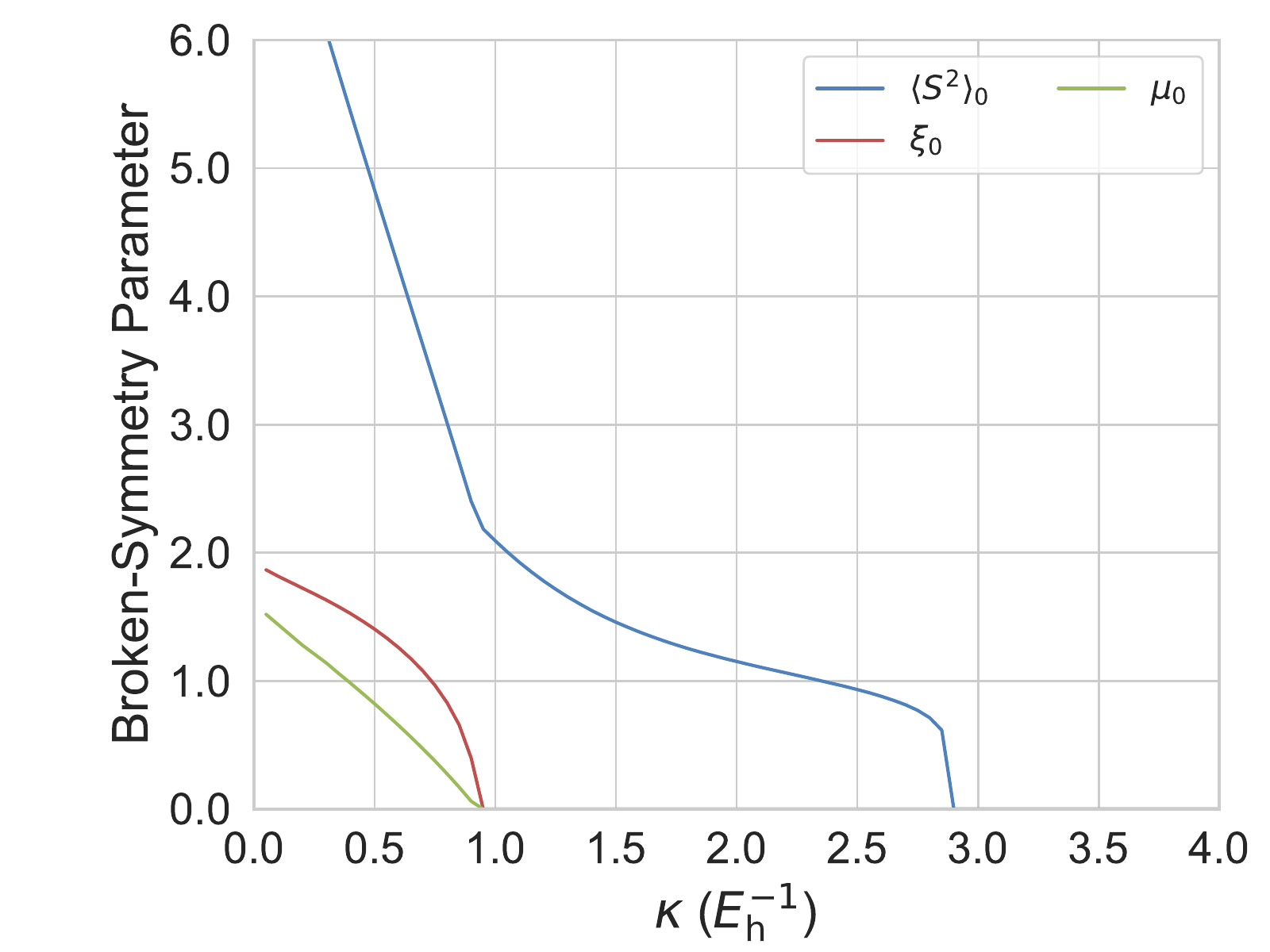}
\caption{\label{fig:c36}
Measures of symmetry breaking ($\langle S^2\rangle_0$, $\xi_0$, and $\mu_0$) as a function of the regularization strength $\kappa$ for \ce{C36} (D$_\text{6h}$). It should be noted that these plots are obtained with a scaled correlation energy variant of $\kappa$-OOMP2 (i.e., $\kappa$-S-OOMP2). These quantities do not include correlation corrections: \insertnew{in other words, the plots characterize symmetry-breaking/restoration in the reference determinant}.
}
\end{figure}
\subsubsection{Symmetry Breaking Landscape}
We computed a landscape of symmetry breaking as a function of $\kappa$ within the 6-31G basis set\cite{Hehre1972} along with the cc-pVDZ auxiliary basis set.\cite{Weigend2002} This may not be quantitatively accurate, but it should be enough to draw qualitative conclusions. Figure \ref{fig:c36} shows that \ce{C36} exhibits multiple classes of one-particle wavefunctions as a function of $\kappa$. As noted by Jim{\'e}nez-Hoyos et al,\cite{Jimenez-Hoyos2014} at the HF level there exists a cGHF solution. This is clearly evident in Figure \ref{fig:c36} for $\kappa < 0.95$, since we have nonzero $\langle S^2\rangle_0$, $\xi_0$, and $\mu_0$. $\xi_0$ and $\mu_0$ vanish at their critical value $\kappa_c = 0.95$ and only unrestricted solutions were obtained for $\kappa \in [0.95,2.85]$. For $\kappa > 2.85$, only restricted solutions were found. This is consistent with St{\"u}ck and Head-Gordon's work which used SOS-OOMP2.\cite{Stuck2011} \insertnew{This landscape has more structure than that of \ce{C60} (see Fig. \ref{fig:c60}) which reflects the increased complexity of the electronic structure of \ce{C36}.}

\replacewith{The optimal $\kappa$ value around 1.5 suggests that both complex generalized and restricted solutions are unphysical.}{The $\kappa$ value within the essential symmetry breaking region exhibits only unrestricted solutions. Therefore, complex generalized and restricted solutions reflect artificial symmetry breaking.} Since \ce{C36} is a well-known singlet biradicaloid, one would expect an unrestricted solution with an $\langle S^2 \rangle_0$ value between 0.0 and 2.0 (i.e., there is only triplet contaminant). Interestingly, this is what was obtained from $\kappa$-S-OOMP2 with $\kappa \approx 1.5$ and also from $\kappa$-OOMP2 with $\kappa=1.45$ (see below). \replacewith{However, this optimal $\kappa$ value was determined based on a crude thermochemistry benchmark so we shall provide more evidence to support this conclusion.}{}\insertnew{Based on the existence of this essential symmetry breaking, we conclude that \ce{C36} is strongly correlated.}
\subsubsection{Singlet-Triplet Gap}
\begin{table}[h!]
  \centering
  \begin{tabular}{c|r|r|r}\hline
Method & \multicolumn{1}{c|}{$\Delta E_\text{S-T}$} & $\langle \hat{S}^2\rangle_{M_S = 0}$ & $\langle \hat{S}^2\rangle_{M_S = 1}$\\\hline
RHF & -19.69 & 0.000 & 2.000\\\hline
RMP2 & 14.46 & 0.000 & 2.000\\\hline
SCS-RMP2 & 9.75 & 0.000 & 2.000\\\hline
SOS-RMP2 & 7.40 & 0.000 & 2.000\\\Xhline{3\arrayrulewidth}
UHF & 32.29 & 7.448 & 8.793 \\\hline
UMP2 & 42.73 & 6.428 & 7.795 \\\Xhline{3\arrayrulewidth}
UOOMP2 & 15.85 & 0.000 & 2.070\\\hline
SCS-UOOMP2 & 30.76 & 0.000 & 1.978\\\hline
SOS-UOOMP2 & 22.91 & 0.000 & 2.002\\\hline
$\kappa$-UOOMP2 & 4.42 & 0.959 & 2.008\\\Xhline{3\arrayrulewidth}
AP+$\kappa$-UOOMP2 & 8.46 &  & \\\Xhline{3\arrayrulewidth}
MRMP2\cite{Varganov2002} & 8.17 &  &
  \end{tabular}
  \caption{
The singlet-triplet gap $\Delta E_\text{S-T} $ (kcal/mol) of \ce{C36} from various methods.
The expectation values of $\langle \hat{S}^2\rangle$ for  $M_S=0$ and $M_S=1$ states are presented as well. 
Note that these values include correlation corrections to $\langle \hat{S}^2\rangle$.
All but MRMP2 results were obtained with the cc-pVTZ basis set.\cite{Dunning1989}
MRMP2 results in ref. \citenum{Varganov2002} were obtained with a D$_\text{6h}$ geometry within the 6-31G(d) basis set. MRMP2 was performed on a CASSCF solution with a (2e, 4o) active space.
  }
  \label{tab:c36}
\end{table}

We will focus on determining whether unscaled $\kappa$-OOMP2 with $\kappa = 1.45$ (i.e., the recommended default $\kappa$-OOMP2 method\cite{Lee2018}) works quantitatively well compared to other MP2 approaches. We observed the same cG to U partial symmetry restoration with $\kappa=1.45$ within the cc-pVDZ basis set. Therefore, due to limited computational resources, for the cc-pVTZ basis set\cite{Dunning1989,Weigend2002} we only ran unrestricted calculations.
In Table \ref{tab:c36}, we present the singlet-triplet gap ($\Delta E_\text{S-T}$) computed with various MP2 approaches. 
Since there is no reliable experimental gap available, we shall compare our results against multi-reference MP2 (MRMP2) results.\cite{Varganov2002} The MRMP2 results were obtained with the 6-31G(d) basis set and a small active space, (2e, 4o), complete active space self-consistent field (CASSCF)  reference state. These might not be a highly accurate reference result, but it can serve as a qualitative answer.

RHF predicts a triplet ground state, a qualitatively wrong result. RMP2, SCS-RMP2,\cite{Grimme2003} and SOS-RMP2\cite{Jung2004} correct the sign of the gap. They are also not too far away from the multi-reference MP2 (MRMP2) result and this is a fortuitous result given the singlet biradicaloid character of the true ground state cannot be captured by doubly occupying restricted orbitals. UHF and UMP2 are heavily spin-contaminated and predict significantly large gaps. We note that the MP2 treatment cannot clean up heavy spin-contamination present at the UHF level. 
As a result, these gaps are qualitatively incorrect as they are too large to support the singlet biradicaloid character. 

Spin-contamination is successfully removed with unregularized OOMP2 methods. The UOOMP2 gap is quantitatively close to RMP2. However, SCS- and SOS-MP2 methods predict much larger gaps with OO than those without OO. This is likely due to overcorrelating the singlet state. \insertnew{This is the case where OOMP2 (or $\kappa$-OOMP2 for $\kappa\rightarrow\infty$) yields too little symmetry breaking as explained in Section \ref{sec:oomp2}.}

$\kappa$-UOOMP2 yields a broken-symmetry solution for $M_S=0$ and $\langle\hat{S}^2\rangle$ is 0.959. This is of a strong singlet biradicaloid character and serves as a good candidate for Yamaguchi's approximate spin-projection (AP)\cite{Yamaguchi1988} to obtain better energetics. The use of AP in conjunction with $\kappa$-UOOMP2 was first discussed in ref. \citenum{Lee2018}. Without AP, the gap was predicted to be 4.42 kcal/mol. This is small enough to conclude that \ce{C36} is a singlet biradicaloid, but the gap is underestimated due to spin-contamination. Applying AP lowers the singlet energy by 4 kcal/mol which yields a gap of 8.46 kcal/mol. This is quite close to the reference MRMP2 energy. However, a more precise benchmark is highly desirable in the future to draw a more definitive conclusion.
\insertnew{
Nevertheless, the result suggests that \ce{C36} is strongly correlated which agrees with the conclusion drawn based on the symmetry breaking landscape.
}
\subsubsection{Correlated 1PDM}

\begin{figure}[h!]
\includegraphics[scale=0.50]{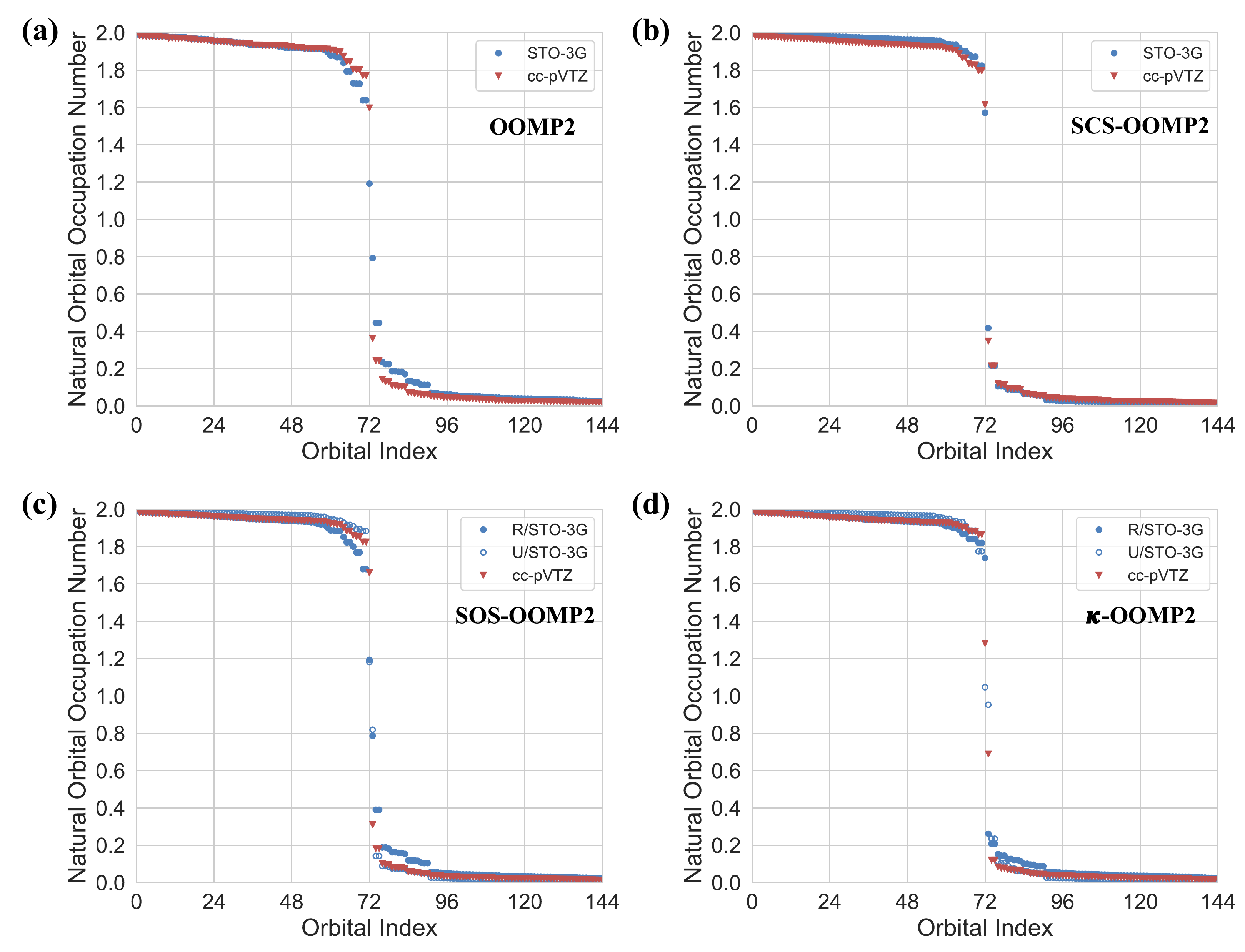}
\caption{Natural orbital occupation numbers of \ce{C36} within the valence space from (a) OOMP2, (b) SCS-OOMP2, (c) SOS-OOMP2, and (d) $\kappa$-OOMP2. Note that there were no unrestricted solutions found for OOMP2 and SCS-OOMP2. Furthermore, the solution from $\kappa$-OOMP2 with cc-pVTZ is unrestricted.
}
\label{fig:c36noon1}
\end{figure}

In Figure \ref{fig:c36noon1}, we discuss NOONs of \ce{C36} from the various OOMP2 methods that were used in the previous sections. For STO-3G, OOMP2, SOS-OOMP2, and $\kappa$-OOMP2 exhibit clear emergent singlet biradicaloid character. \insertnew{It is clear that NOONs exhibit more open-shell character than those of \ce{C60}} Interestingly, we could isolate an unrestricted STO-3G solution with SOS-OOMP2 by reading in an unrestricted $\kappa$-OOMP2 solution. We could not isolate such a solution with the cc-pVTZ basis set. The R to U symmetry breaking of SOS-OOMP2 is interesting in that the unrestricted solution seems to have more closed-shell orbitals than the restricted solution. Moreover, the unrestricted solution is 21 m$E_h$ higher in energy than the restricted one. Within the cc-pVTZ basis set, OOMP2, SCS-OOMP2, and SOS-OOMP2 all predict very similar NOONs and they exhibit only slight singlet biradicaloid character. On the other hand, $\kappa$-OOMP2 exhibits very strong biradicaloid character characterized by an unrestricted solution. HONO and LUNO for this unrestricted solution have NOONs of 1.28 and 0.69, respectively. The true ground state would have less polarized NOONs than $\kappa$-OOMP2.

\begin{figure}[h!]
\includegraphics[scale=0.50]{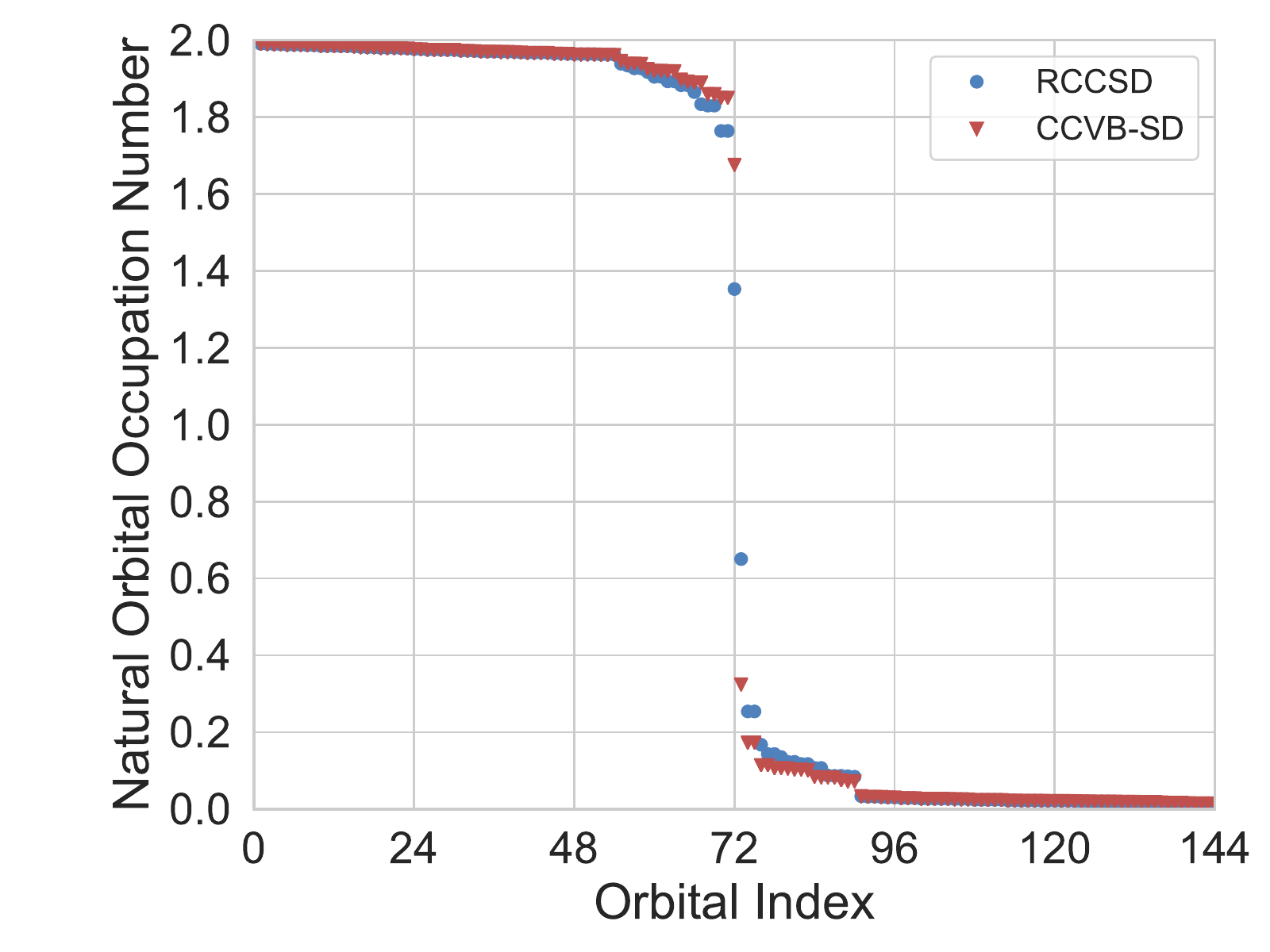}
\caption{Natural orbital occupation numbers of \ce{C36} within the minimal basis set, STO-3G, from RCCSD and CCVB-SD.
}
\label{fig:c36noon2}
\end{figure}
In Figure \ref{fig:c36noon2}, we show NOONs from two CC methods, RCCSD and CCVB-SD. As was noted before, RCCSD and CCVB-SD show qualitative differences when the system is strongly correlated.\cite{Small2012,Lee2017} In particular, RCCSD clearly overcorrelates and likely exhibits non-variationality.\cite{Lee2017} These trends are well reflected in Figure \ref{fig:c36noon2}. The NOONs of the HONO and the LUNO are (1.35, 0.65) and (1.68, 0.32) for RCCSD and CCVB-SD, respectively. These two sets of NOONs show clear differences in that RCCSD exhibits far more open-shell character than does CCVB-SD. In passing we note that the CCVB-SD value (1.68, 0.32) is comparable to NOONs of an acene of length 12 we studied in ref. \citenum{Lee2017}. This is quite substantial strong correlation.

\begin{table}[h!]
  \centering
  \begin{tabular}{c|c|c|c}\hline
Method & STO-3G/R & STO-3G/U & cc-pVTZ\\\hline
OOMP2 & 12.88 (0.36)&N/A & 10.66 (0.30)\\\hline
SCS-OOMP2 & 6.24 (0.17) & N/A& 9.66 (0.27)\\\hline
SOS-OOMP2 & 11.31 (0.31)& 6.17 (0.17)$^1$ & 8.44 (0.23)\\\hline
$\kappa$-OOMP2 & 8.93 (0.25)& 6.95 (0.19)$^1$ & 9.25 (0.26)$^1$\\\hline
RCCSD & 8.23 (0.23) & N/A & \\\hline
CCVB-SD & 6.65 (0.18)& N/A & 
  \end{tabular}
  \caption{
  Number of unpaired electrons (NUE) of \ce{C36} computed from various methods.
  For SOS-OOMP2 and $\kappa$-OOMP2 with STO-3G, we present NUE for both restricted and unrestricted solutions. 
  The first value corresponds to the restricted one and the second corresponds to the unrestricted one.
  The numbers in parentheses are NUE per carbon atom.
  STO-3G/R indicates restricted calculations with STO-3G while STO-3G/U indicates unrestricted calculations with STO-3G.
The cc-pVTZ calculations are done with spin-unrestricted calculations and broken symmetry solutions are indicated by a superscript 1.
  $^1$ Spin-unrestricted solutions.
  }
  \label{tab:nue1}
\end{table}

\replacewith{We have shown that both $\kappa$-OOMP2 and CC methods suggest that \ce{C36} is strongly correlated. As an attempt to quantify such electron correlation, we present Head-Gordon's number of unpaired electrons (NUE),\cite{Head-Gordon2003}
\begin{equation}
\text{NUE} = \sum_i \text{min}(2-n_i, n_i)
\end{equation}
where $i$ is summed over all natural orbitals and $n_i$ is the occupation number of the $i$-th natural orbital.}{}\replacewith{We present this}{We present the NUEs computed from the various methods examined here} in Table \ref{tab:nue1}. SOS-OOMP2 and $\kappa$-OOMP2 exhibit smaller NUE with unrestricted solutions than with restricted solutions. This suggests that the spin polarization occurs within a few orbitals which in turn reduces the global correlation. This is in agreement with Figure \ref{fig:c36noon1}. 
\insertnew{Compared to Table \ref{tab:nue2}, all the NUEs are smaller with \ce{C36} than with \ce{C60}.
However, this is simply due to the fact that there are more electrons in \ce{C60}.
With a proper normalization (i.e., NUE per C atom), it is clear that \ce{C60} exhibits less open-shell character than \ce{C36}.}

For OOMP2 and SOS-OOMP2 (restricted), increasing the basis set size decreases the NUE values. The minimal basis set we used indeed provided more prominent open-shell character than cc-pVTZ in these methods. 
For $\kappa$-OOMP2 (unrestricted) and SCS-OOMP2, the open-shell character increases going from STO-3G to cc-pVTZ. Although HONO and LUNO exhibit less polarization, the other orbitals exhibit more open-shell character with a larger basis set. This can be understood as having more dynamic correlation effects and smaller correlation within a valence space in the cc-pVTZ basis set.

Comparing NUEs of RCCSD and CCVB-SD clearly suggests that RCCSD overcorrelates the system. 
\insertnew{Therefore, this also suggests that \ce{C36} is strongly correlated.}

\subsection{The Nature of Electron Correlation in \ce{C20}} 
\insertnew{
Lastly, we obtain the landscape of symmetry breaking of a smaller fullerene, \ce{C20}. 
We have established that all three probes we used yield a consistent conclusion for \ce{C60} and \ce{C36}.
Therefore, we believe that it is sufficient to use this single probe to obtain an answer to a qualitative question:
is \ce{C20} strongly correlated?
}
\subsubsection{Jahn-Teller distorted structures}
\begin{figure}[h!]
\includegraphics[scale=0.45]{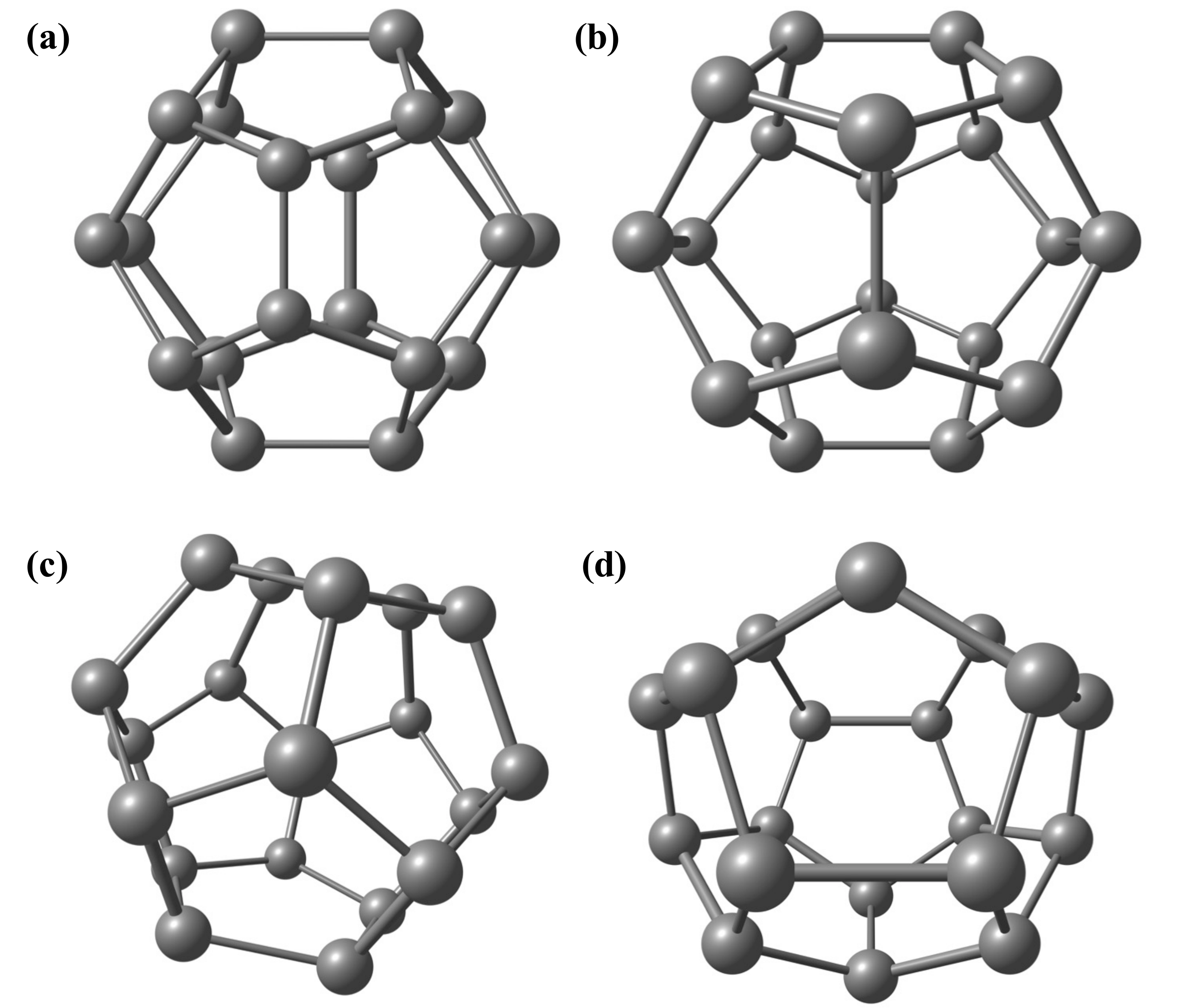}
\caption{Four Jahn-Teller distorted isomers of \ce{C20}: 
(a) \ce{C_{2h}}, (b) \ce{D_{2h}}, (c) \ce{C_{i}}, and (d) \ce{D_{3h}}.
}
\label{geom:c20}
\end{figure}
\ce{C20} has attracted a lot of attention \replacewith{in physical chemistry}{}since it is the smallest possible fullerene suggested by graph-theoretic analyses.\cite{Schwerdtfeger2015} The existence of its cage geometry was controversial for some time,\cite{Parasuk1991,Feyereisen1992,Raghavachari1993,VonHelden1993,Bylaska1996,Martin1996b,Jones1997,Kietzmann1998} but eventually it was experimentally observed in 2000.\cite{Prinzbach2000} Since then, there have been a lot of quantum chemistry studies of \ce{C20} which focus on relative energetics of different conformers such as bowl, cage, and ring.\cite{Sokolova2000,Grimme2002,Lu2003,Paulus2003,An2005,Jin2015,Manna2016} Here we focus on multiple Jahn-Teller distorted conformations of the cage geometry (C$_\text{2h}$, D$_\text{2h}$, C$_\text{i}$, and D$_\text{3h}$).

Manna and Martin carried out a careful study of relative energies among the Jahn-Teller distorted conformers of \ce{C20}. \cite{Manna2016} They used state-of-the-art wavefunction methods in conjunction with high-quality density functional \insertnew{theory} calculations. We took molecular geometries of \ce{C20} from ref. \citenum{Manna2016} which were optimized with PBE0 and the cc-pVTZ basis set.
These geometries are shown in Figure \ref{geom:c20}. Our focus in this section is on the artificial symmetry breaking in these molecules. We studied these within the cc-pVDZ basis set\cite{Dunning1989} along with the appropriate auxiliary basis set.\cite{Weigend2002}
In Table \ref{tab:class1}, we present the classification of HF solutions of these molecules. Since they are commonly thought of as closed-shell molecules, it is striking that RHF is always unstable under spin-symmetry breaking for these molecules.
\begin{table}[h!]
  \centering
  \begin{tabular}{c|c|c|c|c|c|c|c}\hline
Geometry& $\hat{K}$& $\hat{S}^2$& $\hat{S}_\mathbf{n}$& Stuber-Paldus & $\langle S^2\rangle_0$ & $\xi_0$  & $\mu_0$ \\\hline
\ce{C_{2h}}& Broken& Broken& Broken& cGHF    & 5.395 & 1.117 & 0.816\\\hline
\ce{D_{2h}}& Broken& Broken& Broken& cGHF    & 5.397 & 1.626 & 0.818\\\hline
\ce{C_{i}}& Broken& Broken& Conserved& cUHF  & 5.101 & 1.074 & 0.000\\\hline
\ce{D_{3h}}& Broken& Broken& Conserved& cUHF & 5.100 & 1.247 & 0.000\\\hline
  \end{tabular}
  \caption{
Classification of HF solutions of \ce{C20} 
isomers along with 
the corresponding $\langle S^2\rangle_0$, $\xi_0$, and $\mu_0$. The subscript 0 denotes that these properties are computed at the HF level.
  }
  \label{tab:class1}
\end{table}

We will first discuss the cGHF solutions found in C$_\text{2h}$ and D$_\text{2h}$. Although the two geometries differ by 29 mE$_\text{h}$ in terms of the nuclear repulsion energy, the cGHF energies differ only by 10 $\mu$E$_\text{h}$. While this accidental electronic near-degeneracy is surprising, the appearance of non-collinear solutions is also striking since this molecule should be considered closed-shell.\cite{Parasuk1991,Martin1996b,Paulus2003,Adams1993} We will now see how regularized OOMP2 restores this artificial symmetry breaking starting from these broken symmetry solutions as a function of the regularization strength.
\begin{figure}[h!]
\includegraphics[scale=0.5]{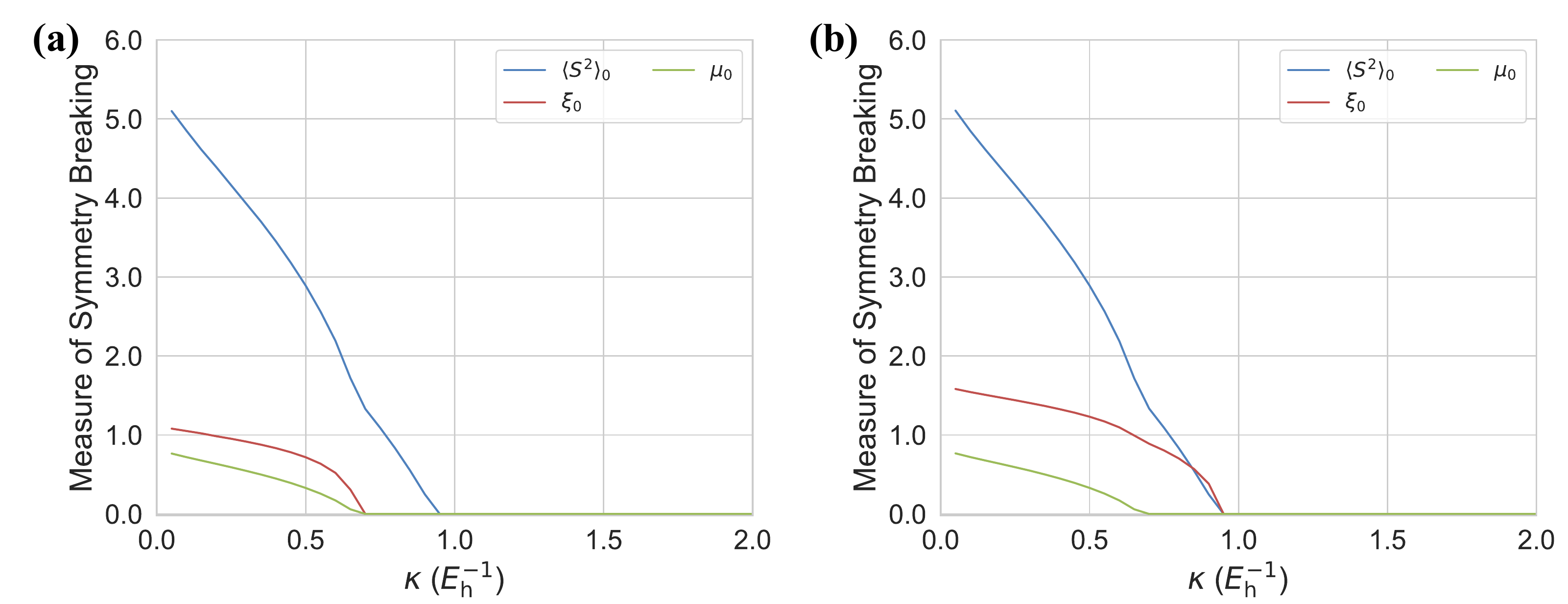}
\caption{\label{fig:c20cghf}
Measures of symmetry breaking ($\langle S^2\rangle_0$, $\xi_0$, and $\mu_0$) as a function of the regularization strength $\kappa$ for (a) \ce{C20} (C$_\text{2h}$) and (b) \ce{C20} (D$_\text{2h}$). It should be noted that these plots are obtained with a scaled correlation energy variant of $\kappa$-OOMP2 (i.e., $\kappa$-S-OOMP2). These quantities do not include correlation corrections: \insertnew{in other words, the plots characterize symmetry-breaking/restoration in the reference determinant}.
}
\end{figure}

In Figure \ref{fig:c20cghf}, $\langle S^2\rangle_0$, $\xi_0$, and $\mu_0$ are plotted as a function of $\kappa$ for C$_\text{2h}$ and D$_\text{2h}$ geometries. 
Although two cGHF solutions exhibit quantitative similarity in $\langle S^2\rangle_0$, $\xi_0$, and $\mu_0$ as shown in Table \ref{tab:class1}, $\kappa_c$ values for each quantity shows a qualitative difference. 

In the case of the C$_\text{2h}$ geometry (Figure \ref{fig:c20cghf} (a)), $\kappa_c$ values are 0.95, 0.70 and 0.70 for $\langle S^2\rangle_0$, $\xi_0$, and $\mu_0$, respectively. This suggests that for $\kappa \in [0.70, 0.95]$ there are unrestricted solutions for this system. Moreover, the symmetry restoration of complex and non-collinearity occurs at the same time. The $\langle S^2\rangle_0$ restoration requires stronger MP2 correlation energies than $\xi_0$ and $\mu_0$. The D$_\text{2h}$ geometry exhibits a different behavior. The $\kappa_c$ values for $\langle S^2\rangle_0$, $\xi_0$, and $\mu_0$ are 0.95, 0.95, and 0.70, respectively. In this case, for $\kappa \in [0.70, 0.95]$ there are complex, unrestricted solutions. The U to G symmetry breaking is easier to restore than complex and unrestriction based on the relative magnitude of $\kappa_c$ values.
\insertnew{These $\kappa_c$ values are in the artificial symmetry breaking range discussed in Section \ref{sec:oomp2}.}
Therefore, we conclude that this symmetry breaking is {\it artificial}.

\begin{figure}[h!]
\includegraphics[scale=0.5]{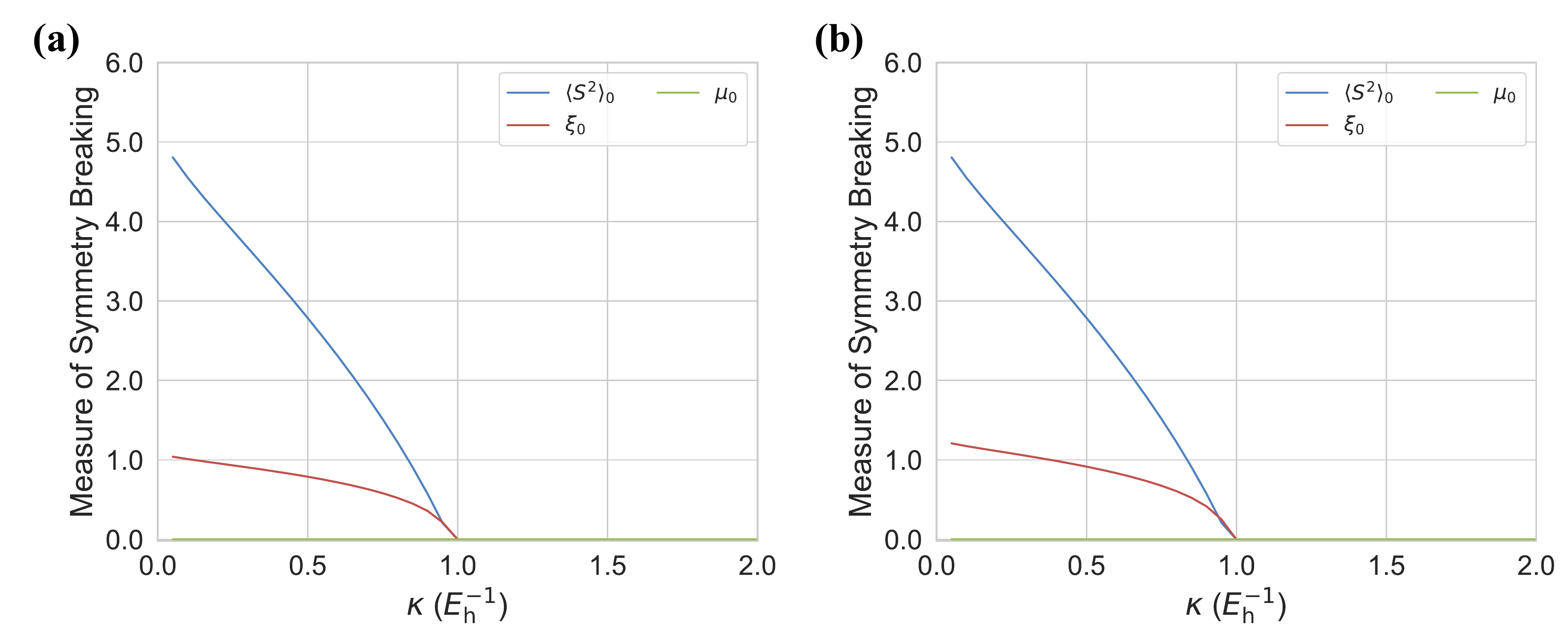}
\caption{\label{fig:c20cuhf}
Measures of symmetry breaking ($\langle S^2\rangle_0$, $\xi_0$, and $\mu_0$) as a function of the regularization strength $\kappa$ for (a) \ce{C20} (C$_\text{i}$) and (b) \ce{C20} (D$_\text{3h}$). It should be noted that these plots are obtained with a scaled correlation energy variant of $\kappa$-OOMP2 (i.e., $\kappa$-S-OOMP2). These quantities do not include correlation corrections: \insertnew{in other words, the plots characterize symmetry-breaking/restoration in the reference determinant}.
}
\end{figure}

There are a total of two cUHF solutions in the C$_\text{i}$ and D$_\text{3h}$ geometries. The nuclear repulsion energies of these molecules differ by 29 m$E_\text{h}$ and the cUHF solutions differ only by 10 $\mu E_\text{h}$. This appearance of electronic degeneracy is similar to the two cGHF solutions of the same molecule. In Figure \ref{fig:c20cuhf}, the three measures of symmetry breaking is shown as a function of $\kappa$ for these two geometries. Clearly, $\mu_0$ is zero at every regularization strength so there is no generalized solution for these geometries. Although there were qualitative differences in $\kappa_c$ between C$_\text{2h}$ and D$_\text{2h}$, C$_\text{i}$ and D$_\text{3h}$ show identical $\kappa_c$ values for $\langle \hat{S}\rangle_0$  and $\xi_0$. The $\kappa_c$ values are 0.95 for both of the symmetries, which is \insertnew{in the artificial symmetry breaking range}. Therefore, this symmetry breaking should also be considered {\it artificial}. 

We emphasize these findings about four different geometries of C$_{20}$ agree with previous studies by others\cite{Parasuk1991,Martin1996b,Paulus2003,Adams1993}: C$_{20}$ is a singlet closed-shell molecule and there is no strong correlation in this molecule as long as the geometry is Jahn-Teller distorted. 

\subsubsection{Dodecahedral (I$_\text{h}$) Structure}
\begin{figure}[h!]
\includegraphics[scale=0.1]{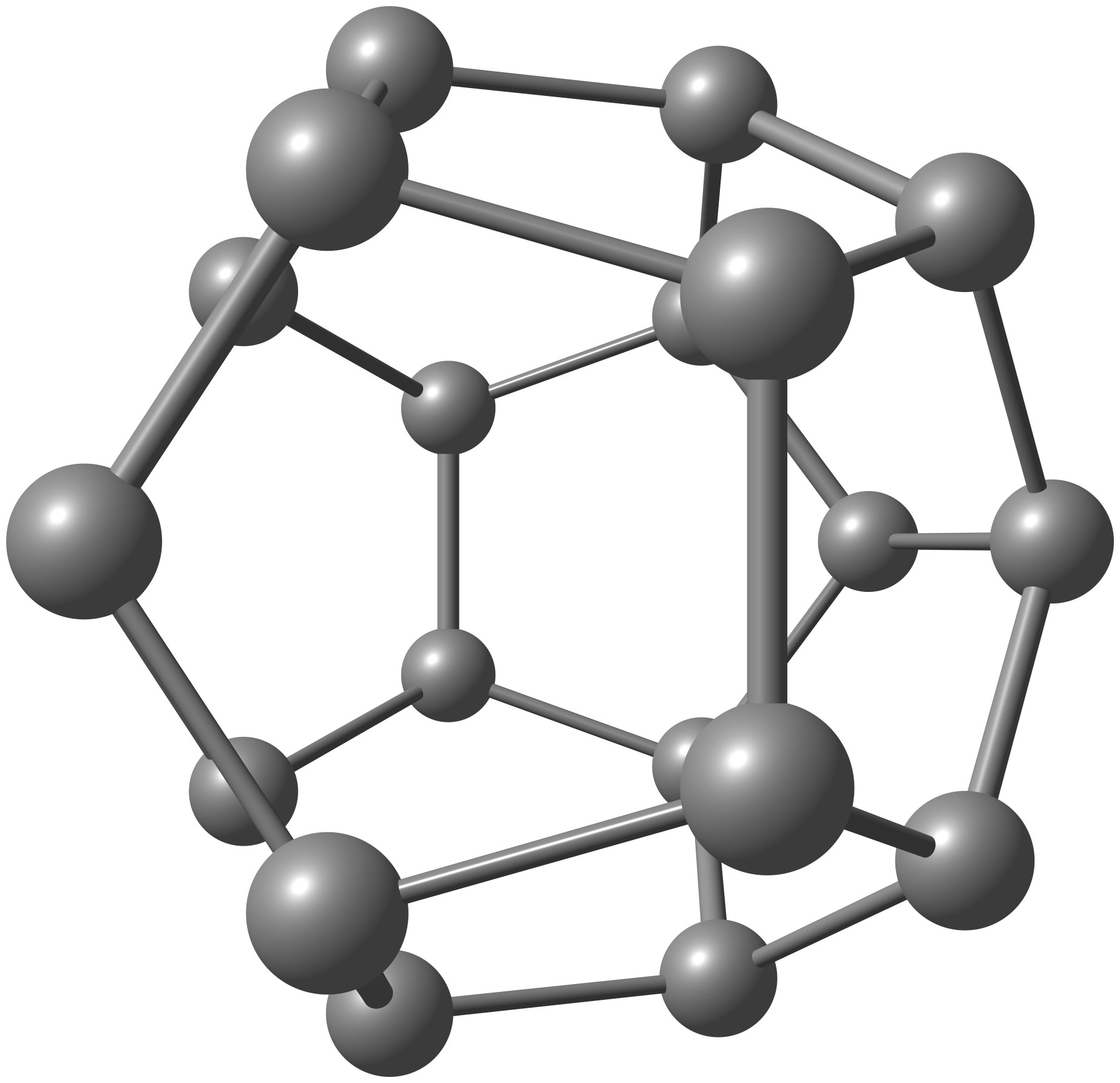}
\caption{The dodecahedral molecular structure of \ce{C20}.
}
\label{geom:c20ih}
\end{figure}
It is possible to force strong correlation by enforcing a higher symmetry.
It is very likely that small fluctuations would be sufficient to break any degeneracies present in a higher symmetry and results into a more stable and lower symmetry geometry. This is the origin of commonly observed Jahn-Teller distortions.
It is very surprising that a dodecahedral geometry (I$_\text{h}$) was found to be the global minimum with cGHF in the work by Jim{\'e}nez-Hoyos et al.\cite{Jimenez-Hoyos2014} We took the geometry from their work (shown in Figure \ref{geom:c20ih}) and ran the same analysis to see how symmetry breaking plays a role in describing the electron correlation of this molecule.

\begin{figure}[h!]
\includegraphics[scale=0.6]{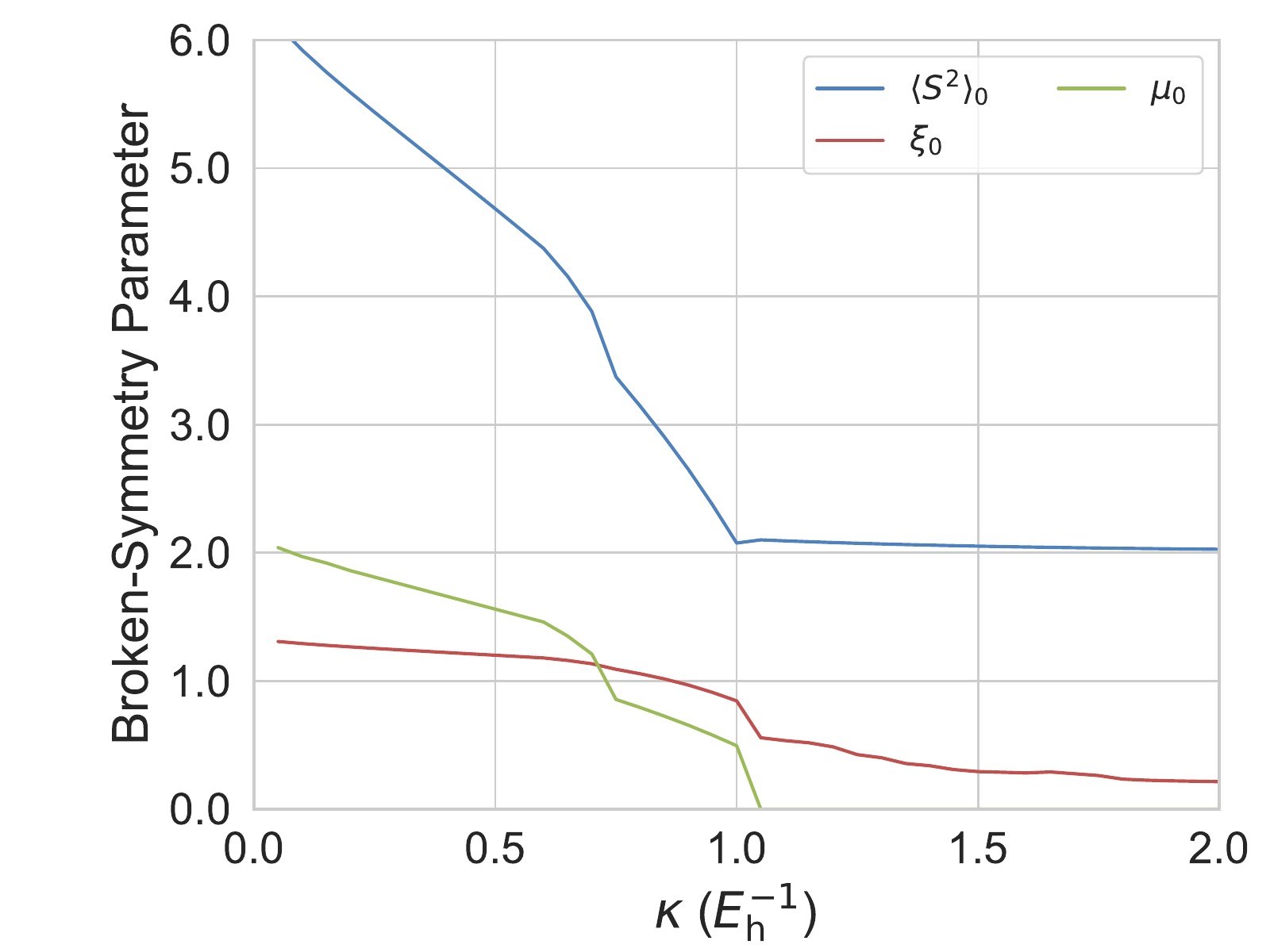}
\caption{\label{fig:c20ih}
Measures of symmetry breaking ($\langle S^2\rangle_0$, $\xi_0$, and $\mu_0$) as a function of the regularization strength $\kappa$ for \ce{C20} with a dodecahedral geometry. It should be noted that these plots are obtained with a scaled correlation energy variant of $\kappa$-OOMP2 (i.e., $\kappa$-S-OOMP2). These quantities do not include correlation corrections: \insertnew{in other words, the plots characterize symmetry-breaking/restoration in the reference determinant}.
}
\end{figure}

In Figure \ref{fig:c20ih}, we present a symmetry breaking landscape of the dodecahedral geometry. It has more structure than other previous cases. There is a discontinuous jump in $\mu_0$ going from $\kappa=1.00$ to $\kappa = 1.05$. This is due to the existence of two distinct low-lying solutions: complex, generalized and complex, unrestricted solutions. The gap between these two competing solutions is controlled by $\kappa$ and around $\kappa = 1.00$ the relative energetics becomes reversed. For $\kappa > 1.00$, cU solutions are found to be the lowest energy solution. 

$\langle \hat{S}\rangle_0$ suggests that a triplet state (with complexity) is the ground state of this system for reasonable $\kappa$ values.
At $\kappa=1.45$, we examined the energy lowering from complexity by comparing the energy between U and cU solutions. We found that these solutions are degenerate as well.

This intricate landscape of symmetry breaking indicates that the system is likely strongly correlated under this geometry. 
Furthermore, a definitive answer to this question was obtained from equation-of-motion spin-flip coupled-cluster with singles and doubles (EOM-SF-CCSD) within the cc-pVDZ basis set starting from an $M_S=1$ $\kappa$-OOMP2 orbitals. The $M_S=1$ unrestricted CCSD calculation has $\langle \hat{S}^2\rangle = 2.02$ which should serve as a good reference state for a subsequent spin-flip calculation. The EOM-SF-CCSD\cite{Krylov2001} calculation yielded a near-exact degeneracy between singlet and triplet states (i.e., near-zero singlet-triplet gap) and this strongly suggests that the system is strongly correlated. This is also consistent with the prediction by molecular orbital theory.\cite{Fan1995}

Therefore, we confirm that the claim by Jim{\'e}nez-Hoyos et al. is correct that \ce{C20} at the I$_\text{h}$ geometry is strongly correlated.
The next question is then whether this geometry is the actual ground state conformation of \ce{C20}. Given the near-zero singlet-triplet gap of the I$_\text{h}$ geometry, \ce{C20} would be quite reactive and unstable. However, experimental findings suggest that \ce{C20} is not as reactive as a pure biradical.\cite{Prinzbach2006} It is likely that the global minimum structure would be one of the Jahn-Teller distorted structures or other structures such as bowl suggested in literature. 
Providing an answer to this question would require geometry optimization with $\kappa$-OOMP2 and this would be an interesting research direction to pursue in the future.

%

\section{Summary}

We have presented an unbiased analysis to determine whether fullerenes, \ce{C20}, \ce{C36}, and \ce{C60}, are strongly correlated. At the Hartree-Fock level, this was already done based on the existence of complex, generalized Hartree-Fock (cGHF) solutions in the work by Jim{\'e}nez-Hoyos et al. As it is common to observe artificial symmetry at the HF level in innocent (i.e., not strongly correlated) systems, we analyzed these solutions beyond the HF level.
This was achieved with three different probes. 

First, we used the recently developed regularized orbital-optimized second-order M{\o}ller-Plesset perturbation theory ($\kappa$-OOMP2) to obtain landscapes of symmetry breaking parameters for spin operators $\hat{S}^2$ and $\hat{S}_\mathbf{n}$ and complex operator $\hat{K}$ as a function of the regularization strength $\kappa$. The critical strength $\kappa_c$ was then used to determine whether a given fullerene is strongly correlated. If $\kappa_c$ is around 1.0, which is far stronger than the optimal $\kappa$ we determined in our previous work, we concluded that the symmetry breaking at the HF level is artificial.

Second, we obtained the singlet-triplet gaps of these fullerenes and quantified strong correlation in them. A singlet-triplet gap measures an energy cost of unpairing an electron pair and this energy cost should be small if the system has strong biradicaloid character. 

Lastly, we studied strong correlation within a minimal basis using two coupled-cluster (CC) methods along with various unregularized OOMP2 methods and $\kappa$-OOMP2. The two methods used in this work are restricted CC with singles and doubles (RCCSD) and CC valence-bond with singles and doubles (CCVB-SD). Based on our previous work,\cite{Lee2017} it is well understood that there is a qualitative difference between RCCSD and CCVB-SD when strong correlation is present. This qualitative difference was probed with natural orbital occupation numbers (NOONs). NOONs from CCVB-SD were in general qualitatively consistent with NOONs from $\kappa$-OOMP2 for the systems we considered in this work. Based on these three independent probes, we reached the following conclusions. 

\insertnew{
\ce{C60} is {\it not strongly correlated} and the symmetry breaking present in its cGHF solution is {\it artificial} based on all three probes. The critical $\kappa$ values for each symmetry breaking is in the range of artificial symmetry breaking. Its singlet-triplet gap is large in both experiments and computations. Furthermore, RCCSD and CCVB-SD show nearly identical behavior. Therefore, the molecule should be described as a closed shell molecule. This is not surprising due to the fact that it is electron paramagnetic resonance silent and found stable in experiments. This then contradicts with the conclusion drawn by Jim{\'e}nez-Hoyos and co-workers. 
\\
\indent On the other hand, \ce{C36} within the D$_\text{6h}$ point group is strongly correlated. In particular, it is a singlet biradicaloid where unrestricted treatment in conjunction with Yamaguchi's approximate projection can be used to obtain a qualitatively (and even quantitatively) correct answer. The cGHF solution found by Jim{\'e}nez-Hoyos et al is likely an artifact due to the limited treatment of electron correlation: the complex and $\langle \hat{S}_\mathbf{n}\rangle$ symmetry breakings are {\it artificial}. However, since breaking $\langle \hat{S}^2 \rangle$ symmetry was found to be {\it essential}, we conclude that this system is strongly correlated. The singlet-triplet gap of this molecule was found to be small and a qualitative difference between the NOONs of RCCSD and CCVB-SD was observed. All three probes indicate that the symmetry breaking in \ce{C36} is {\it essential} and \ce{C36} is strongly correlated.
\\
\indent Lastly, we applied the first probe to the smallest fullerene \ce{C20}. A total of five geometries of \ce{C20} considered in this work all exhibit symmetry breaking. All of the Jahn-Teller distorted geometries (C$_\text{2h}$, D$_\text{2h}$, C$_\text{i}$, and D$_\text{3h}$) were found not to be strongly correlated and the underlying Hartree-Fock symmetry breaking is therefore {\it artificial}. On the other hand, the fully symmetric dodecahedral geometry (I$_\text{h}$) was found to be strongly correlated. In particular, it exhibits a near-zero singlet-triplet gap. 
}

It is the central message of this paper that not every symmetry breaking in a HF solution indicates strong correlation. Many symmetry breakings in Hartree-Fock are simply due to its lack of dynamic correlation, which can be properly recovered by perturbation theory such as MP2. 
\iterate{$\kappa$-OOMP2 emerges as a method that captures dynamic correlation, and attenuates all strong correlation. As a result, $\kappa$-OOMP2 removes artificial symmetry breaking in its reference determinant. However, essential symmetry breaking due to lack of static (or strong) correlation remains.}
The analyses we presented here can be used to probe strong correlation in numerous chemical systems especially when one is unsure about using single-reference methods.

\section{Acknowledgement}
We thank Dave Small for bringing our attention to this problem in the beginning, Eloy Ramos and Luke Bertels for stimulating discussions, and Evgeny Epifanovsky and Xintian Feng for useful discussions on efficient implementations. J.L. thanks Soojin Lee for consistent encouragement. This work was supported by the Director, Office of Science, Office of Basic Energy Sciences of the U.S. Department of Energy under contract No. DE-ACO2-05CH11231.
\section{Appendix}
\beginsupplement
\subsection{Non-Collinearity Test of MP1 Wavefunctions}
In order to perform the non-collinearity test on an MP1 wavefunction, one needs first-order corrections to $\langle\hat{{S}_i}\rangle$ and  $\langle\hat{S}_i\hat{S}_j\rangle$ where $i,j \in \{x,y,z\}$.
The first-order correction to $\langle\hat{O}\rangle$ for an operator $\hat{O}$ is defined as follows:
\begin{equation}
\langle\hat{O}\rangle_1 = \langle{\Psi_1|\hat{O}|\Psi_0}\rangle + \langle{\Psi_0|\hat{O}|\Psi_1}\rangle.
\end{equation}
This can be derived from the derivative with respect to $\lambda$ of the first-order MP energy expression $E^\text{(1)}$ with a modified Hamiltonian, $\hat {H} + \lambda \hat{O}$.
We enumerate the expectation value of each spin operator using this formula.
For $\langle \hat{S}^2 \rangle$, one may use the following identity:
\begin{equation}
\hat{S}^2 = \hat{S}_z + \hat{S}_z^2 + \hat{S}_{-}\hat{S}_{+},
\end{equation}
where
\begin{equation}
\hat{S}_z = \frac12\sum_p \left(\hat{a}_{p_\alpha}^\dagger\hat{a}_{p_\alpha}-\hat{a}_{p_\beta}^\dagger\hat{a}_{p_\beta}\right)
\end{equation}
\begin{equation}
\hat{S}_+ = \hat{S}_x + i \hat{S}_y = \sum_p \hat{a}_{p_\alpha}^\dagger\hat{a}_{p_\beta}
\end{equation}
\begin{equation}
\hat{S}_- = \hat{S}_x - i \hat{S}_y = \sum_p \hat{a}_{p_\beta}^\dagger\hat{a}_{p_\alpha}
\end{equation}
One can evaluate $\langle \hat{S}_i \hat{S}_j \rangle$ for $i, j \in \{x,y\}$ using 
$\langle \hat{S}_i \hat{S}_j \rangle$ for $i, j \in \{+,-\}$. 
We choose to work with these ladder operators for simplicity.

With a cGHF reference, the zeroth order expectation values are as follows:\cite{Small2015,Cassam-Chenai2015}
\begin{align}
\langle
\hat{S}_z
\rangle_0
=&
\frac12
\sum_i
\left(
\langle i_\alpha | i_\alpha \rangle
-\langle i_\beta | i_\beta \rangle
\right)\\ 
\langle
\hat{S_+}
\rangle_0 
=& 
\langle
\hat{S_-}
\rangle_0^*
=
\sum_i \langle i_\alpha | i_\beta \rangle\\
\nonumber
\langle
\hat{S}_z^2
\rangle_0
=&
\frac14
\sum_i
\left(
\langle i_\alpha | i_\alpha \rangle
+\langle i_\beta | i_\beta \rangle
\right)\\ \nonumber
&+
\frac14
\sum_{ij}\sum_{\sigma\in\{\alpha,\beta\}}
\left(
\langle i_\sigma | i_\sigma \rangle \langle j_\sigma | j_\sigma \rangle
- \langle i_\sigma | j_\sigma \rangle \langle j_\sigma | i_\sigma \rangle
\right)
\\
&+\frac14
\sum_{ij}
\left(
\langle i_\beta|j_\beta\rangle \langle j_\alpha|i_\alpha\rangle
-\langle i_\alpha|i_\alpha\rangle \langle j_\beta|j_\beta\rangle
+ \text{h.c.}
\right)
\\
\langle
\hat{S}_-
\hat{S}_+
\rangle_0
=&
\sum_i \langle i_\beta|i_\beta\rangle
+\sum_{ij}
\left(
\langle i_\alpha|i_\beta\rangle \langle j_\beta|j_\alpha\rangle
-\langle i_\beta|j_\alpha\rangle \langle j_\alpha|i_\beta\rangle
\right)
 \\
 \langle
 \hat{S}_+
 \hat{S}_-
 \rangle_0
 =&
 \sum_i \langle i_\alpha|i_\alpha\rangle
 +\sum_{ij}
 \left(
 \langle i_\alpha|i_\beta\rangle \langle j_\beta|j_\alpha\rangle
 -\langle i_\beta|j_\alpha\rangle \langle j_\alpha|i_\beta\rangle
 \right)
\\
\langle
\hat{S}_-
\hat{S}_-
\rangle_0
= \langle
\hat{S}_+
\hat{S}_+
\rangle_0^*
=&
\sum_{ij}
\left(
\langle i_\beta|i_\alpha\rangle \langle j_\beta|j_\alpha\rangle
-\langle j_\beta|i_\alpha\rangle \langle i_\beta|j_\alpha\rangle
\right)
\\ \nonumber
\langle
\hat{S}_+
\hat{S}_z
\rangle_0
=
\langle
\hat{S}_z
\hat{S}_-
\rangle_0^*
=&
-\frac12
\sum_i \langle i_\alpha | i_\beta \rangle
+
\frac12
\sum_{ij}
\left(
\langle i_\alpha | i_\beta \rangle \langle j_\alpha | j_\alpha \rangle
- \langle j_\alpha | i_\beta \rangle \langle i_\alpha | j_\alpha \rangle
\right)\\ 
& -\frac12
\sum_{ij}
\left(
\langle i_\alpha | i_\beta \rangle \langle j_\beta | j_\beta \rangle
- \langle i_\beta | j_\beta \rangle \langle j_\alpha | i_\beta \rangle
\right)
\\ \nonumber
\langle
\hat{S}_-
\hat{S}_z
\rangle_0
=
\langle
\hat{S}_z
\hat{S}_+
\rangle_0^*
=&
\frac12
\sum_i \langle i_\beta | i_\alpha \rangle
+
\frac12
\sum_{ij}
\left(
\langle i_\beta | i_\alpha \rangle \langle j_\alpha | j_\alpha \rangle
- \langle i_\alpha | j_\alpha \rangle \langle j_\beta | i_\alpha \rangle
\right)\\ 
& 
-\frac12
\sum_{ij}
\left(
\langle i_\beta | i_\alpha \rangle \langle j_\beta | j_\beta \rangle
- \langle i_\beta | j_\beta \rangle \langle j_\beta | i_\alpha \rangle
\right)
\end{align}
where
we used the fact that each orbital is of the spinor form in Eq. \eqref{eq:spinor} and
we define
\begin{equation}
\langle p_{\sigma_1} | q_{\sigma_2} \rangle
= \int_\mathbf{r}
(\phi^{\sigma_1}_p(\mathbf{r}))^*
\phi^{\sigma_2}_q(\mathbf{r}).
\label{eq:1eovl}
\end{equation}
We note that there is no spin integration in Eq. \eqref{eq:1eovl}.
These are used to compute the covariance matrix $A_{ij} = \langle \hat{S}_i \hat{S}_j \rangle - \langle \hat{S}_i \rangle\langle \hat{S}_j\rangle$. As noted before, the eigenspectrum of $\mathbf A$ determines whether the GHF wavefunction is genuinely non-collinear. The wavefunction is collinear if and only if there is a zero eigenmode.

Similarly, the first-order corrections to these expectation values can be obtained from
\begin{equation}
\langle\hat{O}\rangle_1
= \frac14 \sum_{ijab}
\left( t_{ij}^{ab}\right)^*
\langle\Psi_{ij}^{ab}|\hat{O}|\Psi_0\rangle 
+ 
\frac14 \sum_{ijab}
\langle\Psi_0|\hat{O}|\Psi_{ij}^{ab}\rangle t_{ij}^{ab}
\end{equation}
This can be easily computed as follows:
\begin{align}
\langle\hat{S}_z\rangle_1
&= 
\langle\hat{S}_+\rangle_1
=
\langle\hat{S}_-\rangle_1
=
0
\end{align}
\begin{align}
\nonumber
\langle\hat{S}_z^2\rangle_1
&=
\frac{1}{4} \sum_{\substack{i<j\\a<b}}  \left( t_{ij}^{ab}\right)^*
\sum_{\sigma\in\{\alpha,\beta\}}
\left(
2\langle a_\sigma | i_\sigma \rangle \langle b_\sigma | j_\sigma \rangle
-2\langle a_\sigma | j_\sigma \rangle \langle b_\sigma | i_\sigma \rangle
\right)\\\nonumber
&+ \frac{1}{4} \sum_{\substack{i<j\\a<b}}  \left( t_{ij}^{ab}\right)^*
(
-2 \langle a_\alpha | i_\alpha \rangle \langle b_\beta | j_\beta\rangle
+2 \langle a_\beta | j_\beta \rangle \langle b_\alpha | i_\alpha\rangle\\
&-2 \langle a_\beta | i_\beta \rangle \langle b_\alpha | j_\alpha\rangle
+2 \langle a_\alpha | j_\alpha \rangle \langle b_\beta | i_\beta\rangle
)
+ \text{h.c.}
\end{align}
\begin{align}\nonumber
\langle\hat{S}_-\hat{S}_+\rangle_1
= \langle\hat{S}_+\hat{S}_-\rangle_1
&=
\sum_{\substack{i<j\\a<b}}
(t_{ij}^{ab})^*
(
\langle a_\beta | i_\alpha \rangle \langle b_\alpha | j_\beta \rangle
-\langle a_\alpha | j_\beta \rangle \langle b_\beta | i_\alpha \rangle\\
&+\langle a_\alpha | i_\beta \rangle \langle b_\beta | j_\alpha \rangle
-\langle a_\beta | j_\alpha \rangle \langle b_\alpha | i_\beta \rangle
) + \text{h.c.}
\end{align}
\begin{align}\nonumber
\langle\hat{S}_-\hat{S}_-\rangle_1  = \langle\hat{S}_+\hat{S}_+\rangle_1 ^*
&= 
\sum_{\substack{i<j\\a<b}}
(t_{ij}^{ab})^*
\left(
2 \langle a_\beta | i_\alpha \rangle \langle b_\beta | j_\alpha \rangle
- 2 \langle a_\beta | j _\alpha \rangle \langle b_\beta | i_\alpha \rangle
\right) \\
&+ 
\sum_{\substack{i<j\\a<b}}
(t_{ij}^{ab})
\left(
2 \langle i_\beta | a_\alpha \rangle \langle j_\beta |b_\alpha\rangle
-2 \langle j_\beta | a_\alpha \rangle \langle i_\beta |b_\alpha\rangle
\right)
\end{align}
\begin{align}\nonumber
\langle\hat{S}_+\hat{S}_z\rangle_1 = \langle\hat{S}_z\hat{S}_-\rangle_1^* &=
\frac12
\sum_{\substack{i<j\\a<b}}
(t_{ij}^{ab})^*
(
\langle a_\alpha| i_\beta\rangle \langle b_\alpha|j_\alpha \rangle 
- \langle a_\alpha | j_\alpha\rangle\langle b_\alpha | i_\beta\rangle\\\nonumber
&+ \langle a_\alpha | i_\alpha\rangle\langle b_\alpha | j_\beta\rangle 
- \langle a_\alpha | j_\beta\rangle\langle b_\alpha | i_\alpha \rangle
- \langle a_\alpha| i_\beta\rangle \langle b_\beta| j_\beta\rangle\\ \nonumber
&+ \langle a_\beta | j_\beta\rangle \langle b_\alpha | i_\beta\rangle
- \langle a_\beta | i_\beta\rangle \langle b_\alpha| j_\beta\rangle
+ \langle a_\alpha| j_\beta\rangle \langle b_\beta | i_\beta\rangle
)\\\nonumber
&+
\frac12
\sum_{\substack{i<j\\a<b}}
t_{ij}^{ab}
(
 \langle i_\alpha| a_\alpha \rangle \langle j_\alpha | b_\beta\rangle
-\langle j_\alpha| a_\beta  \rangle \langle i_\alpha | b_\alpha\rangle\\\nonumber
&+\langle i_\alpha| a_\beta  \rangle \langle j_\alpha | b_\alpha\rangle
-\langle j_\alpha| a_\alpha \rangle \langle i_\alpha | b_\beta\rangle
-\langle i_\beta| a_\beta \rangle \langle j_\alpha | b_\beta\rangle\\
&+\langle j_\alpha| a_\beta  \rangle \langle i_\beta | b_\beta\rangle
-\langle i_\alpha| a_\beta  \rangle \langle j_\beta | b_\beta\rangle
+\langle j_\beta| a_\alpha \rangle \langle i_\alpha | b_\beta\rangle
)
\end{align}
\begin{align}\nonumber
\langle\hat{S}_-\hat{S}_z\rangle_1 = \langle\hat{S}_z\hat{S}_+\rangle_1^* &=
\frac12
\sum_{\substack{i<j\\a<b}}
(t_{ij}^{ab})^*
(
 \langle a_\beta | i_\alpha \rangle \langle b_\alpha | j_\alpha\rangle
-\langle a_\alpha| j_\alpha \rangle \langle b_\beta  | i_\alpha\rangle\\\nonumber
&+\langle a_\alpha| i_\alpha \rangle \langle b_\beta  | j_\alpha\rangle
-\langle a_\beta | j_\alpha \rangle \langle b_\alpha | i_\alpha\rangle
-\langle a_\beta | i_\alpha \rangle \langle b_\beta  | j_\beta \rangle\\ \nonumber
&+\langle a_\beta | j_\beta  \rangle \langle b_\beta  | i_\alpha\rangle
-\langle a_\beta | i_\beta  \rangle \langle b_\beta  | j_\alpha\rangle
+\langle a_\beta | j_\alpha \rangle \langle b_\beta  | i_\beta \rangle
)\\\nonumber
&+
\frac12
\sum_{\substack{i<j\\a<b}}
t_{ij}^{ab}
(
 \langle i_\alpha | a_\alpha \rangle \langle j_\beta | b_\beta\rangle
 -\langle j_\beta | a_\alpha \rangle \langle i_\alpha | b_\alpha\rangle\\\nonumber
 &+\langle i_\beta | a_\alpha \rangle \langle j_\alpha | b_\alpha\rangle
 -\langle j_\alpha | a_\alpha \rangle \langle i_\beta | b_\alpha\rangle
 -\langle i_\beta | a_\beta \rangle \langle j_\beta | b_\alpha\rangle\\
 &+\langle j_\beta | a_\alpha \rangle \langle i_\beta | b_\beta\rangle
 -\langle i_\beta | a_\alpha \rangle \langle j_\beta | b_\beta\rangle
 +\langle j_\beta | a_\beta \rangle \langle i_\beta | b_\alpha\rangle
)
\end{align}
%
%
%
%
\subsection{Complex Generalized HF}
The variation in the energy expression reads
\begin{align}\nonumber
\delta E &= 
\sum_{ia}
\left(-h_{ai}\delta\Theta_{ia} - h_{ia}\delta\Theta^*_{ia}\right)
-\frac12 \sum_{ija}
\left(
\Braket{ij||aj}\delta\Theta_{ia}^*
+\Braket{ij||ia}\delta\Theta_{ja}^*
+\Braket{aj||ij}\delta\Theta_{ia}
+\Braket{ia||ij}\delta\Theta_{ja}
\right)
\end{align}
where $\delta\Theta_{ia}$ is an infinitesimal orbital rotation.
This energy variation can be used to compute orbital gradient and similarly orbital hessian.
\subsubsection{Orbital Gradient}
We compute the gradient of $E$ with respect to the real and imaginary part of $\mat \Theta$,
$$
\frac{\partial E}{\partial \text{Re}(\Theta_{ia})} = -\left[h_{ai} + J_{ai} -K_{ai}\right] + \text{h.c.}
=
-h_{ai}
-\sum_k \langle ak || ik \rangle
-h_{ia}
-\sum_k \langle ik || ak \rangle
$$
$$
\frac{\partial E}{\partial i\text{Im}(\Theta_{ia})} = -\left[h_{ai} + J_{ai} -K_{ai}\right] - \text{h.c.}
=
-h_{ai}
-\sum_k \langle ak || ik \rangle
+h_{ia}
+\sum_k \langle ik || ak \rangle
$$
\subsubsection{Orbital Hessian}
The variation of orbital gradient reads
\begin{align}\nonumber
\delta \frac{\partial E}{\partial \text{Re}(\Theta_{ia})} 
= &
\sum_j 
\left(
-h_{ji}\delta\Theta_{ja}^*
+h_{ab}\delta\Theta_{bi}^*
\right)
+
\sum_{k}\left(
-\sum_j\delta\Theta_{aj}^* \langle jk || ik \rangle
+ \sum_b\delta\Theta_{bi}^* \langle ak || bk \rangle
\right)\\
&+\sum_{kb}\delta \Theta_{bk}^* \langle ak || ib\rangle
+\sum_{kb}\delta \Theta_{bk} \langle ab || ik\rangle
+ \text{h.c.}
\end{align}
and
\begin{align}\nonumber
\delta \frac{\partial E}{\partial i\text{Im}(\Theta_{ia})} 
= &
\sum_j 
\left(
-h_{ji}\delta\Theta_{ja}^*
+h_{ab}\delta\Theta_{bi}^*
\right)
+
\sum_{k}\left(
-\sum_j\delta\Theta_{aj}^* \langle jk || ik \rangle
+ \sum_b\delta\Theta_{bi}^* \langle ak || bk \rangle
\right)\\
&+\sum_{kb} \delta\Theta_{bk}^* \langle ak || ib\rangle
+\sum_{kb}\delta \Theta_{bk} \langle ab || ik\rangle
- \text{h.c.}
\end{align}

These are then used to obtain orbital hessian:
\begin{align}
\frac{\partial^2 E}{\partial \text{Re}(\Theta_{ia})\partial \text{Re}(\Theta_{jb})} 
= &
\left[
-\delta_{ab}\left(
h_{ji} + \sum_{k}\langle jk||ik\rangle 
\right)
+\delta_{ij}
\left(
h_{ab} + \sum_{k}\langle ak||bk\rangle 
\right)
+ \langle aj || ib \rangle
+ \langle ab || ij \rangle
\right]
+ h.c
\end{align}
\begin{align}\nonumber
\frac{\partial^2 E}{\partial i\text{Im}(\Theta_{ia})\partial i\text{Im}(\Theta_{jb})} 
&= 
\left[
\delta_{ab}\left(
h_{ji} + \sum_{k}\langle jk||ik\rangle 
\right)
-\delta_{ij}
\left(
h_{ab} + \sum_{k}\langle ak||bk\rangle 
\right)
- \langle aj || ib \rangle
+ \langle ab || ij \rangle
\right]
+ h.c\\
&=
-\frac{\partial^2 E}{\partial \text{Im}(\Theta_{ia})\partial \text{Im}(\Theta_{jb})} 
\end{align}
\begin{align}
\frac{\partial^2 E}{\partial \text{Im}(\Theta_{ia})\partial \text{Re}(\Theta_{jb})} 
= &
i
\left[
- \langle aj || ib \rangle
+ \langle ab || ij \rangle
- h.c.
\right]
\end{align}
\begin{align}
\frac{\partial^2 E}{\partial \text{Re}(\Theta_{ia})\partial \text{Im}(\Theta_{jb})} 
= &
-i
\left[
- \langle aj || ib \rangle
+ \langle ab || ij \rangle
- h.c.
\right]
\end{align}
\bibliography{goomp2_ms}

\providecommand{\latin}[1]{#1}
\providecommand*\mcitethebibliography{\thebibliography}
\csname @ifundefined\endcsname{endmcitethebibliography}
  {\let\endmcitethebibliography\endthebibliography}{}
\begin{mcitethebibliography}{102}
\providecommand*\natexlab[1]{#1}
\providecommand*\mciteSetBstSublistMode[1]{}
\providecommand*\mciteSetBstMaxWidthForm[2]{}
\providecommand*\mciteBstWouldAddEndPuncttrue
  {\def\EndOfBibitem{\unskip.}}
\providecommand*\mciteBstWouldAddEndPunctfalse
  {\let\EndOfBibitem\relax}
\providecommand*\mciteSetBstMidEndSepPunct[3]{}
\providecommand*\mciteSetBstSublistLabelBeginEnd[3]{}
\providecommand*\EndOfBibitem{}
\mciteSetBstSublistMode{f}
\mciteSetBstMaxWidthForm{subitem}{(\alph{mcitesubitemcount})}
\mciteSetBstSublistLabelBeginEnd
  {\mcitemaxwidthsubitemform\space}
  {\relax}
  {\relax}

\bibitem[Pulay and Hamilton(1988)Pulay, and Hamilton]{Pulay1988}
Pulay,~P.; Hamilton,~T.~P. \emph{J. Chem. Phys.} \textbf{1988}, \emph{88},
  4926--4933\relax
\mciteBstWouldAddEndPuncttrue
\mciteSetBstMidEndSepPunct{\mcitedefaultmidpunct}
{\mcitedefaultendpunct}{\mcitedefaultseppunct}\relax
\EndOfBibitem
\bibitem[Bofill and Pulay(1989)Bofill, and Pulay]{Bofill1989}
Bofill,~J.~M.; Pulay,~P. \emph{J. Chem. Phys.} \textbf{1989}, \emph{90},
  3637--3646\relax
\mciteBstWouldAddEndPuncttrue
\mciteSetBstMidEndSepPunct{\mcitedefaultmidpunct}
{\mcitedefaultendpunct}{\mcitedefaultseppunct}\relax
\EndOfBibitem
\bibitem[Coulson and Fischer(1949)Coulson, and Fischer]{Coulson1949}
Coulson,~C.; Fischer,~I. \emph{Philos. Mag.} \textbf{1949}, \emph{40},
  386--393\relax
\mciteBstWouldAddEndPuncttrue
\mciteSetBstMidEndSepPunct{\mcitedefaultmidpunct}
{\mcitedefaultendpunct}{\mcitedefaultseppunct}\relax
\EndOfBibitem
\bibitem[Jackels and Davidson(1976)Jackels, and Davidson]{Jackels1976}
Jackels,~C.~F.; Davidson,~E.~R. \emph{J. Chem. Phys.} \textbf{1976}, \emph{64},
  2908--2917\relax
\mciteBstWouldAddEndPuncttrue
\mciteSetBstMidEndSepPunct{\mcitedefaultmidpunct}
{\mcitedefaultendpunct}{\mcitedefaultseppunct}\relax
\EndOfBibitem
\bibitem[Davidson and Borden(1983)Davidson, and Borden]{Davidson1983}
Davidson,~E.~R.; Borden,~W.~T. \emph{J. Phys. Chem.} \textbf{1983}, \emph{87},
  4783--4790\relax
\mciteBstWouldAddEndPuncttrue
\mciteSetBstMidEndSepPunct{\mcitedefaultmidpunct}
{\mcitedefaultendpunct}{\mcitedefaultseppunct}\relax
\EndOfBibitem
\bibitem[Andrews \latin{et~al.}(1991)Andrews, Jayatilaka, Bone, Handy, and
  Amos]{Andrews1991}
Andrews,~J.~S.; Jayatilaka,~D.; Bone,~R.~G.; Handy,~N.~C.; Amos,~R.~D.
  \emph{Chem. Phys. Lett.} \textbf{1991}, \emph{183}, 423--431\relax
\mciteBstWouldAddEndPuncttrue
\mciteSetBstMidEndSepPunct{\mcitedefaultmidpunct}
{\mcitedefaultendpunct}{\mcitedefaultseppunct}\relax
\EndOfBibitem
\bibitem[Ayala and Schlegel(1998)Ayala, and Schlegel]{Ayala1998}
Ayala,~P.~Y.; Schlegel,~H.~B. \emph{J. Chem. Phys.} \textbf{1998}, \emph{108},
  7560\relax
\mciteBstWouldAddEndPuncttrue
\mciteSetBstMidEndSepPunct{\mcitedefaultmidpunct}
{\mcitedefaultendpunct}{\mcitedefaultseppunct}\relax
\EndOfBibitem
\bibitem[McLean \latin{et~al.}(1985)McLean, Lengsfield, Pacansky, and
  Ellinger]{McLean1985}
McLean,~A.~D.; Lengsfield,~B.~H.; Pacansky,~J.; Ellinger,~Y. \emph{J. Chem.
  Phys.} \textbf{1985}, \emph{83}, 3567--3576\relax
\mciteBstWouldAddEndPuncttrue
\mciteSetBstMidEndSepPunct{\mcitedefaultmidpunct}
{\mcitedefaultendpunct}{\mcitedefaultseppunct}\relax
\EndOfBibitem
\bibitem[Sherrill \latin{et~al.}(1999)Sherrill, Lee, and
  Head-Gordon]{Sherrill1999}
Sherrill,~C.; Lee,~M.~S.; Head-Gordon,~M. \emph{Chem. Phys. Lett.}
  \textbf{1999}, \emph{302}, 425--430\relax
\mciteBstWouldAddEndPuncttrue
\mciteSetBstMidEndSepPunct{\mcitedefaultmidpunct}
{\mcitedefaultendpunct}{\mcitedefaultseppunct}\relax
\EndOfBibitem
\bibitem[Crawford and Stanton(2000)Crawford, and Stanton]{Crawford2000}
Crawford,~T.~D.; Stanton,~J.~F. \emph{J. Chem. Phys.} \textbf{2000},
  \emph{112}, 7873\relax
\mciteBstWouldAddEndPuncttrue
\mciteSetBstMidEndSepPunct{\mcitedefaultmidpunct}
{\mcitedefaultendpunct}{\mcitedefaultseppunct}\relax
\EndOfBibitem
\bibitem[Paldus and Thiamov{\'{a}}(2007)Paldus, and Thiamov{\'{a}}]{Paldus2007}
Paldus,~J.; Thiamov{\'{a}},~G. \emph{J. Math. Chem.} \textbf{2007}, \emph{44},
  88--120\relax
\mciteBstWouldAddEndPuncttrue
\mciteSetBstMidEndSepPunct{\mcitedefaultmidpunct}
{\mcitedefaultendpunct}{\mcitedefaultseppunct}\relax
\EndOfBibitem
\bibitem[Small \latin{et~al.}(2004)Small, Zaitsev, Jung, Rosokha, Head-Gordon,
  and Kochi]{Small2004}
Small,~D.; Zaitsev,~V.; Jung,~Y.; Rosokha,~S.~V.; Head-Gordon,~M.; Kochi,~J.~K.
  \emph{J. Am. Chem. Soc.} \textbf{2004}, \emph{126}, 13850--13858, PMID:
  15493946\relax
\mciteBstWouldAddEndPuncttrue
\mciteSetBstMidEndSepPunct{\mcitedefaultmidpunct}
{\mcitedefaultendpunct}{\mcitedefaultseppunct}\relax
\EndOfBibitem
\bibitem[Small \latin{et~al.}(2005)Small, Rosokha, Kochi, and
  Head-Gordon]{Small2005}
Small,~D.; Rosokha,~S.~V.; Kochi,~J.~K.; Head-Gordon,~M. \emph{J. Phys. Chem.
  A} \textbf{2005}, \emph{109}, 11261--11267, PMID: 16331910\relax
\mciteBstWouldAddEndPuncttrue
\mciteSetBstMidEndSepPunct{\mcitedefaultmidpunct}
{\mcitedefaultendpunct}{\mcitedefaultseppunct}\relax
\EndOfBibitem
\bibitem[Lochan and Head-Gordon(2007)Lochan, and Head-Gordon]{Lochan2007}
Lochan,~R.~C.; Head-Gordon,~M. \emph{J. Chem. Phys.} \textbf{2007}, \emph{126},
  164101\relax
\mciteBstWouldAddEndPuncttrue
\mciteSetBstMidEndSepPunct{\mcitedefaultmidpunct}
{\mcitedefaultendpunct}{\mcitedefaultseppunct}\relax
\EndOfBibitem
\bibitem[Lykos and Pratt(1963)Lykos, and Pratt]{Lykos1963}
Lykos,~P.; Pratt,~G.~W. \emph{Rev. Mod. Phys.} \textbf{1963}, \emph{35},
  496--501\relax
\mciteBstWouldAddEndPuncttrue
\mciteSetBstMidEndSepPunct{\mcitedefaultmidpunct}
{\mcitedefaultendpunct}{\mcitedefaultseppunct}\relax
\EndOfBibitem
\bibitem[Farnell \latin{et~al.}(1983)Farnell, Pople, and Radom]{Farnell1983}
Farnell,~L.; Pople,~J.~A.; Radom,~L. \emph{J. Phys. Chem.} \textbf{1983},
  \emph{87}, 79--82\relax
\mciteBstWouldAddEndPuncttrue
\mciteSetBstMidEndSepPunct{\mcitedefaultmidpunct}
{\mcitedefaultendpunct}{\mcitedefaultseppunct}\relax
\EndOfBibitem
\bibitem[Nobes \latin{et~al.}(1987)Nobes, Pople, Radom, Handy, and
  Knowles]{Nobes1987}
Nobes,~R.~H.; Pople,~J.~A.; Radom,~L.; Handy,~N.~C.; Knowles,~P.~J. \emph{Chem.
  Phys. Lett.} \textbf{1987}, \emph{138}, 481--485\relax
\mciteBstWouldAddEndPuncttrue
\mciteSetBstMidEndSepPunct{\mcitedefaultmidpunct}
{\mcitedefaultendpunct}{\mcitedefaultseppunct}\relax
\EndOfBibitem
\bibitem[Gill \latin{et~al.}(1988)Gill, Pople, Radom, and Nobes]{Gill1988}
Gill,~P. M.~W.; Pople,~J.~A.; Radom,~L.; Nobes,~R.~H. \emph{J. Chem. Phys.}
  \textbf{1988}, \emph{89}, 7307--7314\relax
\mciteBstWouldAddEndPuncttrue
\mciteSetBstMidEndSepPunct{\mcitedefaultmidpunct}
{\mcitedefaultendpunct}{\mcitedefaultseppunct}\relax
\EndOfBibitem
\bibitem[Jensen(1990)]{Jensen1990}
Jensen,~F. \emph{Chem. Phys. Lett.} \textbf{1990}, \emph{169}, 519--528\relax
\mciteBstWouldAddEndPuncttrue
\mciteSetBstMidEndSepPunct{\mcitedefaultmidpunct}
{\mcitedefaultendpunct}{\mcitedefaultseppunct}\relax
\EndOfBibitem
\bibitem[Yamanaka \latin{et~al.}(1994)Yamanaka, Okumura, Nakano, and
  Yamaguchi]{Yamanaka1994}
Yamanaka,~S.; Okumura,~M.; Nakano,~M.; Yamaguchi,~K. \emph{J. Mol. Struct.}
  \textbf{1994}, \emph{310}, 205--218\relax
\mciteBstWouldAddEndPuncttrue
\mciteSetBstMidEndSepPunct{\mcitedefaultmidpunct}
{\mcitedefaultendpunct}{\mcitedefaultseppunct}\relax
\EndOfBibitem
\bibitem[Meyer(1976)]{Meyer1976}
Meyer,~W. \emph{J. Chem. Phys.} \textbf{1976}, \emph{64}, 2901--2907\relax
\mciteBstWouldAddEndPuncttrue
\mciteSetBstMidEndSepPunct{\mcitedefaultmidpunct}
{\mcitedefaultendpunct}{\mcitedefaultseppunct}\relax
\EndOfBibitem
\bibitem[Shavitt \latin{et~al.}(1976)Shavitt, Rosenberg, and
  Palalikit]{Shavitt1976}
Shavitt,~I.; Rosenberg,~B.~J.; Palalikit,~S. \emph{Int. J. Quantum Chem.}
  \textbf{1976}, \emph{10}, 33--46\relax
\mciteBstWouldAddEndPuncttrue
\mciteSetBstMidEndSepPunct{\mcitedefaultmidpunct}
{\mcitedefaultendpunct}{\mcitedefaultseppunct}\relax
\EndOfBibitem
\bibitem[Dykstra(1977)]{Dykstra1977}
Dykstra,~C.~E. \emph{Chem. Phys. Lett.} \textbf{1977}, \emph{45},
  466--469\relax
\mciteBstWouldAddEndPuncttrue
\mciteSetBstMidEndSepPunct{\mcitedefaultmidpunct}
{\mcitedefaultendpunct}{\mcitedefaultseppunct}\relax
\EndOfBibitem
\bibitem[Handy \latin{et~al.}(1989)Handy, Pople, Head-Gordon, Raghavachari, and
  Trucks]{Handy1989}
Handy,~N.~C.; Pople,~J.~A.; Head-Gordon,~M.; Raghavachari,~K.; Trucks,~G.~W.
  \emph{Chem. Phys. Lett.} \textbf{1989}, \emph{164}, 185--192\relax
\mciteBstWouldAddEndPuncttrue
\mciteSetBstMidEndSepPunct{\mcitedefaultmidpunct}
{\mcitedefaultendpunct}{\mcitedefaultseppunct}\relax
\EndOfBibitem
\bibitem[Brueckner(1954)]{Brueckner1954}
Brueckner,~K.~A. \emph{Phys. Rev.} \textbf{1954}, \emph{96}, 508--516\relax
\mciteBstWouldAddEndPuncttrue
\mciteSetBstMidEndSepPunct{\mcitedefaultmidpunct}
{\mcitedefaultendpunct}{\mcitedefaultseppunct}\relax
\EndOfBibitem
\bibitem[Nesbet(1958)]{Nesbet1958}
Nesbet,~R.~K. \emph{Phys. Rev.} \textbf{1958}, \emph{109}, 1632--1638\relax
\mciteBstWouldAddEndPuncttrue
\mciteSetBstMidEndSepPunct{\mcitedefaultmidpunct}
{\mcitedefaultendpunct}{\mcitedefaultseppunct}\relax
\EndOfBibitem
\bibitem[Sherrill \latin{et~al.}(1998)Sherrill, Krylov, Byrd, and
  Head-Gordon]{Sherrill1998}
Sherrill,~C.~D.; Krylov,~A.~I.; Byrd,~E. F.~C.; Head-Gordon,~M. \emph{J. Chem.
  Phys.} \textbf{1998}, \emph{109}, 4171--4181\relax
\mciteBstWouldAddEndPuncttrue
\mciteSetBstMidEndSepPunct{\mcitedefaultmidpunct}
{\mcitedefaultendpunct}{\mcitedefaultseppunct}\relax
\EndOfBibitem
\bibitem[Krylov \latin{et~al.}(1998)Krylov, Sherrill, Byrd, and
  Head-Gordon]{Krylov1998}
Krylov,~A.~I.; Sherrill,~C.~D.; Byrd,~E. F.~C.; Head-Gordon,~M. \emph{J. Chem.
  Phys.} \textbf{1998}, \emph{109}, 10669\relax
\mciteBstWouldAddEndPuncttrue
\mciteSetBstMidEndSepPunct{\mcitedefaultmidpunct}
{\mcitedefaultendpunct}{\mcitedefaultseppunct}\relax
\EndOfBibitem
\bibitem[St{\"{u}}ck \latin{et~al.}(2011)St{\"{u}}ck, Baker, Zimmerman,
  Kurlancheek, and Head-Gordon]{Stuck2011}
St{\"{u}}ck,~D.; Baker,~T.~A.; Zimmerman,~P.; Kurlancheek,~W.; Head-Gordon,~M.
  \emph{J. Chem. Phys.} \textbf{2011}, \emph{135}, 194306\relax
\mciteBstWouldAddEndPuncttrue
\mciteSetBstMidEndSepPunct{\mcitedefaultmidpunct}
{\mcitedefaultendpunct}{\mcitedefaultseppunct}\relax
\EndOfBibitem
\bibitem[Kurlancheek \latin{et~al.}(2012)Kurlancheek, Lochan, Lawler, and
  Head-Gordon]{Kurlancheek2012}
Kurlancheek,~W.; Lochan,~R.; Lawler,~K.; Head-Gordon,~M. \emph{J. Chem. Phys.}
  \textbf{2012}, \emph{136}, 054113\relax
\mciteBstWouldAddEndPuncttrue
\mciteSetBstMidEndSepPunct{\mcitedefaultmidpunct}
{\mcitedefaultendpunct}{\mcitedefaultseppunct}\relax
\EndOfBibitem
\bibitem[Neese \latin{et~al.}(2009)Neese, Schwabe, Kossmann, Schirmer, and
  Grimme]{Neese2009}
Neese,~F.; Schwabe,~T.; Kossmann,~S.; Schirmer,~B.; Grimme,~S. \emph{J. Chem.
  Theory Comput.} \textbf{2009}, \emph{5}, 3060--3073\relax
\mciteBstWouldAddEndPuncttrue
\mciteSetBstMidEndSepPunct{\mcitedefaultmidpunct}
{\mcitedefaultendpunct}{\mcitedefaultseppunct}\relax
\EndOfBibitem
\bibitem[Kossmann and Neese(2010)Kossmann, and Neese]{Kossmann2010}
Kossmann,~S.; Neese,~F. \emph{J. Phys. Chem. A} \textbf{2010}, \emph{114},
  11768--11781\relax
\mciteBstWouldAddEndPuncttrue
\mciteSetBstMidEndSepPunct{\mcitedefaultmidpunct}
{\mcitedefaultendpunct}{\mcitedefaultseppunct}\relax
\EndOfBibitem
\bibitem[St{\"{u}}ck and Head-Gordon(2013)St{\"{u}}ck, and
  Head-Gordon]{Stuck2013}
St{\"{u}}ck,~D.; Head-Gordon,~M. \emph{J. Chem. Phys.} \textbf{2013},
  \emph{139}, 244109\relax
\mciteBstWouldAddEndPuncttrue
\mciteSetBstMidEndSepPunct{\mcitedefaultmidpunct}
{\mcitedefaultendpunct}{\mcitedefaultseppunct}\relax
\EndOfBibitem
\bibitem[Razban \latin{et~al.}(2017)Razban, St{\"{u}}ck, and
  Head-Gordon]{Razban2017}
Razban,~R.~M.; St{\"{u}}ck,~D.; Head-Gordon,~M. \emph{Mol. Phys.}
  \textbf{2017}, \emph{115}, 2102--2109\relax
\mciteBstWouldAddEndPuncttrue
\mciteSetBstMidEndSepPunct{\mcitedefaultmidpunct}
{\mcitedefaultendpunct}{\mcitedefaultseppunct}\relax
\EndOfBibitem
\bibitem[Lee and Head-Gordon(2018)Lee, and Head-Gordon]{Lee2018}
Lee,~J.; Head-Gordon,~M. \emph{J. Chem. Theory Comput.} \textbf{2018},
  \emph{14}, 5203--5219\relax
\mciteBstWouldAddEndPuncttrue
\mciteSetBstMidEndSepPunct{\mcitedefaultmidpunct}
{\mcitedefaultendpunct}{\mcitedefaultseppunct}\relax
\EndOfBibitem
\bibitem[Fowler and Manolopoulos(2007)Fowler, and Manolopoulos]{Fowler2007}
Fowler,~P.~W.; Manolopoulos,~D.~E. \emph{An Atlas of Fullerenes (Dover Books on
  Chemistry)}; Dover Publications, 2007\relax
\mciteBstWouldAddEndPuncttrue
\mciteSetBstMidEndSepPunct{\mcitedefaultmidpunct}
{\mcitedefaultendpunct}{\mcitedefaultseppunct}\relax
\EndOfBibitem
\bibitem[Prinzbach \latin{et~al.}(2000)Prinzbach, Weiler, Landenberger, Wahl,
  W{\"{o}}rth, Scott, Gelmont, Olevano, and v.~Issendorff]{Prinzbach2000}
Prinzbach,~H.; Weiler,~A.; Landenberger,~P.; Wahl,~F.; W{\"{o}}rth,~J.;
  Scott,~L.~T.; Gelmont,~M.; Olevano,~D.; v.~Issendorff,~B. \emph{Nature}
  \textbf{2000}, \emph{407}, 60--63\relax
\mciteBstWouldAddEndPuncttrue
\mciteSetBstMidEndSepPunct{\mcitedefaultmidpunct}
{\mcitedefaultendpunct}{\mcitedefaultseppunct}\relax
\EndOfBibitem
\bibitem[Jim{\'{e}}nez-Hoyos \latin{et~al.}(2014)Jim{\'{e}}nez-Hoyos,
  Rodr{\'{i}}guez-Guzm{\'{a}}n, and Scuseria]{Jimenez-Hoyos2014}
Jim{\'{e}}nez-Hoyos,~C.~A.; Rodr{\'{i}}guez-Guzm{\'{a}}n,~R.; Scuseria,~G.~E.
  \emph{J. Phys. Chem. A} \textbf{2014}, \emph{118}, 9925--40\relax
\mciteBstWouldAddEndPuncttrue
\mciteSetBstMidEndSepPunct{\mcitedefaultmidpunct}
{\mcitedefaultendpunct}{\mcitedefaultseppunct}\relax
\EndOfBibitem
\bibitem[Small and Head-Gordon(2012)Small, and Head-Gordon]{Small2012}
Small,~D.~W.; Head-Gordon,~M. \emph{J. Chem. Phys.} \textbf{2012}, \emph{137},
  114103\relax
\mciteBstWouldAddEndPuncttrue
\mciteSetBstMidEndSepPunct{\mcitedefaultmidpunct}
{\mcitedefaultendpunct}{\mcitedefaultseppunct}\relax
\EndOfBibitem
\bibitem[Lee \latin{et~al.}(2017)Lee, Small, Epifanovsky, and
  Head-Gordon]{Lee2017}
Lee,~J.; Small,~D.~W.; Epifanovsky,~E.; Head-Gordon,~M. \emph{J. Chem. Theory
  Comput.} \textbf{2017}, \emph{13}, 602--615\relax
\mciteBstWouldAddEndPuncttrue
\mciteSetBstMidEndSepPunct{\mcitedefaultmidpunct}
{\mcitedefaultendpunct}{\mcitedefaultseppunct}\relax
\EndOfBibitem
\bibitem[Fukutome(1981)]{Fukutome1981}
Fukutome,~H. \emph{Int. J. Quantum Chem.} \textbf{1981}, \emph{20},
  955--1065\relax
\mciteBstWouldAddEndPuncttrue
\mciteSetBstMidEndSepPunct{\mcitedefaultmidpunct}
{\mcitedefaultendpunct}{\mcitedefaultseppunct}\relax
\EndOfBibitem
\bibitem[Stuber and Paldus(2003)Stuber, and Paldus]{Stuber2003}
Stuber,~J.~L.; Paldus,~J. In \emph{Fundamental World of Quantum Chemistry: A
  Tribute to the Memory of Per-Olov Löwdin}; Br{\"a}ndas,~E.~J.,
  Kryachko,~E.~S., Eds.; Springer, 2003; Vol.~1; pp 67--139\relax
\mciteBstWouldAddEndPuncttrue
\mciteSetBstMidEndSepPunct{\mcitedefaultmidpunct}
{\mcitedefaultendpunct}{\mcitedefaultseppunct}\relax
\EndOfBibitem
\bibitem[Yamaki \latin{et~al.}(2000)Yamaki, Shigeta, Yamanaka, Nagao, and
  Yamaguchi]{Yamaki2000}
Yamaki,~D.; Shigeta,~Y.; Yamanaka,~S.; Nagao,~H.; Yamaguchi,~K. \emph{Int. J.
  Quantum Chem.} \textbf{2000}, \emph{80}, 701--707\relax
\mciteBstWouldAddEndPuncttrue
\mciteSetBstMidEndSepPunct{\mcitedefaultmidpunct}
{\mcitedefaultendpunct}{\mcitedefaultseppunct}\relax
\EndOfBibitem
\bibitem[Jim{\'{e}}nez-Hoyos \latin{et~al.}(2011)Jim{\'{e}}nez-Hoyos,
  Henderson, and Scuseria]{Jimenez-Hoyos2011}
Jim{\'{e}}nez-Hoyos,~C.~A.; Henderson,~T.~M.; Scuseria,~G.~E. \emph{J. Chem.
  Theory Comput.} \textbf{2011}, \emph{7}, 2667--2674\relax
\mciteBstWouldAddEndPuncttrue
\mciteSetBstMidEndSepPunct{\mcitedefaultmidpunct}
{\mcitedefaultendpunct}{\mcitedefaultseppunct}\relax
\EndOfBibitem
\bibitem[Small \latin{et~al.}(2015)Small, Sundstrom, and
  Head-Gordon]{Small2015}
Small,~D.~W.; Sundstrom,~E.~J.; Head-Gordon,~M. \emph{J. Chem. Phys.}
  \textbf{2015}, \emph{142}, 094112\relax
\mciteBstWouldAddEndPuncttrue
\mciteSetBstMidEndSepPunct{\mcitedefaultmidpunct}
{\mcitedefaultendpunct}{\mcitedefaultseppunct}\relax
\EndOfBibitem
\bibitem[Cassam-Chena{\"{i}}(2015)]{Cassam-Chenai2015}
Cassam-Chena{\"{i}},~P. \emph{Theor. Chem. Acc.} \textbf{2015}, \emph{134},
  125\relax
\mciteBstWouldAddEndPuncttrue
\mciteSetBstMidEndSepPunct{\mcitedefaultmidpunct}
{\mcitedefaultendpunct}{\mcitedefaultseppunct}\relax
\EndOfBibitem
\bibitem[Feyereisen \latin{et~al.}(1993)Feyereisen, Fitzgerald, and
  Komornicki]{Feyereisen1993}
Feyereisen,~M.; Fitzgerald,~G.; Komornicki,~A. \emph{Chem. Phys. Lett.}
  \textbf{1993}, \emph{208}, 359--363\relax
\mciteBstWouldAddEndPuncttrue
\mciteSetBstMidEndSepPunct{\mcitedefaultmidpunct}
{\mcitedefaultendpunct}{\mcitedefaultseppunct}\relax
\EndOfBibitem
\bibitem[Bernholdt and Harrison(1996)Bernholdt, and Harrison]{Bernholdt1996}
Bernholdt,~D.~E.; Harrison,~R.~J. \emph{Chem. Phys. Lett.} \textbf{1996},
  \emph{250}, 477--484\relax
\mciteBstWouldAddEndPuncttrue
\mciteSetBstMidEndSepPunct{\mcitedefaultmidpunct}
{\mcitedefaultendpunct}{\mcitedefaultseppunct}\relax
\EndOfBibitem
\bibitem[Small \latin{et~al.}(2015)Small, Sundstrom, and
  Head-Gordon]{Small2015a}
Small,~D.~W.; Sundstrom,~E.~J.; Head-Gordon,~M. \emph{J. Chem. Phys.}
  \textbf{2015}, \emph{142}, 024104\relax
\mciteBstWouldAddEndPuncttrue
\mciteSetBstMidEndSepPunct{\mcitedefaultmidpunct}
{\mcitedefaultendpunct}{\mcitedefaultseppunct}\relax
\EndOfBibitem
\bibitem[Small and Head-Gordon(2009)Small, and Head-Gordon]{Small2009}
Small,~D.~W.; Head-Gordon,~M. \emph{J. Chem. Phys.} \textbf{2009}, \emph{130},
  084103\relax
\mciteBstWouldAddEndPuncttrue
\mciteSetBstMidEndSepPunct{\mcitedefaultmidpunct}
{\mcitedefaultendpunct}{\mcitedefaultseppunct}\relax
\EndOfBibitem
\bibitem[Small and Head-Gordon(2011)Small, and Head-Gordon]{Small2011}
Small,~D.~W.; Head-Gordon,~M. \emph{Phys. Chem. Chem. Phys.} \textbf{2011},
  \emph{13}, 19285--97\relax
\mciteBstWouldAddEndPuncttrue
\mciteSetBstMidEndSepPunct{\mcitedefaultmidpunct}
{\mcitedefaultendpunct}{\mcitedefaultseppunct}\relax
\EndOfBibitem
\bibitem[Small \latin{et~al.}(2014)Small, Lawler, and Head-Gordon]{Small2014}
Small,~D.~W.; Lawler,~K.~V.; Head-Gordon,~M. \emph{J. Chem. Theory Comput.}
  \textbf{2014}, \emph{10}, 2027--2040\relax
\mciteBstWouldAddEndPuncttrue
\mciteSetBstMidEndSepPunct{\mcitedefaultmidpunct}
{\mcitedefaultendpunct}{\mcitedefaultseppunct}\relax
\EndOfBibitem
\bibitem[Small and Head-Gordon(2017)Small, and Head-Gordon]{Small2017}
Small,~D.~W.; Head-Gordon,~M. \emph{J. Chem. Phys.} \textbf{2017}, \emph{147},
  024107\relax
\mciteBstWouldAddEndPuncttrue
\mciteSetBstMidEndSepPunct{\mcitedefaultmidpunct}
{\mcitedefaultendpunct}{\mcitedefaultseppunct}\relax
\EndOfBibitem
\bibitem[Small and Head-Gordon(2018)Small, and Head-Gordon]{Small2018}
Small,~D.~W.; Head-Gordon,~M. \emph{J. Chem. Phys.} \textbf{2018}, \emph{149},
  144103\relax
\mciteBstWouldAddEndPuncttrue
\mciteSetBstMidEndSepPunct{\mcitedefaultmidpunct}
{\mcitedefaultendpunct}{\mcitedefaultseppunct}\relax
\EndOfBibitem
\bibitem[Lee \latin{et~al.}()Lee, Small, and Head-Gordon]{Lee2018a}
Lee,~J.; Small,~D.~W.; Head-Gordon,~M. \emph{arXiv:1808.06743.} \relax
\mciteBstWouldAddEndPunctfalse
\mciteSetBstMidEndSepPunct{\mcitedefaultmidpunct}
{}{\mcitedefaultseppunct}\relax
\EndOfBibitem
\bibitem[Abe(2013)]{Abe2013}
Abe,~M. \emph{Chem. Rev.} \textbf{2013}, \emph{113}, 7011--7088\relax
\mciteBstWouldAddEndPuncttrue
\mciteSetBstMidEndSepPunct{\mcitedefaultmidpunct}
{\mcitedefaultendpunct}{\mcitedefaultseppunct}\relax
\EndOfBibitem
\bibitem[Head-Gordon(2003)]{Head-Gordon2003}
Head-Gordon,~M. \emph{Chem. Phys. Lett.} \textbf{2003}, \emph{372},
  508--511\relax
\mciteBstWouldAddEndPuncttrue
\mciteSetBstMidEndSepPunct{\mcitedefaultmidpunct}
{\mcitedefaultendpunct}{\mcitedefaultseppunct}\relax
\EndOfBibitem
\bibitem[Shao \latin{et~al.}(2015)Shao, Gan, Epifanovsky, Gilbert, Wormit,
  Kussmann, Lange, Behn, Deng, Feng, Ghosh, Goldey, Horn, Jacobson, Kaliman,
  Khaliullin, Ku{\' s}, Landau, Liu, Proynov, Rhee, Richard, Rohrdanz, Steele,
  Sundstrom, Woodcock, Zimmerman, Zuev, Albrecht, Alguire, Austin, Beran,
  Bernard, Berquist, Brandhorst, Bravaya, Brown, Casanova, Chang, Chen, Chien,
  Closser, Crittenden, Diedenhofen, Distasio, Do, Dutoi, Edgar, Fatehi,
  Fusti-Molnar, Ghysels, Golubeva-Zadorozhnaya, Gomes, Hanson-Heine, Harbach,
  Hauser, Hohenstein, Holden, Jagau, Ji, Kaduk, Khistyaev, Kim, Kim, King,
  Klunzinger, Kosenkov, Kowalczyk, Krauter, Lao, Laurent, Lawler, Levchenko,
  Lin, Liu, Livshits, Lochan, Luenser, Manohar, Manzer, Mao, Mardirossian,
  Marenich, Maurer, Mayhall, Neuscamman, Oana, Olivares-Amaya, Oneill,
  Parkhill, Perrine, Peverati, Prociuk, Rehn, Rosta, Russ, Sharada, Sharma,
  Small, Sodt, Stein, St{\"{u}}ck, Su, Thom, Tsuchimochi, Vanovschi, Vogt,
  Vydrov, Wang, Watson, Wenzel, White, Williams, Yang, Yeganeh, Yost, You,
  Zhang, Zhang, Zhao, Brooks, Chan, Chipman, Cramer, Goddard, Gordon, Hehre,
  Klamt, Schaefer, Schmidt, Sherrill, Truhlar, Warshel, Xu, Aspuru-Guzik, Baer,
  Bell, Besley, Chai, Dreuw, Dunietz, Furlani, Gwaltney, Hsu, Jung, Kong,
  Lambrecht, Liang, Ochsenfeld, Rassolov, Slipchenko, Subotnik, {Van Voorhis},
  Herbert, Krylov, Gill, and Head-Gordon]{Shao2015}
Shao,~Y.; Gan,~Z.; Epifanovsky,~E.; Gilbert,~A.~T.; Wormit,~M.; Kussmann,~J.;
  Lange,~A.~W.; Behn,~A.; Deng,~J.; Feng,~X.; Ghosh,~D.; Goldey,~M.;
  Horn,~P.~R.; Jacobson,~L.~D.; Kaliman,~I.; Khaliullin,~R.~Z.; Ku{\' s},~T.;
  Landau,~A.; Liu,~J.; Proynov,~E.~I.; Rhee,~Y.~M.; Richard,~R.~M.;
  Rohrdanz,~M.~A.; Steele,~R.~P.; Sundstrom,~E.~J.; Woodcock,~H.~L.;
  Zimmerman,~P.~M.; Zuev,~D.; Albrecht,~B.; Alguire,~E.; Austin,~B.;
  Beran,~G.~J.; Bernard,~Y.~A.; Berquist,~E.; Brandhorst,~K.; Bravaya,~K.~B.;
  Brown,~S.~T.; Casanova,~D.; Chang,~C.~M.; Chen,~Y.; Chien,~S.~H.;
  Closser,~K.~D.; Crittenden,~D.~L.; Diedenhofen,~M.; Distasio,~R.~A.; Do,~H.;
  Dutoi,~A.~D.; Edgar,~R.~G.; Fatehi,~S.; Fusti-Molnar,~L.; Ghysels,~A.;
  Golubeva-Zadorozhnaya,~A.; Gomes,~J.; Hanson-Heine,~M.~W.; Harbach,~P.~H.;
  Hauser,~A.~W.; Hohenstein,~E.~G.; Holden,~Z.~C.; Jagau,~T.~C.; Ji,~H.;
  Kaduk,~B.; Khistyaev,~K.; Kim,~J.; Kim,~J.; King,~R.~A.; Klunzinger,~P.;
  Kosenkov,~D.; Kowalczyk,~T.; Krauter,~C.~M.; Lao,~K.~U.; Laurent,~A.~D.;
  Lawler,~K.~V.; Levchenko,~S.~V.; Lin,~C.~Y.; Liu,~F.; Livshits,~E.;
  Lochan,~R.~C.; Luenser,~A.; Manohar,~P.; Manzer,~S.~F.; Mao,~S.~P.;
  Mardirossian,~N.; Marenich,~A.~V.; Maurer,~S.~A.; Mayhall,~N.~J.;
  Neuscamman,~E.; Oana,~C.~M.; Olivares-Amaya,~R.; Oneill,~D.~P.;
  Parkhill,~J.~A.; Perrine,~T.~M.; Peverati,~R.; Prociuk,~A.; Rehn,~D.~R.;
  Rosta,~E.; Russ,~N.~J.; Sharada,~S.~M.; Sharma,~S.; Small,~D.~W.; Sodt,~A.;
  Stein,~T.; St{\"{u}}ck,~D.; Su,~Y.~C.; Thom,~A.~J.; Tsuchimochi,~T.;
  Vanovschi,~V.; Vogt,~L.; Vydrov,~O.; Wang,~T.; Watson,~M.~A.; Wenzel,~J.;
  White,~A.; Williams,~C.~F.; Yang,~J.; Yeganeh,~S.; Yost,~S.~R.; You,~Z.~Q.;
  Zhang,~I.~Y.; Zhang,~X.; Zhao,~Y.; Brooks,~B.~R.; Chan,~G.~K.;
  Chipman,~D.~M.; Cramer,~C.~J.; Goddard,~W.~A.; Gordon,~M.~S.; Hehre,~W.~J.;
  Klamt,~A.; Schaefer,~H.~F.; Schmidt,~M.~W.; Sherrill,~C.~D.; Truhlar,~D.~G.;
  Warshel,~A.; Xu,~X.; Aspuru-Guzik,~A.; Baer,~R.; Bell,~A.~T.; Besley,~N.~A.;
  Chai,~J.~D.; Dreuw,~A.; Dunietz,~B.~D.; Furlani,~T.~R.; Gwaltney,~S.~R.;
  Hsu,~C.~P.; Jung,~Y.; Kong,~J.; Lambrecht,~D.~S.; Liang,~W.; Ochsenfeld,~C.;
  Rassolov,~V.~A.; Slipchenko,~L.~V.; Subotnik,~J.~E.; {Van Voorhis},~T.;
  Herbert,~J.~M.; Krylov,~A.~I.; Gill,~P.~M.; Head-Gordon,~M. \emph{Mol. Phys.}
  \textbf{2015}, \emph{113}, 184--215\relax
\mciteBstWouldAddEndPuncttrue
\mciteSetBstMidEndSepPunct{\mcitedefaultmidpunct}
{\mcitedefaultendpunct}{\mcitedefaultseppunct}\relax
\EndOfBibitem
\bibitem[Hunter(2007)]{Hunter2007}
Hunter,~J.~D. \emph{Comput. Sci. Eng.} \textbf{2007}, \emph{9}, 90--95\relax
\mciteBstWouldAddEndPuncttrue
\mciteSetBstMidEndSepPunct{\mcitedefaultmidpunct}
{\mcitedefaultendpunct}{\mcitedefaultseppunct}\relax
\EndOfBibitem
\bibitem[Che()]{Chemcraft}
$\texttt{Chemcraft}$. \url{https://www.chemcraftprog.com}, Accessed:
  2017-10-31\relax
\mciteBstWouldAddEndPuncttrue
\mciteSetBstMidEndSepPunct{\mcitedefaultmidpunct}
{\mcitedefaultendpunct}{\mcitedefaultseppunct}\relax
\EndOfBibitem
\bibitem[Paul \latin{et~al.}(2002)Paul, Kim, Sun, Boyd, and Reed]{Paul2002}
Paul,~P.; Kim,~K.-C.; Sun,~D.; Boyd,~P. D.~W.; Reed,~C.~A. \emph{J. Am. Chem.
  Soc.} \textbf{2002}, \emph{124}, 4394--4401\relax
\mciteBstWouldAddEndPuncttrue
\mciteSetBstMidEndSepPunct{\mcitedefaultmidpunct}
{\mcitedefaultendpunct}{\mcitedefaultseppunct}\relax
\EndOfBibitem
\bibitem[Tomita \latin{et~al.}(2003)Tomita, Andersen, Hansen, and
  Hvelplund]{Tomita2003}
Tomita,~S.; Andersen,~J.; Hansen,~K.; Hvelplund,~P. \emph{Chem. Phys. Lett.}
  \textbf{2003}, \emph{382}, 120--125\relax
\mciteBstWouldAddEndPuncttrue
\mciteSetBstMidEndSepPunct{\mcitedefaultmidpunct}
{\mcitedefaultendpunct}{\mcitedefaultseppunct}\relax
\EndOfBibitem
\bibitem[Hehre \latin{et~al.}(1969)Hehre, Stewart, and Pople]{Hehre1969}
Hehre,~W.~J.; Stewart,~R.~F.; Pople,~J.~A. \emph{J. Chem. Phys.} \textbf{1969},
  \emph{51}, 2657--2664\relax
\mciteBstWouldAddEndPuncttrue
\mciteSetBstMidEndSepPunct{\mcitedefaultmidpunct}
{\mcitedefaultendpunct}{\mcitedefaultseppunct}\relax
\EndOfBibitem
\bibitem[Dunning(1989)]{Dunning1989}
Dunning,~T.~H. \emph{J. Chem. Phys.} \textbf{1989}, \emph{90}, 1007--1023\relax
\mciteBstWouldAddEndPuncttrue
\mciteSetBstMidEndSepPunct{\mcitedefaultmidpunct}
{\mcitedefaultendpunct}{\mcitedefaultseppunct}\relax
\EndOfBibitem
\bibitem[Sassara \latin{et~al.}(1996)Sassara, Zerza, and Chergui]{Sassara1996}
Sassara,~A.; Zerza,~G.; Chergui,~M. \emph{Chem. Phys. Lett.} \textbf{1996},
  \emph{261}, 213--220\relax
\mciteBstWouldAddEndPuncttrue
\mciteSetBstMidEndSepPunct{\mcitedefaultmidpunct}
{\mcitedefaultendpunct}{\mcitedefaultseppunct}\relax
\EndOfBibitem
\bibitem[Weigend \latin{et~al.}(2002)Weigend, K{\"{o}}hn, and
  H{\"{a}}ttig]{Weigend2002}
Weigend,~F.; K{\"{o}}hn,~A.; H{\"{a}}ttig,~C. \emph{J. Chem. Phys.}
  \textbf{2002}, \emph{116}, 3175--3183\relax
\mciteBstWouldAddEndPuncttrue
\mciteSetBstMidEndSepPunct{\mcitedefaultmidpunct}
{\mcitedefaultendpunct}{\mcitedefaultseppunct}\relax
\EndOfBibitem
\bibitem[Varganov \latin{et~al.}(2002)Varganov, Avramov, Ovchinnikov, and
  Gordon]{Varganov2002}
Varganov,~S.~A.; Avramov,~P.~V.; Ovchinnikov,~S.~G.; Gordon,~M.~S. \emph{Chem.
  Phys. Lett.} \textbf{2002}, \emph{362}, 380--386\relax
\mciteBstWouldAddEndPuncttrue
\mciteSetBstMidEndSepPunct{\mcitedefaultmidpunct}
{\mcitedefaultendpunct}{\mcitedefaultseppunct}\relax
\EndOfBibitem
\bibitem[Piskoti \latin{et~al.}(1998)Piskoti, Yarger, and Zettl]{Piskoti1998}
Piskoti,~C.; Yarger,~J.; Zettl,~A. \emph{Nature} \textbf{1998}, \emph{393},
  771--774\relax
\mciteBstWouldAddEndPuncttrue
\mciteSetBstMidEndSepPunct{\mcitedefaultmidpunct}
{\mcitedefaultendpunct}{\mcitedefaultseppunct}\relax
\EndOfBibitem
\bibitem[Fowler \latin{et~al.}(1999)Fowler, Heine, Rogers, Sandall, Seifert,
  and Zerbetto]{Fowler1999}
Fowler,~P.; Heine,~T.; Rogers,~K.; Sandall,~J.; Seifert,~G.; Zerbetto,~F.
  \emph{Chem. Phys. Lett.} \textbf{1999}, \emph{300}, 369--378\relax
\mciteBstWouldAddEndPuncttrue
\mciteSetBstMidEndSepPunct{\mcitedefaultmidpunct}
{\mcitedefaultendpunct}{\mcitedefaultseppunct}\relax
\EndOfBibitem
\bibitem[Fowler \latin{et~al.}(1999)Fowler, Mitchell, and
  Zerbetto]{Fowler1999a}
Fowler,~P.~W.; Mitchell,~D.; Zerbetto,~F. \emph{Journal of the American
  Chemical Society} \textbf{1999}, \emph{121}, 3218--3219\relax
\mciteBstWouldAddEndPuncttrue
\mciteSetBstMidEndSepPunct{\mcitedefaultmidpunct}
{\mcitedefaultendpunct}{\mcitedefaultseppunct}\relax
\EndOfBibitem
\bibitem[Jagadeesh and Chandrasekhar(1999)Jagadeesh, and
  Chandrasekhar]{Jagadeesh1999}
Jagadeesh,~M.~N.; Chandrasekhar,~J. \emph{Chem. Phys. Lett.} \textbf{1999},
  \emph{305}, 298--302\relax
\mciteBstWouldAddEndPuncttrue
\mciteSetBstMidEndSepPunct{\mcitedefaultmidpunct}
{\mcitedefaultendpunct}{\mcitedefaultseppunct}\relax
\EndOfBibitem
\bibitem[Ito \latin{et~al.}(2000)Ito, Monobe, Yoshii, and Tanaka]{Ito2000}
Ito,~A.; Monobe,~T.; Yoshii,~T.; Tanaka,~K. \emph{Chem. Phys. Lett.}
  \textbf{2000}, \emph{328}, 32--38\relax
\mciteBstWouldAddEndPuncttrue
\mciteSetBstMidEndSepPunct{\mcitedefaultmidpunct}
{\mcitedefaultendpunct}{\mcitedefaultseppunct}\relax
\EndOfBibitem
\bibitem[Slanina \latin{et~al.}(2000)Slanina, Uhl\'{i}k, Zhao, and
  \={O}sawa]{Slanina2000}
Slanina,~Z.; Uhl\'{i}k,~F.; Zhao,~X.; \={O}sawa,~E. \emph{J. Chem. Phys.}
  \textbf{2000}, \emph{113}, 4933\relax
\mciteBstWouldAddEndPuncttrue
\mciteSetBstMidEndSepPunct{\mcitedefaultmidpunct}
{\mcitedefaultendpunct}{\mcitedefaultseppunct}\relax
\EndOfBibitem
\bibitem[Yuan \latin{et~al.}(2000)Yuan, Yang, Deng, and Zhu]{Lan-FengYuan2000}
Yuan,~L.-F.; Yang,~J.; Deng,~K.; Zhu,~Q.-S. \emph{J. Phys. Chem. A}
  \textbf{2000}, \emph{104}, 6666--6671\relax
\mciteBstWouldAddEndPuncttrue
\mciteSetBstMidEndSepPunct{\mcitedefaultmidpunct}
{\mcitedefaultendpunct}{\mcitedefaultseppunct}\relax
\EndOfBibitem
\bibitem[Paulus(2003)]{Paulus2003}
Paulus,~B. \emph{Phys. Chem. Chem. Phys.} \textbf{2003}, \emph{5}, 3364\relax
\mciteBstWouldAddEndPuncttrue
\mciteSetBstMidEndSepPunct{\mcitedefaultmidpunct}
{\mcitedefaultendpunct}{\mcitedefaultseppunct}\relax
\EndOfBibitem
\bibitem[Becke(1988)]{Becke1988}
Becke,~A.~D. \emph{Phys. Rev. A} \textbf{1988}, \emph{38}, 3098--3100\relax
\mciteBstWouldAddEndPuncttrue
\mciteSetBstMidEndSepPunct{\mcitedefaultmidpunct}
{\mcitedefaultendpunct}{\mcitedefaultseppunct}\relax
\EndOfBibitem
\bibitem[Lee \latin{et~al.}(1988)Lee, Yang, and Parr]{Lee1988}
Lee,~C.; Yang,~W.; Parr,~R.~G. \emph{Phys. Rev. B} \textbf{1988}, \emph{37},
  785--789\relax
\mciteBstWouldAddEndPuncttrue
\mciteSetBstMidEndSepPunct{\mcitedefaultmidpunct}
{\mcitedefaultendpunct}{\mcitedefaultseppunct}\relax
\EndOfBibitem
\bibitem[Hariharan and Pople(1973)Hariharan, and Pople]{Hariharan1973}
Hariharan,~P.~C.; Pople,~J.~A. \emph{Theor. Chim. Acta.} \textbf{1973},
  \emph{28}, 213--222\relax
\mciteBstWouldAddEndPuncttrue
\mciteSetBstMidEndSepPunct{\mcitedefaultmidpunct}
{\mcitedefaultendpunct}{\mcitedefaultseppunct}\relax
\EndOfBibitem
\bibitem[Hehre \latin{et~al.}(1972)Hehre, Ditchfield, and Pople]{Hehre1972}
Hehre,~W.~J.; Ditchfield,~R.; Pople,~J.~A. \emph{J. Chem. Phys.} \textbf{1972},
  \emph{56}, 2257--2261\relax
\mciteBstWouldAddEndPuncttrue
\mciteSetBstMidEndSepPunct{\mcitedefaultmidpunct}
{\mcitedefaultendpunct}{\mcitedefaultseppunct}\relax
\EndOfBibitem
\bibitem[Grimme(2003)]{Grimme2003}
Grimme,~S. \emph{J. Chem. Phys.} \textbf{2003}, \emph{118}, 9095--9102\relax
\mciteBstWouldAddEndPuncttrue
\mciteSetBstMidEndSepPunct{\mcitedefaultmidpunct}
{\mcitedefaultendpunct}{\mcitedefaultseppunct}\relax
\EndOfBibitem
\bibitem[Jung \latin{et~al.}(2004)Jung, Lochan, Dutoi, and
  Head-Gordon]{Jung2004}
Jung,~Y.; Lochan,~R.~C.; Dutoi,~A.~D.; Head-Gordon,~M. \emph{J. Chem. Phys.}
  \textbf{2004}, \emph{121}, 9793--9802\relax
\mciteBstWouldAddEndPuncttrue
\mciteSetBstMidEndSepPunct{\mcitedefaultmidpunct}
{\mcitedefaultendpunct}{\mcitedefaultseppunct}\relax
\EndOfBibitem
\bibitem[Yamaguchi \latin{et~al.}(1988)Yamaguchi, Jensen, Dorigo, and
  Houk]{Yamaguchi1988}
Yamaguchi,~K.; Jensen,~F.; Dorigo,~A.; Houk,~K.~N. \emph{Chem. Phys. Lett.}
  \textbf{1988}, \emph{149}, 537--542\relax
\mciteBstWouldAddEndPuncttrue
\mciteSetBstMidEndSepPunct{\mcitedefaultmidpunct}
{\mcitedefaultendpunct}{\mcitedefaultseppunct}\relax
\EndOfBibitem
\bibitem[Schwerdtfeger \latin{et~al.}(2015)Schwerdtfeger, Wirz, and
  Avery]{Schwerdtfeger2015}
Schwerdtfeger,~P.; Wirz,~L.~N.; Avery,~J. \emph{WIRES: Comput. Mol. Sci.}
  \textbf{2015}, \emph{5}, 96--145\relax
\mciteBstWouldAddEndPuncttrue
\mciteSetBstMidEndSepPunct{\mcitedefaultmidpunct}
{\mcitedefaultendpunct}{\mcitedefaultseppunct}\relax
\EndOfBibitem
\bibitem[Parasuk and Alml{\"{o}}f(1991)Parasuk, and Alml{\"{o}}f]{Parasuk1991}
Parasuk,~V.; Alml{\"{o}}f,~J. \emph{Chem. Phys. Lett.} \textbf{1991},
  \emph{184}, 187--190\relax
\mciteBstWouldAddEndPuncttrue
\mciteSetBstMidEndSepPunct{\mcitedefaultmidpunct}
{\mcitedefaultendpunct}{\mcitedefaultseppunct}\relax
\EndOfBibitem
\bibitem[Feyereisen \latin{et~al.}(1992)Feyereisen, Gutowski, Simons, and
  Alml{\"{o}}f]{Feyereisen1992}
Feyereisen,~M.; Gutowski,~M.; Simons,~J.; Alml{\"{o}}f,~J. \emph{J. Chem.
  Phys.} \textbf{1992}, \emph{96}, 2926--2932\relax
\mciteBstWouldAddEndPuncttrue
\mciteSetBstMidEndSepPunct{\mcitedefaultmidpunct}
{\mcitedefaultendpunct}{\mcitedefaultseppunct}\relax
\EndOfBibitem
\bibitem[Raghavachari \latin{et~al.}(1993)Raghavachari, Strout, Odom, Scuseria,
  Pople, Johnson, and Gill]{Raghavachari1993}
Raghavachari,~K.; Strout,~D.; Odom,~G.; Scuseria,~G.; Pople,~J.; Johnson,~B.;
  Gill,~P. \emph{Chem. Phys. Lett.} \textbf{1993}, \emph{214}, 357--361\relax
\mciteBstWouldAddEndPuncttrue
\mciteSetBstMidEndSepPunct{\mcitedefaultmidpunct}
{\mcitedefaultendpunct}{\mcitedefaultseppunct}\relax
\EndOfBibitem
\bibitem[von Helden \latin{et~al.}(1993)von Helden, Hsu, Gotts, Kemper, and
  Bowers]{VonHelden1993}
von Helden,~G.; Hsu,~M.; Gotts,~N.; Kemper,~P.; Bowers,~M. \emph{Chem. Phys.
  Lett.} \textbf{1993}, \emph{204}, 15--22\relax
\mciteBstWouldAddEndPuncttrue
\mciteSetBstMidEndSepPunct{\mcitedefaultmidpunct}
{\mcitedefaultendpunct}{\mcitedefaultseppunct}\relax
\EndOfBibitem
\bibitem[Bylaska \latin{et~al.}(1996)Bylaska, Taylor, Kawai, and
  Weare]{Bylaska1996}
Bylaska,~E.~J.; Taylor,~P.~R.; Kawai,~R.; Weare,~J.~H. \emph{J. Phys. Chem.}
  \textbf{1996}, \emph{100}, 6966--6972\relax
\mciteBstWouldAddEndPuncttrue
\mciteSetBstMidEndSepPunct{\mcitedefaultmidpunct}
{\mcitedefaultendpunct}{\mcitedefaultseppunct}\relax
\EndOfBibitem
\bibitem[Martin \latin{et~al.}(1996)Martin, El-Yazal, and
  Fran{\c{c}}ois]{Martin1996b}
Martin,~J.~M.; El-Yazal,~J.; Fran{\c{c}}ois,~J.-P. \emph{Chem. Phys. Lett.}
  \textbf{1996}, \emph{248}, 345--352\relax
\mciteBstWouldAddEndPuncttrue
\mciteSetBstMidEndSepPunct{\mcitedefaultmidpunct}
{\mcitedefaultendpunct}{\mcitedefaultseppunct}\relax
\EndOfBibitem
\bibitem[Jones and Seifert(1997)Jones, and Seifert]{Jones1997}
Jones,~R.~O.; Seifert,~G. \emph{Phys. Rev. Lett.} \textbf{1997}, \emph{79},
  443--446\relax
\mciteBstWouldAddEndPuncttrue
\mciteSetBstMidEndSepPunct{\mcitedefaultmidpunct}
{\mcitedefaultendpunct}{\mcitedefaultseppunct}\relax
\EndOfBibitem
\bibitem[Kietzmann \latin{et~al.}(1998)Kietzmann, Rochow, Gantef\"{o}r,
  Eberhardt, Vietze, Seifert, and Fowler]{Kietzmann1998}
Kietzmann,~H.; Rochow,~R.; Gantef\"{o}r,~G.; Eberhardt,~W.; Vietze,~K.;
  Seifert,~G.; Fowler,~P.~W. \emph{Phys. Rev. Lett.} \textbf{1998}, \emph{81},
  5378--5381\relax
\mciteBstWouldAddEndPuncttrue
\mciteSetBstMidEndSepPunct{\mcitedefaultmidpunct}
{\mcitedefaultendpunct}{\mcitedefaultseppunct}\relax
\EndOfBibitem
\bibitem[Sokolova \latin{et~al.}(2000)Sokolova, L{\"{u}}chow, and
  Anderson]{Sokolova2000}
Sokolova,~S.; L{\"{u}}chow,~A.; Anderson,~J.~B. \emph{Chem. Phys. Lett.}
  \textbf{2000}, \emph{323}, 229--233\relax
\mciteBstWouldAddEndPuncttrue
\mciteSetBstMidEndSepPunct{\mcitedefaultmidpunct}
{\mcitedefaultendpunct}{\mcitedefaultseppunct}\relax
\EndOfBibitem
\bibitem[Grimme and M{\"{u}}ck-Lichtenfeld(2002)Grimme, and
  M{\"{u}}ck-Lichtenfeld]{Grimme2002}
Grimme,~S.; M{\"{u}}ck-Lichtenfeld,~C. \emph{ChemPhysChem} \textbf{2002},
  \emph{3}, 207--209\relax
\mciteBstWouldAddEndPuncttrue
\mciteSetBstMidEndSepPunct{\mcitedefaultmidpunct}
{\mcitedefaultendpunct}{\mcitedefaultseppunct}\relax
\EndOfBibitem
\bibitem[Lu \latin{et~al.}(2003)Lu, Re, Choe, Nagase, Zhou, Han, Peng, Zhang,
  and Zhao]{Lu2003}
Lu,~J.; Re,~S.; Choe,~Y.-k.; Nagase,~S.; Zhou,~Y.; Han,~R.; Peng,~L.;
  Zhang,~X.; Zhao,~X. \emph{Phys. Rev. B} \textbf{2003}, \emph{67},
  125415\relax
\mciteBstWouldAddEndPuncttrue
\mciteSetBstMidEndSepPunct{\mcitedefaultmidpunct}
{\mcitedefaultendpunct}{\mcitedefaultseppunct}\relax
\EndOfBibitem
\bibitem[An \latin{et~al.}(2005)An, Gao, Bulusu, and Zeng]{An2005}
An,~W.; Gao,~Y.; Bulusu,~S.; Zeng,~X.~C. \emph{J. Chem. Phys.} \textbf{2005},
  \emph{122}, 204109\relax
\mciteBstWouldAddEndPuncttrue
\mciteSetBstMidEndSepPunct{\mcitedefaultmidpunct}
{\mcitedefaultendpunct}{\mcitedefaultseppunct}\relax
\EndOfBibitem
\bibitem[Jin \latin{et~al.}(2015)Jin, Perera, Lotrich, and Bartlett]{Jin2015}
Jin,~Y.; Perera,~A.; Lotrich,~V.~F.; Bartlett,~R.~J. \emph{Chem. Phys. Lett.}
  \textbf{2015}, \emph{629}, 76--80\relax
\mciteBstWouldAddEndPuncttrue
\mciteSetBstMidEndSepPunct{\mcitedefaultmidpunct}
{\mcitedefaultendpunct}{\mcitedefaultseppunct}\relax
\EndOfBibitem
\bibitem[Manna and Martin(2016)Manna, and Martin]{Manna2016}
Manna,~D.; Martin,~J. M.~L. \emph{J. Phys. Chem. A} \textbf{2016}, \emph{120},
  153--160\relax
\mciteBstWouldAddEndPuncttrue
\mciteSetBstMidEndSepPunct{\mcitedefaultmidpunct}
{\mcitedefaultendpunct}{\mcitedefaultseppunct}\relax
\EndOfBibitem
\bibitem[Adams \latin{et~al.}(1993)Adams, Sankey, Page, and
  O'Keeffe]{Adams1993}
Adams,~G.~B.; Sankey,~O.~F.; Page,~J.~B.; O'Keeffe,~M. \emph{Chem. Phys.}
  \textbf{1993}, \emph{176}, 61--66\relax
\mciteBstWouldAddEndPuncttrue
\mciteSetBstMidEndSepPunct{\mcitedefaultmidpunct}
{\mcitedefaultendpunct}{\mcitedefaultseppunct}\relax
\EndOfBibitem
\bibitem[Krylov(2001)]{Krylov2001}
Krylov,~A.~I. \emph{Chem. Phys. Lett.} \textbf{2001}, \emph{338},
  375--384\relax
\mciteBstWouldAddEndPuncttrue
\mciteSetBstMidEndSepPunct{\mcitedefaultmidpunct}
{\mcitedefaultendpunct}{\mcitedefaultseppunct}\relax
\EndOfBibitem
\bibitem[Fan \latin{et~al.}(1995)Fan, Lin, and Yang]{Fan1995}
Fan,~M.-F.; Lin,~Z.; Yang,~S. \emph{J. Mol. Struc-THEOCHEM} \textbf{1995},
  \emph{337}, 231--240\relax
\mciteBstWouldAddEndPuncttrue
\mciteSetBstMidEndSepPunct{\mcitedefaultmidpunct}
{\mcitedefaultendpunct}{\mcitedefaultseppunct}\relax
\EndOfBibitem
\bibitem[Prinzbach \latin{et~al.}(2006)Prinzbach, Wahl, Weiler, Landenberger,
  W{\"{o}}rth, Scott, Gelmont, Olevano, Sommer, and von
  Issendorff]{Prinzbach2006}
Prinzbach,~H.; Wahl,~F.; Weiler,~A.; Landenberger,~P.; W{\"{o}}rth,~J.;
  Scott,~L.~T.; Gelmont,~M.; Olevano,~D.; Sommer,~F.; von Issendorff,~B.
  \emph{Chem.-Eur. J.} \textbf{2006}, \emph{12}, 6268--6280\relax
\mciteBstWouldAddEndPuncttrue
\mciteSetBstMidEndSepPunct{\mcitedefaultmidpunct}
{\mcitedefaultendpunct}{\mcitedefaultseppunct}\relax
\EndOfBibitem
\end{mcitethebibliography}
\bibliographystyle{achemso}
\end{document}